\documentclass[journal]{IEEEtran}

\usepackage{verbatim}
\usepackage{amsfonts}
\usepackage{amssymb}
\usepackage{stfloats}
\usepackage{cite}
\usepackage{graphicx}
\usepackage{psfrag}
\usepackage{subfigure}
\usepackage{amsmath}
\usepackage{array}
\usepackage{epstopdf}
\usepackage{authblk}
\usepackage{graphicx} 
\usepackage{amsthm} 
\usepackage{lipsum}
\usepackage{verbatim} 
\usepackage{authblk}
\usepackage{mathtools}
\usepackage{cuted}
\usepackage{booktabs}
\usepackage{cases}
\usepackage{multirow}
\usepackage{bbding}
\usepackage{pifont}
\usepackage{makecell}
\usepackage{threeparttable}
\usepackage[bookmarks=true, colorlinks=true, breaklinks=true]{hyperref}

\newtheorem{theorem}{Theorem}

\newtheorem{lemma}{Lemma}
\newtheorem{corollary}{Corollary}

\newtheorem{assumption}{Assumption}

\newtheorem{remark}{\bf Remark}

\def\({\left(}
\def\){\right)}

\def\b0{{\mathbf{0}}}

\newcommand{\tr}{\mathrm{tr}}

\usepackage{algorithmic}
\usepackage{algorithm}

\usepackage{setspace}

\usepackage{xcolor}

\renewcommand{\IEEEQED}{\IEEEQEDopen}

\makeatletter

\newcommand{\Rmnum}[1]{\expandafter\@slowromancap\romannumeral #1@}
\makeatother

\begin{document}
	\bstctlcite{ref:BSTcontrol}
	
	\title{{Improving Convergence for Semi-Federated Learning: An Energy-Efficient Approach by Manipulating Over-the-Air Distortion}}
	
	\author{
		Jingheng~Zheng,~Hui~Tian,~\IEEEmembership{Senior~Member,~IEEE,}~Wanli~Ni,~\IEEEmembership{Member,~IEEE,}
		Yang~Tian,~and~Ping~Zhang,~\IEEEmembership{Fellow,~IEEE}
		\thanks{Jingheng Zheng, Hui Tian and Ping Zhang are with the State Key Laboratory of Networking and Switching Technology, Beijing University of Posts and Telecommunications, Beijing 100876, China (e-mail: zhengjh@bupt.edu.cn; tianhui@bupt.edu.cn; pzhang@bupt.edu.cn).}
		\thanks{Wanli Ni is with the Department of Electronic Engineering, Tsinghua University, Beijing 100084, China (e-mail: niwanli@tsinghua.edu.cn).}
		\thanks{Yang Tian is with the School of Information and Communication Engineering, Beijing Information Science and Technology University, Beijing 102206, China (e-mail: tianyang9108@163.com).}
		\vspace{-1.0 cm}
	}
	
	\maketitle

	\begin{abstract}
	In this paper, we propose a hybrid learning framework that combines federated and split learning, termed semi-federated learning (SemiFL), in which over-the-air computation is utilized for gradient aggregation.
	A key idea is to strategically adjust the learning rate by manipulating over-the-air distortion for improving SemiFL's convergence.
	Specifically, we intentionally amplify amplitude distortion to increase the learning rate in the non-stable region, thereby accelerating convergence and reducing communication energy consumption.
	In the stable region, we suppress noise perturbation to maintain a small learning rate for improving SemiFL's final convergence.
	Theoretical results demonstrate the antagonistic effects of over-the-air distortion in different regions, under both independent and identically distributed (IID) and non-IID data settings.
	Then, we formulate two energy consumption minimization problems, one for each region, which implements a two-region mean square error threshold configuration scheme.
	Accordingly, we propose two resource allocation algorithms with closed-form solutions. 
	Simulation results show that under different network and data distribution conditions, strategically manipulating over-the-air distortion can efficiently adjust the learning rate to improve SemiFL's convergence.
	Moreover, energy consumption can be reduced by using the proposed algorithms.
\end{abstract}


\vspace{-0.6 cm}
\section{Introduction}
\label{introduction}
Past few years have witnessed the thriving emergence of artificial intelligence (AI)-enabled applications, such as unmanned vehicles, smart healthcare, and AI-empowered Internet of Things~\cite{Du2020Federated,Nguyen2021Federated}.
When deploying these advanced applications in the sixth-generation (6G) wireless network, a critical challenge is how to efficiently train high-quality AI models~\cite{Le2024Applications}.
The conventional centralized learning (CL) framework necessitates collecting all available data from devices to train a global model~\cite{Liu2024Federated}.
Nonetheless, this centralized approach raises privacy leakage issues owing to the transmission of massive raw data.
Due to its ability to preserve privacy, federated learning (FL) has garnered substantial research interests~\cite{Duan2023Combining}.
As a distributed framework, FL deployed in wireless networks allows devices collaboratively train a shared global model by aggregating local gradients at the base station (BS), thereby preserving data privacy.
Recently, the data collection and storage capabilities of local devices have been significantly enhanced~\cite{Deng2020Edge}.
Substantial available data impose overwhelming local training burdens, whereas the BS's strong computational capabilities remain underutilized in conventional FL schemes.
Hence, it is necessary to design a new model training framework that utilizes resources rationally to balance the workload between the BS and devices.


\vspace{-0.2 cm}
\subsection{Motivations}
\label{motivation}
By synthesizing FL and CL into an implement-efficient framework, the existing semi-federated learning (SemiFL) enables the BS to capitalize on its powerful computational capability to execute CL concurrently with FL across devices~\cite{Ni2023SemiFL,Elbir2022Ahybrid,Zheng2023Semi,Feng2023Hybrid}.
Although SemiFL alleviates workload on local devices, partially uploading local datasets still poses privacy leakage risks.
In the literature, techniques such as mixup~\cite{Oh2020Mix2FLD} and fully homomorphic encryption~\cite{Zhang2016Privacy} can mitigate privacy leakage. 
However, the mixup technique simply superposes local data using a set of normalized weight coefficients, offering privacy protection to only a limited extent.
As for fully homomorphic encryption, it requires considerable local computation due to the complicated encryption and decryption operations, limiting its implementations on resource-constrained devices.
In view of the drawbacks of existing methods, another key question is that: \textit{How can we ameliorate the SemiFL framework to ensure efficient and privacy-preserving data uploading}?

Over-the-air computation (AirComp) has been widely adopted for aggregating local gradients within SemiFL frameworks~\cite{Sahin2023Asurvey,Zhu2024Over}.
As an analog transmission scheme, AirComp exploits the superposition property of wireless channels to directly aggregate devices' gradients during concurrent transmissions over the same time-frequency resources~\cite{Zhu2019MIMO}.
However, noise and fading in wireless channels cause over-the-air distortion, deviating the aggregated signal from its ideal form.
Regarding this issue, prevailing approaches primarily adhere to the distortion-suppressing criterion that aims to minimize over-the-air distortion, typically requiring substantial transmit power to achieve~\cite{Li2024Energy,Wang2024Digital,Yao2024Wireless}.
Nevertheless, recent techniques such as gradient sparsification~\cite{Oh2024Communication} and gradient compression~\cite{Xu2024Federated} imply the robustness of gradients, suggesting that precise aggregation of local gradients may be overly conservative.
Meanwhile, the work in~\cite{Neelakantan2015Adding} reports that adding artificial noise to gradients can improve the learning performance of AI models.
Motivated by these observations, an intriguing yet critical question rises: \textit{Is it possible to accelerate the convergence of SemiFL by intentionally leveraging over-the-air distortion during gradient aggregation}?

Another critical challenge for SemiFL is the joint optimization involving multiple categories of resources~\cite{Huang2023Wireless}.
Through adjusting gradient and data transmissions, along with training workloads, resource allocation significantly affects the learning performance and energy consumption of SemiFL~\cite{Han2024Semi}.
Unlike conventional FL, where resource allocation involves only devices' local resources, SemiFL expands the range of optimizable resources.
They typically include the amount of uploaded data, the transmit power for gradient and data uploading, and the computational capability of the BS~\cite{Zheng2023Convergence}.
Regretfully, a comprehensive allocation scheme involving these unique resources for SemiFL remains unexplored in the existing literature.
Hence, it becomes indeterminate that: \textit{How to design efficient resource allocation scheme to facilitate the convergence acceleration of SemiFL}?

\subsection{Related Work}
\label{related_work}
{
	There has been growing interest in hybrid learning, as seen in recent works like~\cite{Chen2024Role,Chen2025Completion,Liu2024FedCD}.
	The authors of~\cite{Chen2024Role} proposed a hybrid training framework in which the BS first pre-trained a model centrally using public data, and then fine-tuned it with local data via an FL framework.
	The authors of~\cite{Chen2025Completion} proposed an unmanned aerial vehicle (UAV) assisted semi-decentralized hybrid FL scheme, which enables asynchronous model training across clusters.
	Devices within the same cluster perform local model training and aggregation via device-to-device communication, while a randomly selected cluster uploads its intra-cluster aggregated model to update the global model at the UAV.
	The authors of~\cite{Liu2024FedCD} split a device's local model into two parts: one was shared with neighboring devices for local aggregation, and the other was uploaded to the BS for global aggregation.
	The two parts of layers were sent back to the device for combination.
	However, none of these works have considered leveraging over-the-air distortion to accelerate learning convergence, let alone exploiting the intermediate outputs of a model’s shallow layers to protect data privacy.}

AirComp has attracted attention in the field of both FL and SemiFL, such as the work in~\cite{Wang2024Digital,Yao2024Wireless,Zheng2023Semi,Ren2024Convergence}.
Specifically, the authors of~\cite{Wang2024Digital} mitigated wireless fading for FL model aggregation by integrating digital modulation with AirComp.
In~\cite{Yao2024Wireless}, the authors provided quantitative comparisons between a digital transmission scheme and AirComp, revealing the superior spectrum utilization of AirComp but its greater vulnerability to aggregation errors. 
As for SemiFL, the authors of~\cite{Zheng2023Semi} suppressed the over-the-air distortion of aggregated gradients by keeping the corresponding mean squared error (MSE) below a specific threshold.
In~\cite{Ren2024Convergence}, the authors aimed to minimize the latency of AirComp-based SemiFL, whereas over-the-air distortion was still suppressed by imposing a constraint on MSE.
Although the above work uniformly dedicated to restricting over-the-air distortion, the authors of~\cite{Yang2022Revisiting,Chen2023Edge,Zhang2022Turning} have suggested that the noise of wireless channel could potentially ameliorate the convergence of FL.
However, \cite{Yang2022Revisiting,Chen2023Edge,Zhang2022Turning} focus solely on utilizing the noise perturbation, whereas the amplitude distortion has not been investigated or leveraged.
In view of this, the utilization of over-the-air distortion can be further extended.
Thus, \cite{Yang2022Revisiting,Chen2023Edge,Zhang2022Turning} adopt a relatively conservative approach to leveraging over-the-air distortion.

Other studies have been devoted to the joint allocation of multi-type resources in the SemiFL framework, including the work in~\cite{Feng2023Hybrid,Huang2023Wireless,Han2024Semi}.
Specifically, by jointly optimizing the communication and computation resources of a hybrid learning framework, the authors of~\cite{Feng2023Hybrid} balanced model accuracy loss, latency, and energy consumption.
Subject to restricted energy budget, the authors of~\cite{Huang2023Wireless} minimized the training latency of a hybrid learning framework by jointly optimizing device selection, the number of local and global iterations, and bandwidth allocation.
The authors of~\cite{Han2024Semi} minimized the latency of a SemiFL framework comprising computing-heterogeneous devices by optimizing transmit power and the receive factor.
Though the above work addresses the resource allocation issue to a limited degree, its scope has yet to encompass the idea of accelerating convergence by leveraging over-the-air distortion.
In addition, the category of optimizable resources in the SemiFL framework can be further expanded to include more aspects such as the data allocation between devices and the BS.

\begin{figure*}[t]
	\centering
	\includegraphics[width=0.9 \textwidth]{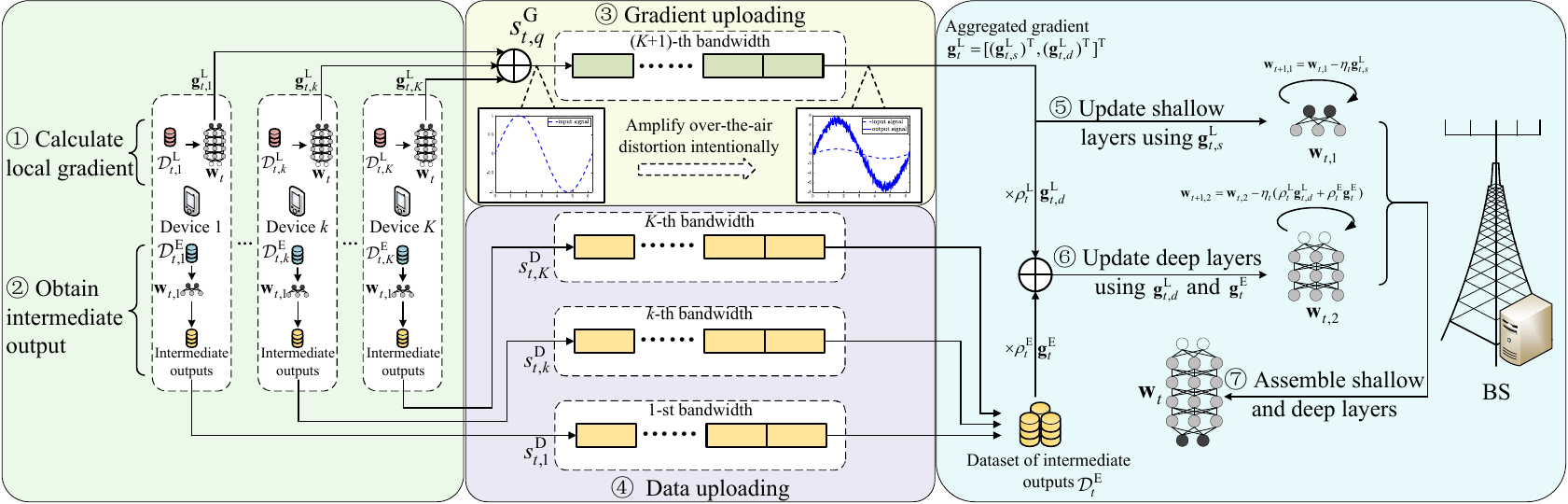}
	\vspace{-0.2 cm}
	{\caption{Illustration of over-the-air distortion accelerated SemiFL. Devices upload local gradients and intermediate outputs, leveraging intentionally amplified over-the-air distortion to accelerate convergence. The BS update shallow layers ${\bf{w}}_{t,1}$ using the aggregated gradient ${\bf{g}}^{\rm L}_{t}$, while update deep layers ${\bf{w}}_{t,2}$ by combining ${\bf{g}}^{\rm L}_{t}$ with the edge gradient ${\bf{g}}^{\rm E}_{t}$. The entire updated model is obtained by assembling the shallow layers ${\bf{w}}_{t,1}$ and the deep layers ${\bf{w}}_{t,2}$.}
		\label{system_model_picture}}
	\vspace{-0.4 cm}
\end{figure*}

\subsection{Contributions and Organization}
\label{contribution_and_organization}
To address the aforementioned key problems, we develop a new SemiFL framework in this paper. 
Unlike existing SemiFL works~\cite{Ni2023SemiFL,Elbir2022Ahybrid,Zheng2023Semi,Feng2023Hybrid}, our framework replaces CL with split learning (SL)~\cite{Liu2023Wireless}, where devices process raw data locally with shallow layers and upload intermediate outputs to the BS for deep-layer training, thereby enhancing privacy preservation.
Additionally, our SemiFL framework intentionally amplifies over-the-air distortion during local gradient aggregation, accelerating convergence in the non-stable region.
The main contributions of this paper are summarized as follows:
\begin{itemize}
	\item We propose a new SemiFL framework that integrates FL and SL.
	Through incorporating SL, SemiFL enhances privacy protection by uploading only the intermediate outputs of local model's shallow layers to the BS.
	\item We propose a new approach to adjust learning rate of the FL part in SemiFL by manipulating over-the-air distortion during gradient aggregation.
	By amplifying amplitude distortion, we achieve a larger learning rate in an energy-efficient manner, accelerating SemiFL's in the non-stable region.
	In the stable region, we propose to cancel amplitude distortion while suppressing noise perturbation, ensuring stable final convergence.
	\item We derive closed-form theoretical results to confirm the acceleration effect of over-the-distortion while revealing the adverse effect of amplified over-the-air distortion to final convergence, under independent and identically distributed (IID) data conditions.
	Under non-IID data conditions, both convergence acceleration and final convergence are degraded by data heterogeneity.
	\item We formulate two separate problems for the non-stable and stable regions, aiming to minimize energy consumption while adhering to latency, amplitude, MSE, and resource constraints.
	We decompose each of these problems and propose two algorithms to solve them, enabling energy-efficient resource allocation that supports our joint design of wireless communication and learning approach for SemiFL.
\end{itemize}
We evaluate the proposed framework and algorithms by using SemiFL to train a multilayer perceptron (MLP), a convolutional neural network (CNN), and a residual network (ResNet) on the Fashion-MNIST, CIFAR-10, and CIFAR-100 datasets, respectively.
Simulation results validate that over-the-air distortion effectively accelerates the convergence of SemiFL, and the energy consumption can be significantly reduced by using the proposed algorithms.

The remainder of this paper is organized as follows.
In Section~\ref{system_model}, we describe the system model of SemiFL.
In Section~\ref{convergence_analysis_and_problem_formulation}, we present the theoretical analysis results and formulate two problems to reduce energy consumption.
In Section~\ref{proposed_algorithms}, we present proposed algorithms, followed by simulation results in~\ref{simulation_results}.
The paper is concluded in Section~\ref{conclusion}.

\section{System Model}
\label{system_model}
As shown in Fig.~\ref{system_model_picture}, we consider a SemiFL framework consisting of an $N_r$-antenna BS and $K$ single-antenna devices.
The devices are indexed by the set $\mathcal{K}=\{1,2,\ldots,K\}$.
The $k$-th device possess a local dataset, denoted by $\mathcal{D}_k$, containing $|\mathcal{D}_k|=D, \forall k \in \mathcal{K}$ data samples for training a shared global model ${\bf{w}} \in \mathbb{R}^{Q}$.
Here, $|\cdot|$ denotes set cardinality, $\mathbb{R}$ denotes the set of real numbers, and $Q$ denotes the dimensions of ${\bf{w}}$.

\vspace{-0.4 cm}
\subsection{SemiFL Framework}
\label{SemiFL_framework}
Consider a $T$-round SemiFL where the round indexes are collected by the set $\mathcal{T}=\{1,2,\ldots,T\}$.
The goal is to minimize the global loss function $F({\bf{w}})$, defined by
\begin{align}
	\label{overall_goal}
	F({\bf{w}})=\sum\nolimits_{k=1}^{K} \sum\nolimits_{n=1}^{D} f({\bf{w}}; {\bf{\Omega}}_{n,k}),
\end{align}
where $f({\bf{w}}; {\bf{\Omega}}_{n,k})$ denotes the loss function regarding a data sample ${\bf{\Omega}}_{n,k}$.
To this end, in the $t$-th round, the $k$-th device divides the local dataset $\mathcal{D}_k$ into two disjoint subsets, i.e., the edge dataset $\mathcal{D}^{\rm E}_{t,k}$ for SL at the BS containing $\theta_{t,k}D$ data samples and the local dataset $\mathcal{D}^{\rm L}_{t,k}$ consisting of $(1-\theta_{t,k})D$ data samples for local training, where $\theta_{t,k} \in (0,1)$ denotes the ratio of SL data.

From the perspective of FL, the local dataset $\mathcal{D}^{\rm L}_{t,k}$ are retained locally for calculating the local gradient ${\bf{g}}^{\rm L}_{t,k}=[g^{\rm L}_{t,k,1},\ldots,g^{\rm L}_{t,k,Q}]^{\rm T} \in \mathbb{R}^{Q}$, defined by
\begin{align}
	\label{FL_gradient}
	{\bf{g}}^{\rm L}_{t,k}=\sum\nolimits_{n\in \mathcal{D}^{\rm L}_{t,k}} {\nabla f({\bf{w}}_t;{\bf{\Omega}}_{n,k})}, \forall k \in \mathcal{K},
\end{align}
where ${\bf{w}}_t$ denotes the global model in the $t$-th round and $\nabla$ denotes the gradient operator.
Then, all devices upload their local gradients $\{{\bf{g}}^{\rm L}_{t,k}\}$ to the BS for aggregation.
To facilitate SL, we rewrite the global model ${\bf{w}}_t$ as a combination of shallow layers ${\bf{w}}_{t,1} \in \mathbb{R}^{Q_1}$ and deep layers ${\bf{w}}_{t,2} \in \mathbb{R}^{Q_2}$, i.e., ${\bf{w}}_t=[{\bf{w}}_{t,1}^{\rm T},{\bf{w}}_{t,2}^{\rm T}]^{\rm T} \in \mathbb{R}^{Q}$ with $Q_1+Q_2=Q$.
As a result, the aggregated gradient can be denoted by ${\bf{g}}^{\rm L}_{t}=[({\bf{g}}_{t,s}^{\rm L})^{\rm T},({\bf{g}}_{t,d}^{\rm L})^{\rm T}]^{\rm T}=[g^{\rm L}_{1},\ldots,g^{\rm L}_{Q_1},g^{\rm L}_{Q_1+1},\ldots,g^{\rm L}_{Q}]^{\rm T} \in \mathbb{R}^{Q}$, where ${\bf{g}}^{\rm L}_{t,s} \in \mathbb{R}^{Q_1}$ and ${\bf{g}}^{\rm L}_{t,d} \in \mathbb{R}^{Q_2}$ denote the aggregated gradients of shallow layers and deep layers, respectively.

To enable SL, the $k$-th device uploads the edge dataset $\mathcal{D}^{\rm E}_{t,k}$ to the BS.
To preserve data privacy, the $k$-th device first inputs the raw data in $\mathcal{D}^{\rm E}_{t,k}$ into the shallow layers ${\bf{w}}_{t,1}$, and replace them with the resultant intermediate outputs.
Then, the $k$-th device uploads these intermediate outputs in $\mathcal{D}^{\rm E}_{t,k}$ to the BS.
The BS collects the intermediate outputs in a dataset $\mathcal{D}^{\rm E}_{t}$ and uses it to calculate the gradient of the deep layers ${\bf{w}}_{t,2}$.
Concretely, the edge gradient, ${\bf{g}}^{\rm E}_{t} \in \mathbb{R}^{Q_2}$, is defined by
\begin{align}
	\label{CL_gradient}
	{\bf{g}}^{\rm E}_{t}=\sum\nolimits_{n \in \mathcal{D}^{\rm E}_{t}} {\nabla f({\bf{w}}_{t,2};{\bf{\Omega}}_{n})},
\end{align}
where ${\bf{\Omega}}_{n}$ denotes an intermediate output in $\mathcal{D}^{\rm E}_{t}$.

Using the aggregated gradient ${\bf{g}}^{\rm L}_{t}$ and the edge gradient ${\bf{g}}^{\rm E}_{t}$, the shallow layers ${\bf{w}}_{t,1}$ and deep layers ${\bf{w}}_{t,2}$ of global model ${\bf{w}}_{t}$ are updated by (\ref{global_update_shallow_layer}) and (\ref{global_update_deep_layer}), respectively, given by
\vspace{-0.2 cm}
\begin{align}
	\label{global_update_shallow_layer}
	&{\bf{w}}_{t+1,1}={\bf{w}}_{t,1}-\eta_t{\bf{g}}^{\rm L}_{t,s}, \\
	\label{global_update_deep_layer}
	&{\bf{w}}_{t+1,2}={\bf{w}}_{t,2}-\eta_t(\rho^{\rm L}_t{\bf{g}}^{\rm L}_{t,d}+\rho^{\rm E}_t{\bf{g}}^{\rm E}_{t}),
\end{align}
where $\eta_t$ denotes the learning rate.
Here, $\rho^{\rm L}_t=1-(1/K)\sum\nolimits_{k=1}^{K} \theta_{t,k}$ and $\rho^{\rm E}_t=(1/K)\sum\nolimits_{k=1}^{K} \theta_{t,k}$ denote the FL and SL weight coefficients, respectively.
Next, the BS broadcasts the updated global model ${\bf{w}}_{t+1}=[{\bf{w}}_{t+1,1}^{\rm T},{\bf{w}}_{t+1,2}^{\rm T}]^{\rm T}$ to all devices for the next round of SemiFL.

\vspace{-0.4 cm}
\subsection{Over-the-Air Gradient Aggregation and Data Uploading}
\label{over-the-air_gradient_aggregation_and_data_uploading}
As depicted in Fig.~\ref{system_model_picture}, $K+1$ orthogonal bandwidths are employed to aggregate local gradients $\{{\bf{g}}^{\rm L}_{t,k}\}$ over the air while uploading datasets $\{\mathcal{D}^{\rm E}_{t,k}\}$.
%
%

%
\subsubsection{Over-the-Air Gradient Aggregation}
\label{over-the-air_gradient_aggregation}
Let ${\hat{\bf{g}}}^{\rm L}_{t,k}=[\hat{g}^{\rm L}_{t,k,1},\ldots,\hat{g}^{\rm L}_{t,k,Q}] \in \mathbb{R}^{\rm Q}$ denote the normalized gradient signal of the local gradient ${\bf{g}}^{\rm L}_{t,k}$, where $\mathbb{E}[\hat{g}^{\rm L}_{t,k,q}]=0$ and $\mathbb{E}[|\hat{g}^{\rm L}_{t,k,q}|^2]=1, \forall q$.
{
	Each entry in $\hat{\bf{g}}_{t,k}^{\rm L}$ corresponds to an analog symbol obtained by mapping the respective entry of the local gradient ${\bf{g}}_{t,k}^{\rm L}$.
	To be transmitted on the wireless channel, $\hat{\bf{g}}_{t,k}^{\rm L}$ is further mapped onto single-carrier waveforms, exemplified by single-carrier frequency-division multiple access (SC-FDMA).
}
Devices concurrently send gradient signals $\{{\hat{\bf{g}}}^{\rm L}_{t,k}\}$ over the shared bandwidth entry by entry.
Then, the $q$-th signal of the gradient aggregated is given by
\vspace{-0.2 cm}
\begin{align}
	\label{aggregated_signal_FL_gradient}
	s^{\rm G}_{t,q}=\frac{{\bf{b}}^{\rm H}_{t}}{K\sqrt{\nu_t}}\left(\sum\nolimits_{k=1}^{K} {\bf{h}}^{\rm G}_{t,k} p^{\rm G}_{t,k} \hat{g}^{\rm L}_{t,k,q}+{\bf{n}}^{\rm G}_{t} \right), \forall q,
\end{align}
where ${\bf{b}}_t \in \mathbb{C}^{N_r}$ satisfying $\|{\bf{b}}_t\|=1$ and $\nu_t > 0$ denote the receive beamformer and the normalizing factor for gradient aggregation, respectively, ${\bf{h}}^{\rm G}_{t,k} \in \mathbb{C}^{N_r}$ denotes the channel coefficient vector between the $k$-th device and the BS for gradient aggregation, $p^{\rm G}_{t,k}$ denotes the transmit power coefficient of the $k$-th device for gradient uploading, and ${\bf{n}}^{\rm G}_{t} \in \mathbb{C}^{N_r}$ yielding $\mathcal{CN}({\bf{0}},\sigma^2 {\bf{I}})$ denotes noise. 
Here, $(\cdot)^{\rm H}$, $\|\cdot\|$, $\sigma^2$, and ${\bf{I}}$ denote conjugate transpose, vector $2$-norm, the noise power, and an identity matrix, respectively.
We set the transmit power coefficient of the $k$-th device for gradient uploading to $p^{\rm G}_{t,k}=\sqrt{\omega_t} ({\bf{b}}^{\rm H}_t {\bf{h}}^{\rm G}_{t,k})^{\rm H}/{|{\bf{b}}^{\rm H}_t {\bf{h}}^{\rm G}_{t,k}|^2}$, where $\omega_t >0$ denotes the power scaling factor.
Note that coupling $p^{\rm G}_{t,k}$ with the receive beamformer ${\bf{b}}_t$ helps eliminate the influence of the channel ${\bf{h}}^{\rm G}_{t,k}$.
Moreover, introducing the power scaling factor $\omega_t$ offers an efficient approach to generate over-the-air distortion in coordination with $\nu_t$.
As a result, the $q$-th entry of the aggregated gradient ${\bf{g}}^{\rm L}_t$ can be obtained by
%
\begin{align}
	\label{signal_aggregated_gradient}
	g^{\rm L}_{t,q} ={\rm Re}\{s^{\rm G}_{t,q}\} =\frac{\sqrt{\omega_t}}{\sqrt{\nu_t}} \frac{1}{K} \sum\nolimits_{k=1}^{K} {\hat{g}^{\rm L}_{t,k,q}} + \hat{n}^{\rm G}_{t,q}, \forall q,
\end{align}
where ${\rm Re}\{\cdot\}$ takes the real part and $\hat{n}^{\rm G}_{t,q}\!=\!{{\rm Re}\{{\bf{b}}^{\rm H}_t{\bf{n}}^{\rm G}_t\!\}}/\!{\sqrt{\nu_t}}$, yielding $\mathcal{N}(0,\frac{\sigma^2}{2\nu_t})$.
Since gradient descent for AI model training typically requires an aggregated gradient with real-value elements, we thus extract the real part of $s^{\rm G}_{t,q}$.

Denote the ideally aggregated gradient by $\hat{\bf{g}}^{\rm L}_t=[\hat{g}^{\rm L}_{t,1},\ldots,\hat{g}^{\rm L}_{t,Q}] \in \mathbb{R}^{Q}$, where $\hat{g}^{\rm L}_{t,q}=(\sum\nolimits_{k=1}^{K} \hat{g}^{\rm L}_{t,k,q})/K, \forall q$.
Then, by assembling the received entries, the aggregated gradient ${\bf{g}}^{\rm L}_t$ can be expressed by
\vspace{-0.2 cm}
\begin{align}
	\label{aggregated_FL_gradient}
	{\bf{g}}^{\rm L}_t=\frac{\sqrt{\omega_t}}{\sqrt{\nu_t}}\hat{\bf{g}}^{\rm L}_t + \hat{\bf{n}}^{\rm G}_t,
\end{align}
where $\hat{\bf{n}}^{\rm G}_t=[\hat{n}^{\rm G}_{t,1},\ldots,\hat{n}^{\rm G}_{t,Q}] \in \mathbb{R}^{\rm Q}$, yielding $\mathcal{N}({\bf{0}},\frac{\sigma^2}{2\nu_t}{\bf{I}})$.
Meanwhile, the MSE between $g^{\rm L}_{t,q}$ and $\hat{g}^{\rm L}_{t,q}$, i.e., ${\rm MSE}_{t}=\mathbb{E}[|g^{\rm L}_{t,q}-\hat{g}^{\rm L}_{t,q}|^2]$, is used as a metric to quantify the over-the-air distortion, as derived by
\vspace{-0.2 cm}
\begin{align}
	\label{MSE}
	{\rm MSE}_{t}=\frac{1}{K} \left( \frac{\sqrt{\omega_t}}{\sqrt{\nu_t}} - 1 \right)^2  + \frac{\sigma^2}{2 \nu_t}.
\end{align}
Based on the formulas of the aggregated gradient, ${\bf{g}}^{\rm L}_t$, and the MSE that measures aggregation distortion, ${\rm MSE}_t$, it is seen that over-the-air distortion arises from two aspects: amplitude distortion $\frac{\sqrt{\omega_t}}{\sqrt{\nu_t}}$ and noise perturbation $\hat{\bf{n}}^{\rm G}_{t}$.
The former occurs when the amplitude of the ideally aggregated gradient is modified, while the latter is due to the perturbation of aggregation noise.
The works in~\cite{Zhang2022Turning} and~\cite{Ge2015Escaping} only make use of the noise perturbation to escape from saddle points, whereas this paper jointly adjusts both amplitude distortion and noise perturbation to enlarge the learning rate in the non-stable region.
Our approach accelerates SemiFL's convergence while efficiently reducing energy consumption during gradient aggregation.

\begin{figure}[t]
	\centering
	\includegraphics[width=0.35 \textwidth]{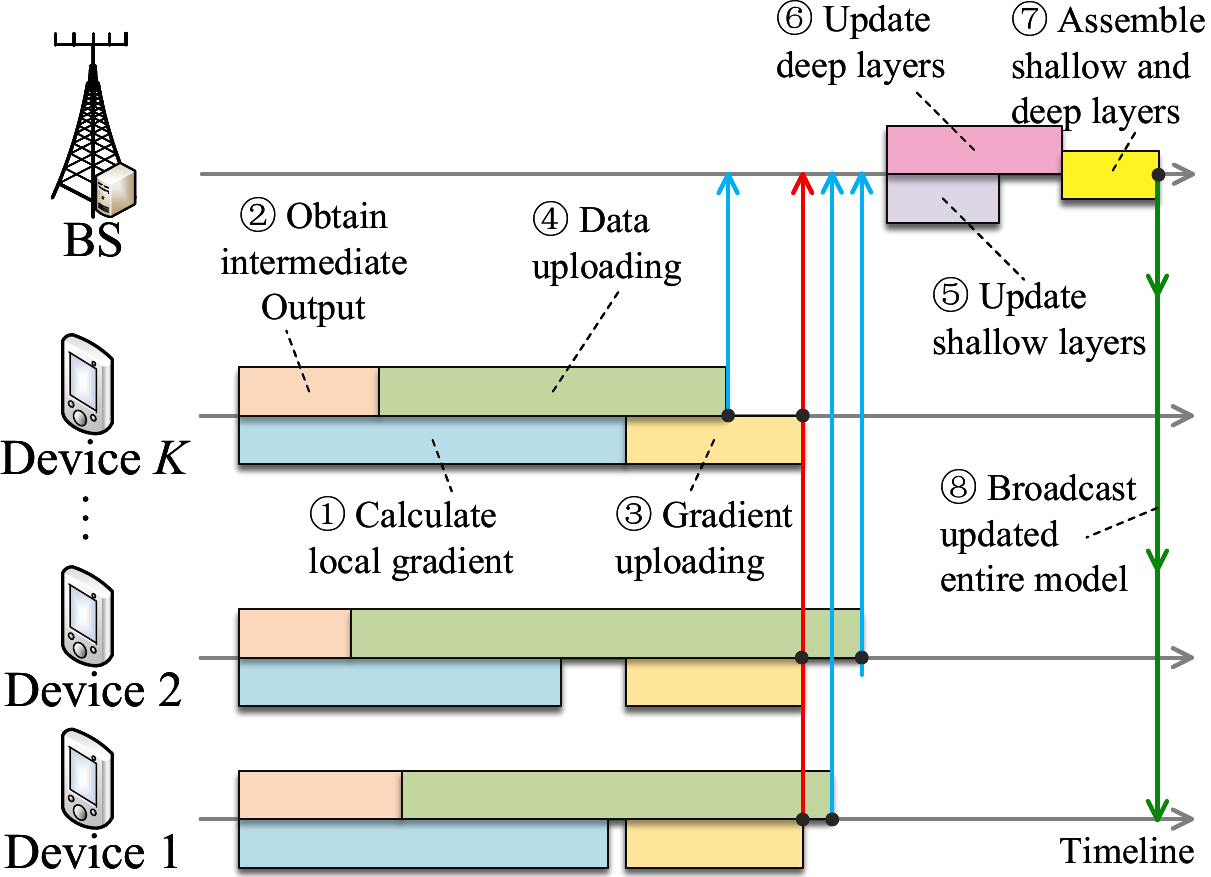}
	\vspace{-0.4 cm}
	{\caption{A workflow illustration of the proposed over-the-air distortion accelerated SemiFL framework.}
		\label{system_workflow}}
	\vspace{-0.6 cm}
\end{figure}

%
\subsubsection{Data Uploading}
\label{data_uploading}
Apart from gradient aggregation, the $k$-th device also sends the intermediate outputs in $\mathcal{D}^{\rm E}_{t,k}$ to the BS on its dedicated bandwidth.
{
	We consider a binary phase shift keying modulation scheme. First, the intermediate outputs in $\mathcal{D}_{t,k}^{\rm E}$ are mapped to $D\theta_{t,k}\bar{C}$ binary transmission symbols. Then, these symbols are carried by single-carrier waveforms such as SC-FDMA.}
Denote the signal of the intermediate outputs from the $k$-th device by $d_{t,k}$, where $\mathbb{E}[d_{t,k}]=0$ and $\mathbb{E}[|d_{t,k}|^2]=1$.
Then, the data signal of the $k$-th device received by the BS, $s^{\rm D}_{t,k}$, can be given by
\vspace{-0.2 cm}
\begin{align}
	\label{data_signal}
	s^{\rm D}_{t,k}=\frac{{\bf{v}}^{\rm H}_{t,k}}{\sqrt{\zeta_{t,k}}} \left({\bf{h}}^{\rm D}_{t,k} p^{\rm D}_{t,k} d_{t,k} + {\bf{n}}^{\rm D}_t\right), \forall k \in \mathcal{K},
\end{align}
where ${\bf{v}}^{\rm H}_{t,k}$ satisfying $\|{\bf{v}}^{\rm H}_{t,k}\|=1$ and $\zeta_{t,k} > 0$ denote the receive beamformer and the normalizing factor for data uploading, respectively, ${\bf{h}}^{\rm D}_{t,k}$ denotes the channel coefficient vector between the $k$-th device and the BS for data uploading, $p^{\rm D}_{t,k}$ denotes the transmit power of the $k$-th device for data uploading, and ${\bf{n}}^{\rm D}_t$ yielding $\mathcal{N}({\bf{0}},\sigma^2{\bf{I}})$ denotes the noise.
We set the transmit power of the $k$-th deice for data uploading to $p^{\rm D}_{t,k}=\sqrt{\zeta_{t,k}}({\bf{v}}^{\rm H}_{t,k}{\bf{h}}^{\rm D}_{t,k})^{\rm H}/|{\bf{v}}^{\rm H}_{t,k}{\bf{h}}^{\rm D}_{t,k}|^2$.
Then, the data signal $s^{\rm D}_{t,k}$ is reduced to
\vspace{-0.2 cm}
\begin{align}
	\label{reduced_data_signal}
	s^{\rm D}_{t,k}=d_{t,k}+\frac{{\bf{v}}^{\rm H}_{t,k}{\bf{n}}^{\rm D}_t}{\sqrt{\zeta_{t,k}}}, \forall k \in \mathcal{K}.
\end{align}
Based on (\ref{reduced_data_signal}), one can calculate the signal-to-noise ratio (SNR) and achievable data rate of the $k$-th device via (\ref{SNR_data_signal}) and (\ref{data_rate}), respectively, given by
\vspace{-0.2 cm}
\begin{align}
	\label{SNR_data_signal}
	&{\rm SNR}_{t,k}=\frac{\zeta_{t,k}}{\|{\bf{v}}_{t,k}\|^2\sigma^2}=\frac{\zeta_{t,k}}{\sigma^2}, \forall k \in \mathcal{K}, \\
	\label{data_rate}
	&R_{t,k}=B\log_2\left(1+\frac{\zeta_{t,k}}{\sigma^2}\right), \forall k \in \mathcal{K},
\end{align}
where $B$ denotes the bandwidth of each spectrum segment for data uploading and $\log_2(\cdot)$ denotes the base-$2$ logarithm.
Based on (\ref{SNR_data_signal}), one can see that increasing the normalizing factor $\zeta_{t,k}$ increases SNR, which increases the data rate.

\vspace{-0.3 cm}
\subsection{Latency and Energy Consumption}
\label{latency_and_energy_consumption}
The $t$-th round of the SemiFL process is mainly composed of four phases, including the gradient uploading phase, the data uploading phase, the local computing phase, and the edge computing phase.
{
	In Fig.~\ref{system_workflow}, it is seen that each device computes its local gradient and generates intermediate outputs simultaneously.
	Once the intermediate outputs are ready, each device uploads them over a dedicated orthogonal spectrum segment.
	Meanwhile, all devices wait until the last device completes its local gradient computation before simultaneously uploading their gradients via AirComp using the same time–frequency resources.
	Then, the BS updates the shallow and deep layers in parallel.
	Eventually, the updated shallow and deep layers are assembled to attain the updated entire model, which is broadcast to all devices for the next round of SemiFL.}
Unlike~\cite{Zheng2024Retransmission} that suppresses over-the-air distortion throughout SemiFL, we strategically amplify it during local gradient aggregation for convergence acceleration in this paper, despite using latency and energy calculation formulas similar to those in~\cite{Zheng2024Retransmission}.
{
	We note that gradient uploading and data uploading are allocated dedicated, non-overlapping spectral bands, which prevents mismatch during transmission.}
\noindent The latency and energy consumption of each phase are modeled as follows:

\subsubsection{Gradient Uploading}
\label{gradient_uploading_latency_energy}
Suppose each AirComp block contains $M$ gradient signals.
The latency of gradient uploading, $T^{\rm G}_t$, is given by
\vspace{-0.2 cm}
\begin{align}
	\label{latency_gradient_uploading}
	T^{\rm G}_t=\left\lceil\frac{Q}{M}\right\rceil T_s,
\end{align}
where $T_s$ denotes the duration of an AirComp block and $\lceil \cdot \rceil$ denotes the ceiling function.
Note that since AirComp employs analog transmission, it is inappropriate to use Shannon's formula to calculate its transmission latency.
We account for this latency by multiplying transmitted blocks $\lceil \frac{Q}{M} \rceil$ by block duration $T_s$.
The energy consumption of the $k$-th device for uploading gradient can be given by
\begin{align}
	\label{energy_consumption_gradient_uploading}
	E^{\rm G}_{t,k}=|p^{\rm G}_{t,k}|^2 T^{\rm G}_t=\frac{\omega_t\lceil\frac{Q}{M}\rceil T_s}{|{\bf{b}}^{\rm H}_t {\bf{h}}^{\rm G}_{t,k}|^2}, \forall k \in \mathcal{K}.
\end{align}

\subsubsection{Data Uploading}
\label{data_uploading_latency_energy}
Suppose an intermediate output is presented by $\bar{C}$ bits.
Then, the data uploading latency of the $k$-th device, $T^{\rm D}_{t,k}$, can be calculated by
\vspace{-0.2 cm}
\begin{align}
	\label{latency_data_uploading}
	T^{\rm D}_{t,k}=\frac{D\theta_{t,k}\bar{C}}{R_{t,k}}, \forall k \in \mathcal{K}.
\end{align}
Consequently, the energy consumption of the $k$-th device for data uploading can be given by
\begin{align}
	\label{energy_consumption_data_uploading}
	E^{\rm D}_{t,k}=|p^{\rm D}_{t,k}|^2T^{\rm D}_{t,k}=\frac{\zeta_{t,k}D\theta_{t,k}\bar{C}}{|{\bf{v}}^{\rm H}_{t,k}{\bf{h}}^{\rm D}_{t,k}|^2R_{t,k}}, \forall k \in \mathcal{K}.
\end{align}

\subsubsection{Local Computing}
\label{FL_computing_latency_energy}
Denote the CPU frequency of the $k$-th device by $\hat{f}_{t,k}$.
Then, the local computing latency of the $k$-th device, $T^{\rm L}_{t,k}$, is given by
\begin{align}
	\label{latency_FL_computing}
	T^{\rm F}_{t,k}=\frac{D(1-\theta_{t,k})\hat{C}_k}{\hat{f}_{t,k}}, \forall k \in \mathcal{K},
\end{align}
where $\hat{C}_k$ denotes the number of CPU circles required for a data sample.
{The latency for obtaining SL data is hidden by local computing.
	This is because the entire local model has substantially more layers than the shallow layers, leading to much longer forward and backward propagation latency.}
The energy consumption of the $k$-th device for local computing is given by~\cite{Yang2021Energy}
\vspace{-0.2 cm}
\begin{align}
	\label{energy_consumption_FL_computing}
	E^{\rm F}_{t,k}=D(1-\theta_{t,k})\hat{C}_k\hat{\kappa}\hat{f}^2_{t,k},\forall k \in \mathcal{K},
\end{align}
where $\hat{\kappa}$ denotes the effective switched capacitance of devices.

\subsubsection{Edge Computing}
\label{CL_computing_latency_energy}
Denoting the CPU frequency of the BS by $\tilde{f}_t$, the edge computing latency of the BS is given by
\vspace{-0.2 cm}
\begin{align}
	\label{latency_CL_computing}
	T^{\rm E}_t=\frac{D(\sum\nolimits_{k=1}^{K} \theta_{t,k})\tilde{C}}{\tilde{f}_t},
\end{align}
where $D(\sum\nolimits_{k=1}^{K} \theta_{t,k})$ denotes the number of intermediate outputs in $\mathcal{D}^{\rm E}_t$, and $\tilde{C}$ denotes the number of CPU circles for processing an intermediate output.
The energy consumption of the BS for edge computing is given by
\vspace{-0.2 cm}
\begin{align}
	\label{energy_consumption_CL_computing}
	E^{\rm E}_t=D\left(\sum\nolimits_{k=1}^{K} \theta_{t,k}\right)\tilde{C}\tilde{\kappa}\tilde{f}^2_t,
\end{align}
where $\tilde{\kappa}$ denotes the effective switched capacitance of the BS.

\subsubsection{Overall Latency and Energy Consumption}
\label{overall_latency_and_energy_consumption}
Based on (\ref{latency_gradient_uploading})--(\ref{energy_consumption_CL_computing}), the overall latency of the $t$-th round is given by
\vspace{-0.2 cm}
\begin{align}
	\label{overall_latency}
	T^{\rm ALL}_t=\max\{& T^{\rm D}_{t,1}+T^{\rm E}_t, \ldots, T^{\rm D}_{t,K}+T^{\rm E}_t, \notag \\ 
	&T^{\rm F}_{t,1}+T^{\rm G}_t, \ldots, T^{\rm F}_{t,K}+T^{\rm G}_t\}.
\end{align}
In addition, the overall energy consumption of the $t$-th round can be calculated by
\vspace{-0.2 cm}
\begin{align}
	\label{overall_energy_consumption}
	E^{\rm ALL}_t=\sum\nolimits_{k=1}^{K} (E^{\rm G}_{t,k}+E^{\rm D}_{t,k}+E^{\rm F}_{t,k})+E^{\rm E}_t.
\end{align}

\vspace{-0.3 cm}
\section{\small  Convergence Analysis and Problem Formulation}
\label{convergence_analysis_and_problem_formulation}
\begin{figure}[t]
	\centering
	\includegraphics[width=0.4 \textwidth]{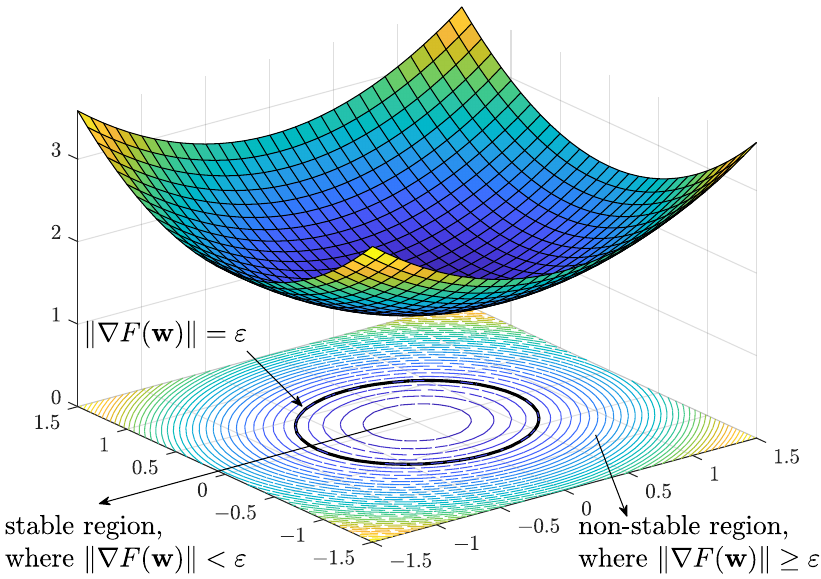}
	\vspace{-0.2 cm}
	\caption{An illustration of the non-stable and stable regions of SemiFL.}
	\label{surface_non_stable_stable_picture}
	\vspace{-0.6 cm}
\end{figure}

In this section, as shown in Fig.~\ref{surface_non_stable_stable_picture}, the SemiFL process is categorized into two types of regions: the non-stable region $\mathcal{R}^{\rm NS}$ and the stable region $\mathcal{R}^{\rm S}$, which are defined by $\mathcal{R}^{\rm NS}=\{{\bf{w}}_t |~ \|\nabla F({\bf{w}}_t)\| \ge \varepsilon, \forall t \in \mathcal{T} \}$ and $\mathcal{R}^{\rm S}=\{{\bf{w}}_t |~ \|\nabla F({\bf{w}}_t)\| < \varepsilon, \forall t \in \mathcal{T} \}$~\cite{Zhang2022Turning}, respectively, where the constant $\varepsilon > 0$.
Next, we analyze the convergence of SemiFL, and formulate two distinct problems for each type of region to minimize energy consumption.

\vspace{-0.4 cm}
\subsection{Convergence Analysis}
\label{convergence_analysis}
\begin{assumption}
	\label{assumption_1}
	The global loss function $F({\bf{w}})$ is $L$-smooth regarding a constant $L>0$.
	For any ${\bf{w}}$, ${\bf{w}}' \in \mathbb{R}^{Q}$, we have
	\vspace{-0.2 cm}
	\begin{align}
		\label{L_smooth}
		F({\bf{w}}) \le F({\bf{w}}') \!+\! {\nabla F({\bf{w}}')}^{\rm T}\left({\bf{w}} \!-\! {\bf{w}}' \right) \!+\! \frac{L}{2} \|{\bf{w}} \!-\! {\bf{w}}'\|^2.
	\end{align}
\end{assumption}

\begin{assumption}
	\label{assumption_2}
	The global loss function $F({\bf{w}})$ is $\mu$-strongly convex.
	For any ${\bf{w}}$, ${\bf{w}}' \in \mathbb{R}^{Q}$ and a constant $\mu>0$, we have
	\vspace{-0.2 cm}
	\begin{align}
		\label{mu_strongly_convex}
		F({\bf{w}}) \ge F({\bf{w}}') \!+\! {\nabla F({\bf{w}}')}^{\rm T} \left({\bf{w}} \!-\! {\bf{w}}' \right) \!+\! \frac{\mu}{2} \|{\bf{w}} \!-\! {\bf{w}}'\|^2.
	\end{align}
\end{assumption}

\begin{assumption}
	\label{assumption_3}
	The expected squared $2$-norm of the global gradient $\nabla F({\bf{w}}_t)$ is bounded by a constant $A^2 > 0$.
	For any ${\bf{w}}_t \in \mathbb{R}^{Q}$, we have
	\vspace{-0.2 cm}
	\begin{align}
		\label{bounded_gradient}
		\mathbb{E}[\|\nabla F({\bf{w}}_t)\|^2] \le A^2.
	\end{align}
\end{assumption}

\begin{assumption}
	\label{assumption_4}
	The ideally aggregated gradient $\hat{\bf{g}}^{\rm L}_{t}$ and the edge gradient ${\bf{g}}^{\rm E}_t$ are unbiased estimations of the global gradient $\nabla F({\bf{w}}_t)$.
	Thus, we have
	\vspace{-0.2 cm}
	\begin{align}
		\label{unbiased_gradient_estimate_FL}
		&\mathbb{E}[\hat{\bf{g}}^{\rm L}_{t}] = \nabla F({\bf{w}}_t), \\
		\label{unbiased_gradient_estimate_CL}
		&\mathbb{E}[{\bf{g}}^{\rm E}_t] = \nabla F({\bf{w}}_t).
	\end{align}
\end{assumption}

We are now in position to characterize the convergence behavior of SemiFL in the non-stable region $\mathcal{R}^{\rm NS}$. 
Specifically, we derive a closed-form lower bound for the expected loss function reduction between two consecutive rounds in $\mathcal{R}^{\rm NS}$, as presented in Theorem~\ref{theorem_1}.

\vspace{-0.2 cm}
\begin{theorem}
	\label{theorem_1}
	Given Assumptions~\ref{assumption_2}~--~\ref{assumption_4}, for global models ${\bf{w}}_t$, ${\bf{w}}_{t+1} \in \mathcal{R}^{\rm NS}$ and $\varepsilon\ge A/\sqrt{2 \mu}$, the expected global loss function reduction between two consecutive rounds in the non-stable region $\mathcal{R}^{\rm NS}$ is lower bounded by
	\vspace{-0.2 cm}
	\begin{align}
		\label{non_stable_lower_bound}
		\mathbb{E}[F({\bf{w}}_{t})\!-\!F({\bf{w}}_{t+1})] \ge &\frac{\eta_t^2}{4}(2\mu\varepsilon^2\!-\! A^2) \!\! \left[1 \! +\!\! \left(\frac{\sqrt{\omega_t}}{\sqrt{\nu_t}}\!-\!1\right) \! \rho^{\rm L}_{t}\right]^2\notag \\&-A^2+\frac{\mu \sigma^2 Q \eta_t^2}{4\nu_t}(\rho^{\rm L}_{t})^2.
	\end{align}
\end{theorem}
%
\begin{IEEEproof}
	Please refer to Appendix~\ref{proof_of_theorem_1} in~\cite{Zheng2024Appendix}.
\end{IEEEproof}

\vspace{-0.2 cm}
\begin{remark}
	\label{remark_1}
	Based on the lower bound (\ref{non_stable_lower_bound}) in Theorem~\ref{theorem_1}, we have the following two key insights in terms of the convergence acceleration effect of over-the-air distortion for SemiFL:
	\begin{itemize}
		\item Increasing the amplitude distortion $\frac{\sqrt{\omega_t}}{\sqrt{\nu_t}}$ increases the lower bound of $\mathbb{E}[F({\bf{w}}_t)-F({\bf{w}}_{t+1})]$ by enlarging the learning rate of the FL component.
		This indicates that even in the worst case, the expected training loss descent $\mathbb{E}[F({\bf{w}}_t)-F({\bf{w}}_{t+1})]$ is increased by the amplitude distortion $\frac{\sqrt{\omega_t}}{\sqrt{\nu_t}}$ at least.
		Hence, the convergence of SemiFL in the non-stable region can be accelerated by amplifying the amplitude distortion $\frac{\sqrt{\omega_t}}{\sqrt{\nu_t}}$.
		\item For $\frac{\sqrt{\omega_t}}{\sqrt{\nu_t}} \ge 1$, it is observed in (\ref{non_stable_lower_bound}) that the lower bound of $\mathbb{E}[F({\bf{w}}_{t})-F({\bf{w}}_{t+1})]$ increases with the FL weight coefficient $\rho^{\rm L}_t$.
		This suggests another insight: the convergence acceleration effect of over-the-air distortion can be enhanced by uploading fewer data to the BS in the non-stable region $\mathcal{R}^{\rm NS}$, as $\rho^{\rm L}_{t}$ can be increased to boost the loss descent in SemiFL.
		\item The aggregated FL gradient is transmitted over wireless channels whereas the edge gradient corresponding to SL, ${\bf{g}}^{\rm E}_{t}$, is computed directly at the BS.
		Hence, the amplitude distortion $\frac{\sqrt{\omega_t}}{\sqrt{\nu_t}}$ affects only the FL gradients, while having no impact on SL.
	\end{itemize}
\end{remark}

\vspace{-0.4 cm}
\begin{remark}
	\label{remark_2}
	In our work, we aim to achieve a native joint design of wireless communication and learning, particularly by manipulating over-the-air distortion to enlarging the learning rate of the FL part in the non-stable region of SemiFL.
	In essence, our proposed approach enables a joint design of communication and learning that accelerates convergence while reducing energy consumption.
	This is because our approach turns down the power scaling factor $\omega_t$ while using a normalizing factor $\nu_t$ much smaller than $\omega_t$, which reduces transmit power of devices and increases amplitude distortion to increase the learning rate.
	{Mathematically, in conventional learning rate tuning schemes, a common form of the learning rate $\eta_t$ is an inverse proportional decaying scheme~\cite{Zhu2024Over,Guo2021Analog}, i.e., $\eta_t = \frac{\eta_0}{t+\Lambda}$, where $\eta_0$ is the initial learning rate and $\Lambda \ge 0$.
		However, we initialize the learning rate $\eta_t$ to a sufficiently small constant, and intentionally amplifies the amplitude distortion $\frac{\sqrt{\omega_t}}{\sqrt{\nu_t}}\gg1$ to equivalently apply an enlarged learning rate, i.e., $\frac{\sqrt{\omega_t}}{\sqrt{\nu_t}}\eta_t$.}
	\vspace{-0.2 cm}
\end{remark}

To explore the impact of data heterogeneity, consider a classification task for the proposed SemiFL which contains $C$ categories of data in total.
The local gradient of the $k$-th device can be rewritten as
\begin{align}
	{\bf{g}}^{\rm L}_{t,k}=\sum\nolimits_{c=1}^{C} p_{t,k,c} {\bf{g}}_{t,k,c},\forall k \in \mathcal{K},
\end{align}
where $p_{t,k,c}$ denotes the proportion of the $c$-th category of data for the $k$-th device, and ${\bf{g}}_{t,k,c}$ denotes the local gradient of the $k$-th device calculated using the $c$-th category of data.
In the case of non-IID data, we have the following corollary:
\vspace{-0.2 cm}
\begin{corollary}
	\label{corollary_1}
	Given Assumptions 2 -- 4 and non-IID data, the expected global loss function reduction between two consecutive rounds in the non-stable region is lower bounded by
	\vspace{-0.4 cm}
	\begin{align}
		\label{lower_bound_non_iid}
		\hspace{-0.2 cm}&\mathbb{E}[F({\bf{w}}_t)\!\!-\!\! F({\bf{w}}_{t+1})] \!\ge\! \frac{\mu^2 \eta_t^2}{2} (\rho^{\rm L}_t \frac{\sqrt{\omega_t}}{\sqrt{\nu_t}}\!+\!\rho^{\rm E}_t)^2\varepsilon^2 \!\!+\!\! \frac{\mu \sigma Q \eta_t^2}{4\nu_t}(\rho_t^{\rm L})^2 \!\notag \\
		\hspace{-0.2 cm}&\! - \!\frac{A^2 C}{K}(\rho^{\rm L}_t)^2\Delta d_t \!-\![\frac{\eta_t^2\omega_t}{4\nu_t}\!+\!\frac{\eta_t^2}{4}(\rho^{\rm L}_t\frac{\sqrt{\omega_t}}{\sqrt{\nu_t}}\!+\!\rho^{\rm E}_t)^2\!+\!1]A^2,
	\end{align}
	where $\Delta d_t=\sum\nolimits_{k=1}^{K}\sum\nolimits_{c=1}^{C}(p_{t,k,c}-\frac{1}{C})^2$ measures the data heterogeneity.
\end{corollary}
\begin{IEEEproof}
	Please refer to Appendix~\ref{proof_of_corollary_1} in~\cite{Zheng2024Appendix}.
\end{IEEEproof}

In Corollary~\ref{corollary_1}, we have the following findings:
\begin{itemize}
	\item The lower bound decreases as $\Delta d_t$ increases. 
	This means data heterogeneity slows down the convergence rate in non-stable region. 
	When $p_{t,k,c}=1/C$, $\Delta d_t$ vanishes, which actually reduces the lower bound to a special case where the data are IID across devices.
	\item The negative impact of non-IID data decreases as the coefficient of FL $\rho^{\rm L}_t$ decreases.
	This indicates that increasing the amount of SL data in SemiFL can address data heterogeneity to some extents, which also elaborates the motivation of investigating SemiFL.
	\item A trade-off exists between accelerating convergence and mitigating data heterogeneity since increasing $\rho^{\rm L}_t$ amplifies the positive impact of over-the-air distortion $\frac{\sqrt{\omega_t}}{\sqrt{\nu_t}}$ while simultaneously exacerbating the negative effect of data heterogeneity $\Delta d_t$.
\end{itemize}

{
	When the global loss function $F({\bf{w}})$ is non-convex, we consider a $\delta$-nonconvex assumption as follows:
	\vspace{-0.2 cm}
	\begin{assumption}
		The global loss function $F({\bf{w}})$ is $\delta$-nonconvex.
		This means all eigenvalues of $\nabla^2 F({\bf{w}})$ lie within the interval $[-\delta,L]$, where $\delta \in (0,L]$, $L \ge 0$ denotes the $L$-smooth constant, and $\nabla^2 F({\bf{w}})$ denotes the Hessian of $F({\bf{w}})$.
	\end{assumption}
	On the one hand, we address the convexity as follows:
	%
	\begin{align}
		\label{convexify_non_convex_loss}
		\hat{F}({\bf{w}})=F({\bf{w}}) + \delta\|{\bf{w}}\|^2.
	\end{align}
	Note that $\hat{F}({\bf{w}})$ is $(L+2\delta)$-smooth $\delta$-strongly convex.
	One the other hand, we plug (\ref{convexify_non_convex_loss}) into Theorem~\ref{theorem_1} to derive the following corollary:
	\vspace{-0.2 cm}
	\begin{corollary}
		\label{corollary_2}
		When $F({\bf{w}})$ is $\delta$-nonconvex, define $\hat{F}({\bf{w}})$ as (\ref{convexify_non_convex_loss}).
		For ${\bf{w}}_t \in \mathcal{R}^{\rm NS}$, one can have
		\vspace{-0.2 cm}
		\begin{align}
			\label{stable_convex_bound}
			&F({\bf{w}}_{t})\!-\!F({\bf{w}}_{t+1})\ge\frac{\eta_t^2}{4}(2\mu\varepsilon^2\!-\! A^2) \!\! \left[1 \! +\!\! \left(\frac{\sqrt{\omega_t}}{\sqrt{\nu_t}}\!-\!1\right) \! \rho^{\rm L}_{t}\right]^2\notag \\
			&-A^2+\frac{\mu \sigma^2 Q \eta_t^2}{4\nu_t}(\rho^{\rm L}_{t})^2 - \frac{\delta+\mu}{2} \|{\bf{w}}_{t}-{\bf{w}}_{t+1}\|^2.
		\end{align}
	\end{corollary}
	\begin{IEEEproof}
		Please refer to Appendix~\ref{proof_of_corollary_2} in~\cite{Zheng2024Appendix}.
	\end{IEEEproof}
	\noindent Compared with Theorem~1, it is seen that the last negative term, $- \frac{\delta+\mu}{2} \|{\bf{w}}_{t}-{\bf{w}}_{t+1}\|^2$, decreases the expected training loss descent between two consecutive rounds.
	This indicates that the non-convex $F({\bf{w}})$ compromises the convergence acceleration effect of over-the-air distortion in the non-stable region.}

Then, in the stable region $\mathcal{R}^{\rm S}$, we derive an upper bound for the expected optimality gap of the global loss function $F({\bf{w}}_t)$ as $t$ approaches infinity, which is detailed in Theorem~\ref{theorem_2}.

\vspace{-0.2 cm}
\begin{theorem}
	\label{theorem_2}
	Given Assumptions~\ref{assumption_1}~--~\ref{assumption_4}, set $\eta_t=\frac{1}{\mu}$, $\frac{\sqrt{\omega_t}}{\sqrt{\nu_t}}=1$, and $\nu_t=\nu, \forall t \in \mathcal{T}$.
	For a global model ${\bf{w}}_t \in \mathcal{R}^{\rm S}$ , as $t \rightarrow \infty$, the expected optimality gap of the global loss function can be upper bounded by
	\vspace{-0.2 cm}
	\begin{align}
		\label{stable_upper_bound}
		\lim_{t \rightarrow \infty} \mathbb{E}[F({\bf{w}}_t)-F({\bf{w}}^{*})] &\le \frac{L}{\mu} \frac{1}{4\mu-L}\left(A^2+\frac{\sigma^2 Q}{2 \nu}\right),\\
		&\overset{\triangle}{=}{\psi}(\nu), \notag
	\end{align}
	where ${\bf{w}}^{*}$ denotes the optimal global model.
\end{theorem}
\begin{IEEEproof}
	Please refer to Appendix~\ref{proof_of_theorem_2} in~\cite{Zheng2024Appendix}.
\end{IEEEproof}

\vspace{-0.2 cm}
\begin{remark}
	\label{remark_3}
	%
	%
	{
		Upon entering the stable region $\mathcal{R}^{\rm S}$, we eliminate amplitude distortion to guarantee $\frac{\sqrt{\omega_t}}{\sqrt{\nu_t}}=1$ while suppressing the noise perturbation.
		This ensures that the learning rate $\eta_t$ can be maintained at a sufficiently small constant to achieve stable final convergence.}
	Moreover, based on (\ref{stable_upper_bound}) and (\ref{aggregated_FL_gradient}), we observe that both the expected optimality gap $\psi(\nu)$ and the noise $\hat{\bf{n}}_{t}^{\rm G}$ decrease as $\nu$ increases.
	This demonstrates that increasing $\nu$ improves final convergence by suppressing noise perturbation.
\end{remark}

\vspace{-0.2 cm}
In the stable region, we have the following corollary under non-IID data conditions:
\vspace{-0.2 cm}
\begin{corollary}
	\label{corollary_3}
	Given Assumptions 1 -- 4 and non-IID data, set $\eta_t=\frac{1}{\mu}$, $\frac{\sqrt{\omega_t}}{\sqrt{\nu_t}}=1$, and $\nu_t=\nu, \forall t \in \mathcal{T}$.
	For ${\bf{w}}_t \in \mathcal{R}^{\rm S}$, as $t \rightarrow \infty$, the expected optimality gap of the global loss function can be upper bounded by
	\begin{align}
		\label{upper_bound_non_iid}
		&\lim_{t \rightarrow \infty} \mathbb{E}[F({\bf{w}}_t)-F({\bf{w}}^{*})] \le \frac{L}{\mu} \frac{1}{2\mu \!-\! L} \left(A^2\!\!+\!\frac{\sigma^2Q}{2\nu}\right) \notag \\
		&+ \frac{L A^2 C}{\mu^2 K} \! \lim_{t \rightarrow \infty} \sum\limits_{\tau=1}^{t-1} \! \xi^{t-1-\tau} (\rho^{\rm L}_t)^2 \Delta d_{\tau}.
	\end{align}
\end{corollary}
\begin{IEEEproof}
	Please refer to Appendix~\ref{proof_of_corollary_3} in~\cite{Zheng2024Appendix}.
\end{IEEEproof}
In Corollary~\ref{corollary_3}, we see that non-IID data increases the upper bound by accumulating $\Delta d_{t}$, which compromises the final convergence of SemiFL.
However, this can also be mitigated by decreasing the FL coefficient $\rho^{\rm L}_t$.

{
	Considering a non-convex global loss function $F({\bf{w}})$, we have the following corollary:
	\vspace{-0.2 cm}
	\begin{corollary}
		\label{corollary_4}
		When $F({\bf{w}})$ is $\delta$-nonconvex, define $\hat{F}({\bf{w}})$ as (\ref{convexify_non_convex_loss}).
		By setting $\eta=\frac{1}{\delta}$, $\delta=\mu$, and $\nu_t=\nu$, for ${\bf{w}}_t \in \mathcal{R}^{\rm S}$, as $T \rightarrow \infty$, one can have
		\vspace{-0.2 cm}
		\begin{align}
			\label{stable_non_convex_bound}
			\lim_{T \rightarrow \infty}\mathbb{E}[F({\bf{w}}_{T})\!-\! F({\bf{w}}^{*})]\le&
			\frac{L}{\mu}\frac{1}{2\mu-L}(A^2+\frac{\sigma^2Q}{2\nu}) \notag \\
			&+  \lim_{T \rightarrow \infty}\sum\limits_{\tau=1}^{T-1}(\xi')^{T-1-\tau}\Delta_{\tau},
		\end{align}
		where $\xi'=\frac{L}{2\mu}$ and $\Delta_{t-1}=\delta\|{\bf{w}}^{*}\|^2+3\delta\|{\bf{w}}_{t-1}\|^2$.
	\end{corollary}
	\begin{IEEEproof}
		Please refer to Appendix~\ref{proof_of_corollary_4} in~\cite{Zheng2024Appendix}.
	\end{IEEEproof}
	\noindent Comparing to Theorem~2, an additional non-negative term $\lim\limits_{T \rightarrow \infty}^{~}\sum\nolimits_{\tau=1}^{T-1}(\xi')^{T-1-\tau}\Delta_{\tau}$ has been introduced.
	Moreover, one can obtain that $\frac{1}{2\mu-L}>\frac{1}{4\mu-L}$.
	Therefore, in the stable region, the non-convexity of $F({\bf{w}})$ incurs an additional gap in contrast to the convex case..
}

%
The opposing effects of over-the-air distortion motivate the exploration of a two-region MSE threshold configuration scheme to leverage advantages from both sides, as elaborated in Remark~\ref{remark_4}.

\vspace{-0.2 cm}
\begin{remark}[Two-region MSE threshold configuration.]
	\label{remark_4}
	In the non-stable region, we set a high MSE threshold for model aggregation, such that the amplitude distortion $\frac{\sqrt{\omega_t}}{\sqrt{\nu_t}}$ can be increased to enlarge the learning rate for accelerating convergence of SemiFL.
	To achieve this, one can assign a small value to the normalizing factor $\nu_t$, i.e., $\nu_t=\nu^{\rm low},\forall t \in \{t|~{\bf{w}}_t \in \mathcal{R}^{\rm NS}\}$.
	In the stable region, we set a low MSE threshold for model aggregation, which aims to suppress noise perturbation for attaining improved final convergence of SemiFL.
	This can be achieved by assigning a large value to the normalizing factor $\nu_t$, i.e., $\nu_t=\nu^{\rm high},\forall t \in \{t|~{\bf{w}}_t \in \mathcal{R}^{\rm S}\}$.
\end{remark}

\vspace{-0.2 cm}
We demonstrate that with the above two-region MSE threshold configuration scheme, the adverse effects of amplified over-the-air distortion on the final convergence can be gradually mitigated as the SemiFL training progresses.
Specifically, suppose SemiFL reaches the stable region $\mathcal{R}^{\rm S}$ in the $T'$-th round.
For $t \ge T'$, we have
\vspace{-0.2 cm}
\begin{align}
	\label{two_region_convergence}
	&\lim_{t \rightarrow \infty} \mathbb{E}[F({\bf{w}}_t)\!-\!F({\bf{w}}^{*})] \notag \\ 
	\le& \lim_{t \rightarrow \infty} \left\{ \xi^{t-1} \mathbb{E}[F({\bf{w}}_1)\!\!-\!\!F({\bf{w}}^{*})] \right. \notag \\
	&\left.+ \frac{L}{2\mu^2} \! \sum\nolimits_{\tau=T'}^{t-1} \xi^{-\tau}(A^2\!+\!\frac{\sigma^2Q}{2\nu^{\rm high}})\right\} \notag \\
	& \underbrace{+\lim_{t \rightarrow \infty} \frac{L}{2\mu^2}\sum\nolimits_{\tau=1}^{T'-1} \! \xi^{\rm -\tau} (A^2\!+\!\frac{\sigma^2Q}{2\nu^{\rm low}})}_{\text{negative effect of amplified over-the-air distortion in $\mathcal{R}^{\rm S}$}} \notag \\
	=&\frac{L}{\mu}\frac{1}{4\mu-L}(A^2+\frac{\sigma^2Q}{2\nu^{\rm high}}).
\end{align}
One can see that the last term, which reflects the adverse affects of amplified over-the-air distortion in the stable region, asymptotically approaches zero as $t \rightarrow \infty$.
Moreover, (\ref{two_region_convergence}) demonstrates that large values of $\nu_t^{\rm high}$ attenuate this result, thereby resulting in improved final convergence.

\subsection{Problem Formulation}
\label{probelm_formulation}
Following the two-region MSE threshold configuration scheme, we aim to minimize the per-round energy consumption in different types of regions by jointly optimizing the normalizing factors, power scaling factor, receive beamformers, CPU frequencies, and data allocation.
Experimentally, one can set a slope threshold for the accuracy or loss curves. 
When the slope falls below this threshold consistently, it is considered to have entered the stable region.

\subsubsection{Non-Stable Region}
\label{non_stable_region}
The optimization problem of the $t$-th round, where $t \in \{t|~{\bf{w}}_t \in \mathcal{R}^{\rm NS}\}$, is formulated as
%
	\begin{subequations}
		\label{p_1}
		\begin{eqnarray}
			\label{objective_p_1}
			\hspace{-0.4 cm}&\mathop {\min }\limits_{\{{\zeta}_{k}\},\nu,\omega,{\bf{b}},\{{\bf{v}}_{k}\},\atop{\{\hat{f}_{k}\}, \tilde{f},\{\theta_{k}\}}} & E^{\rm ALL} \\
			\label{latency_contraint_p_1}
			\hspace{-0.4 cm}&{\rm s.t.}& T^{\rm ALL} \le T_{\max}, \\
			\label{omega_nu_ratio_constraint_p_1}
			\hspace{-0.4 cm}&{}& \epsilon_1 = \frac{\sqrt{\omega}}{\sqrt{\nu}}, \\
			\label{MSE_constraint_p_1}
			\hspace{-0.4 cm}&{}& {\rm MSE} \le \epsilon_2, \\
			\label{zeta_contraint_p_1}
			\hspace{-0.4 cm}&{}&0 \le \zeta_{k} \le p_{\max} |{\bf{v}}_{k}^{\rm H} {\bf{h}}_{k}^{\rm D}|^2,\forall k \in \mathcal{K}, \\
			\label{omega_constraint_p_1}
			\hspace{-0.4 cm}&{}&0 \le \omega \le p_{\max} |{\bf{b}}^{\rm H} {\bf{h}}_{k}^{\rm G}|^2,\forall k \in \mathcal{K}, \\
			\label{theta_constraint_p_1}
			\hspace{-0.4 cm}&{}&0 \le \theta_k \le \theta_{\max}, \forall k \in \mathcal{K}, \\
			\label{device_CPU_constraint_p_1}
			\hspace{-0.4 cm}&{}&0 \le \hat{f}_{k} \le \hat{f}_{\max}, \forall k \in \mathcal{K}, \\
			\label{BS_CPU_constraint_p_1}
			\hspace{-0.4 cm}&{}&0 \le \tilde{f} \le \tilde{f}_{\max}, \\
			\label{receive_beamformer_v_constraint_p_1}
			\hspace{-0.4 cm}&{}& \|{\bf{v}}_{k}\|=1, \forall k \in \mathcal{K}, \\
			\label{receive_beamformer_b_constraint_p_1}
			\hspace{-0.4 cm}&{}& \|{\bf{b}}\|=1,
		\end{eqnarray}
	\end{subequations}
%
\noindent where $T_{\max}$ denotes the maximum allowable latency per round, $\epsilon_1 > 1$ denotes the minimum power scaling factor-to-normalizing factor ratio in $\mathcal{R}^{\rm NS}$, $\epsilon_2 > 0$ denotes the MSE threshold in $\mathcal{R}^{\rm NS}$, $p_{\max}$ denotes the maximum transmit power of devices, $\theta_{\max}$ denotes the maximum ratio of SL data, $\hat{f}_{\max}$ and $\tilde{f}_{\max}$ denote the maximum CPU frequencies of devices and the BS, respectively.
The subscript $t$ is omitted.
As we intend to accelerate the convergence of SemiFL in the non-stable region $\mathcal{R}^{\rm NS}$ by amplifying over-the-air distortion, the ratio $\frac{\sqrt{\omega}}{\sqrt{\nu}}$ is forced to be greater than $1$ in constraint (\ref{omega_nu_ratio_constraint_p_1}).
Correspondingly, we choose a high MSE threshold $\epsilon_2$ in constraint (\ref{MSE_constraint_p_1}).
This aligns with the discussions in Remark~\ref{remark_1}.
Moreover, we guarantee a large FL weight coefficient $\rho^{\rm L}_t$ by intentionally imposing a maximum allowable ratio of SL data, $\theta_{\max}$, so as to better utilize the acceleration effect of over-the-air distortion, as also discussed in Remark~\ref{remark_1}.

\subsubsection{Stable Region}
\label{stable_region}
The optimization problem of the $t$-th round, where $t \in \{t|~{\bf{w}}_t \in \mathcal{R}^{\rm S}\}$, is formulated by
%
	\begin{subequations}
		\label{p_2}
		\begin{eqnarray}
			\label{objective_p_2}
			{\hspace{-0.9 cm}}&\mathop {\min }\limits_{\{{\zeta}_{k}\},\nu,\omega,{\bf{b}},\{{\bf{v}}_{k}\},\atop{\{\hat{f}_{k}\}, \tilde{f},\{\theta_{k}\}}} & E^{\rm ALL} \\
			\label{optimality_gap_contraint_p_2}
			{\hspace{-0.9 cm}}&{\rm s.t.}& \psi(\nu) \le \epsilon_3, \\
			\label{omega_nu_ratio_constraint_p_2}
			{\hspace{-0.9 cm}}&{}& \frac{\sqrt{\omega}}{\sqrt{\nu}}=1, \\
			\label{MSE_constraint_p_2}
			{\hspace{-0.9 cm}}&{}& {\rm MSE} \le \epsilon_4, \\
			\label{theta_constraint_p_2}
			{\hspace{-0.9 cm}}&{}&\theta_{\min} \le \theta_k \le 1, \forall k \in \mathcal{K}, \\
			\label{other_constraints_p_2}
			{\hspace{-0.9 cm}}&{}&\text{(\ref{latency_contraint_p_1}),~(\ref{zeta_contraint_p_1}),~(\ref{omega_constraint_p_1}),~and~(\ref{device_CPU_constraint_p_1})~--~(\ref{receive_beamformer_b_constraint_p_1}),}
		\end{eqnarray}
	\end{subequations}
%
\noindent where $\epsilon_3$ denotes the maximum expected optimality gap, $\epsilon_4$ denotes the MSE threshold in $\mathcal{R}^{\rm S}$, $\theta_{\min}$ denotes the minimum ratio of SL data. 
The subscript $t$ is omitted.
Note that we restrict the expected optimality gap to be below $\epsilon_3$ in constraint (\ref{optimality_gap_contraint_p_2}) to guarantee the final convergence of SemiFL.
To mitigate the negative effect of over-the-air distortion $\mathcal{R}^{\rm S}$, we impose a low MSE threshold $\epsilon_4 \ll \epsilon_2$ in constraint (\ref{MSE_constraint_p_2}), as analyzed in Remark~\ref{remark_4}.
In addition, imposing constraint (\ref{omega_nu_ratio_constraint_p_2}) helps suppress over-the-air distortion by forcing the first term of MSE to be zero according to (\ref{MSE}).

Due to the indefinite Hessian matrices of $E^{\rm ALL}$, $T^{\rm ALL}$, and ${\rm MSE}$, problems (\ref{p_1}) and (\ref{p_2}) are non-convex and intractable.
In the next section, we propose two resource allocation optimization algorithms to solve the formulated two problems in each type of region.

\section{Proposed Algorithms}
\label{proposed_algorithms}
In this section, we decompose problem (\ref{p_1}) of the non-stable region into four subproblems. By solving each problem iteratively, we obtain a solution to problem (\ref{p_1}).
Note that closed-form solutions are derived for some subproblems. 
Then, we solve problem (\ref{p_2}) corresponding to the stable region in a similar manner.

\vspace{-0.2 cm}
\subsection{Algorithm for Non-Stable Region}
\label{algorithm_for_non_stable_region}
We first decompose problem (\ref{p_1}) in the non-stable region $\mathcal{R}^{\rm NS}$ into four subproblems and conquer them one by one, as presented below:
\subsubsection{Normalizing and Power Scaling Factors}
\label{normalizing_and_power_scaling_factors_NS}
Given receive beamformers, CPU frequencies, and data allocation, the problem of normalizing and power scaling factors is reduced to$\!\!\!\!$
\vspace{-0.2 cm}
\begin{spacing}{1.0}
	\begin{subequations}
		\label{p_3}
		\begin{eqnarray}
			\label{objective_p_3}
			{\hspace{-0.8 cm}}&\mathop {\min }\limits_{\{{\zeta}_{k}\},\nu,\omega,\tau_1} & \sum\nolimits_{k=1}^{K} \left[C_{1,k} \frac{\zeta_k}{\log(1+\frac{\zeta_k}{\sigma^2})} + C_{2,k} \omega \right] \\
			\label{1_constraint_p_3}
			{\hspace{-0.8 cm}}&{\rm s.t.}& \tau_1 - T_{\max} \le 0, \\
			\label{2_constraint_p_3}
			{\hspace{-0.8 cm}}&{}&  \frac{C_{3,k}}{\log(1+\frac{\zeta_k}{\sigma^2})} + T^{\rm E} - \tau_1 \le 0, \forall k \in \mathcal{K}, \\
			\label{3_constraint_p_3}
			{\hspace{-0.8 cm}}&{}& \max_{k \in \mathcal{K}} \{T^{\rm L}_{k}\} + T^{\rm G} - \tau_1 \le 0, \\
			\label{4_constraint_p_3}
			{\hspace{-0.8 cm}}&{}& \zeta_{k} - C_{4,k} \le 0, \forall k \in \mathcal{K}, \\
			\label{5_constraint_p_3}
			{\hspace{-0.8 cm}}&{}& \omega - C_{5} \le 0, \\
			\label{6_constraint_p_3}
			{\hspace{-0.8 cm}}&{}& C_{6} \sqrt{\nu} - \sqrt{\omega} = 0, \\
			\label{7_constraint_p_3}
			{\hspace{-0.8 cm}}&{}& C_{7} \nu -2 \sqrt{\omega}\sqrt{\nu} + \omega + C_{8} \le 0,
		\end{eqnarray}
	\end{subequations}
\end{spacing}
\noindent where $C_{1,k}=D\bar{C}\theta_{k}/(|{\bf{v}}^{\rm H}_{k} {\bf{h}}^{\rm D}_{k}|^2 B), \forall k \in \mathcal{K}$, $C_{2,k}=\lceil Q/M \rceil T_s/|{\bf{b}}^{\rm H} {\bf{h}}^{\rm G}_{k}|^2, \forall k \in \mathcal{K}$, $C_{3,k}=D \bar{C} \theta_{k}/B, \forall k \in \mathcal{K}$, $C_{4,k}=p_{\max}|{\bf{v}}^{\rm H}_{k} {\bf{h}}^{\rm D}_{k}|^2, \forall k \in \mathcal{K}$, $C_{5}=p_{\max} \min_{k \in \mathcal{K}} |{\bf{b}}^{\rm H} {\bf{h}}^{\rm G}_{k}|^2$, $C_{6}=\epsilon_1$, $C_{7}=1-K\epsilon_2$, $C_{8}=K\sigma^2/2$, and $\tau_1\ge T^{\rm ALL}$ is an auxiliary variable.
Problem (\ref{p_3}) is non-convex because of the existence of the concave term ${\zeta_k}/{\log(1+{\zeta_k}/{\sigma^2})}$.
To make problem (\ref{p_3}) tractable, we further decompose it into two subproblems: one regarding the normalizing factors $\{\zeta_{k}\}$ and $\tau_1$, and the other regarding the normalizing factors $\nu$ and $\omega$.

By plugging in $\sqrt{\omega}=C_{6}\sqrt{\nu}$ based on constraint (\ref{6_constraint_p_3}), the subproblem regarding $\nu$ and $\omega$ is reduced to
\begin{subequations}
	\label{p_4}
	\begin{eqnarray}
		\label{objective_p_4}
		&\mathop {\min }\limits_{\nu}&  \left(\sum\nolimits_{k=1}^{K} C_{2,k}\right)C^2_{6} \nu \\
		\label{1_constraint_p_4}
		{\hspace{-0.8 cm}}&{\rm s.t.}& - \frac{C_{8}}{C_{7}-2C_{6}+C^2_{6}} \le \nu \le \frac{C_{5}}{C_{6}}.
	\end{eqnarray}
\end{subequations}
Since $(\sum_{k=1}^{K} C_{2,k})C^2_{6} \ge 0$, the closed-form solution to problem (\ref{p_4}) is given by
\vspace{-0.2 cm}
\begin{align}
	\label{solution_nu}
	\nu^{*}=-\frac{C_{8}}{C_{7}-2C_{6}+C^2_{6}}.
\end{align}
Correspondingly, the closed-form solution to $\omega$ is given by
\begin{align}
	\label{solution_omega}
	\omega^{*}=-\frac{C_{6}^2 C_{8}}{C_{7}-2C_{6}+C^2_{6}}.
\end{align}

On the other hand, for notation convenience, define two functions: $h_{1,k}(\zeta_k)={\zeta_k}/\log(1+\zeta_k/\sigma^2),\forall k \in \mathcal{K}$ and $h_{2,k}(\tau_1)=\sigma^2 (2^{C_{3,k}/(\tau_1-T^{\rm E})}-1),\forall k \in \mathcal{K}$.
Then, the subproblem regarding $\{\zeta_{k}\}$ and $\tau_1$ is given by
\vspace{-0.2 cm}
\begin{subequations}
	\label{p_5}
	\begin{eqnarray}
		\label{objective_p_5}
		&\mathop {\min }\limits_{\{\zeta_{k}\},\tau_1}&  \sum\nolimits_{k=1}^{K} C_{1,k} h_{1,k}(\zeta_k) \\
		\label{1_constraint_p_5}
		{\hspace{-0.8 cm}}&{\rm s.t.}&\max_{k \in \mathcal{K}}\{T^{\rm L}_{k}\} + T^{\rm G} \le \tau_1 \le T_{\max}, \\
		\label{2_constraint_p_5}
		{\hspace{-0.8 cm}}&{}& h_{2,k}(\tau_1) \le \zeta_k \le C_{4,k}, \forall k \in \mathcal{K},
	\end{eqnarray}
\end{subequations}
which is non-convex because of the non-convexity of (\ref{objective_p_5}).
However, it is noticed that $h_{1,k}(\zeta_k)$ is a monotonically increasing univariate concave function of $\zeta_k$ whose minimum can be attained at the boundary of the feasible region~\cite{Boyd2004Convex}.
Furthermore, since $h_{2,k}(\tau_1)$ decreases as $\tau_1$ increases, $\tau_1$ should be maximized within the feasible region of problem (\ref{p_5}) to minimize $h_{2,k}(\tau_1)$.
Thus, the optimal $\tau_1$ is given by
\vspace{-0.2 cm}
\begin{align}
	\label{solution_tau_1}
	&\tau_1^{*} = T_{\max}.
\end{align}
Then, to minimize (\ref{objective_p_5}), one should minimize $\zeta_k$ within the feasible region.
Hence, the closed-form optimal normalizing factors $\{\zeta_k\}$ are given by
\vspace{-0.2 cm}
\begin{align}
	\label{solution_zeta}
	&\zeta^{*}_k = h_{2,k}(T_{\max}), \forall k \in \mathcal{K}.
\end{align}

\begin{figure*}
	\begin{align}
		\label{define_H_DU}
		&{\bf{H}}^{\rm D}_{k} = \left[ {\begin{array}{*{20}{c}}
				{{\rm Re}\{{\bf{h}}_{k}^{{\rm{D}}}\} {\rm Re}\{{\bf{h}}_{k}^{{\rm{D}}}\}^{\rm T}+{\rm Im}\{{\bf{h}}_{k}^{{\rm{D}}}\} {\rm Im}\{{\bf{h}}_{k}^{{\rm{D}}}\}^{\rm T} } 
				&{{\rm Re}\{{\bf{h}}_{k}^{{\rm{D}}}\}{\rm Im}\{{\bf{h}}_{k}^{{\rm{D}}}\}^{\rm T}-{\rm Im}\{{\bf{h}}_{k}^{{\rm{D}}}\} {\rm Re}\{{\bf{h}}_{k}^{{\rm{D}}}\}^{\rm T} }\\
				{{\rm Im}\{{\bf{h}}_{k}^{{\rm{D}}}\}{\rm Re}\{{\bf{h}}_{k}^{{\rm{D}}}\}^{\rm T}-{\rm Re}\{{\bf{h}}_{k}^{{\rm{D}}}\} {\rm Im}\{{\bf{h}}_{k}^{{\rm{D}}}\}^{\rm T} } 
				&{{\rm Re}\{{\bf{h}}_{k}^{{\rm{D}}}\} {\rm Re}\{{\bf{h}}_{k}^{{\rm{D}}}\}^{\rm T}+{\rm Im}\{{\bf{h}}_{k}^{{\rm{D}}}\} {\rm Im}\{{\bf{h}}_{k}^{{\rm{D}}}\}^{\rm T} }
		\end{array}} \right], \forall k \in \mathcal{K}, \\
		\label{define_H_GU}
		&{\bf{H}}^{\rm G}_{k} = \left[  {\begin{array}{*{20}{c}}
				{{\rm Re}\{{\bf{h}}_{k}^{{\rm{G}}}\} {\rm Re}\{{\bf{h}}^{\rm G}_{k}\}^{\rm T}+{\rm Im}\{{\bf{h}}^{\rm G}_{k}\} {\rm Im}\{{\bf{h}}^{\rm G}_{k}\}^{\rm T} } 
				&{{\rm Re}\{{\bf{h}}_{k}^{{\rm{G}}}\} {\rm Im}\{{\bf{h}}^{\rm G}_{k}\}^{\rm T}-{\rm Im}\{{\bf{h}}^{\rm G}_{k}\} {\rm Re}\{{\bf{h}}^{\rm G}_{k}\}^{\rm T} }\\
				{{\rm Im}\{{\bf{h}}_{k}^{{\rm{G}}}\} {\rm Re}\{{\bf{h}}^{\rm G}_{k}\}^{\rm T}-{\rm Re}\{{\bf{h}}^{\rm G}_{k}\} {\rm Im}\{{\bf{h}}^{\rm G}_{k}\}^{\rm T} } 
				&{{\rm Re}\{{\bf{h}}_{k}^{{\rm{G}}}\} {\rm Re}\{{\bf{h}}^{\rm G}_{k}\}^{\rm T}+{\rm Im}\{{\bf{h}}^{\rm G}_{k}\} {\rm Im}\{{\bf{h}}^{\rm G}_{k}\}^{\rm T} }
		\end{array}} \right], \forall k \in \mathcal{K}.
	\end{align}
	\hrule
	\vspace{-0.4 cm}
\end{figure*}	

\subsubsection{Receive Beamformers}
\label{receive_beamformers_NS}
For notation convenience, define two rank-one matrices: ${\bf{V}}_k=[{\rm Re}\{{\bf{v}}_{k}\}^{\rm T},{\rm Im}\{{\bf{v}}_{k}\}^{\rm T}]^{\rm T}[{\rm Re}\{{\bf{v}}_{k}\}^{\rm T},{\rm Im}\{{\bf{v}}_{k}\}^{\rm T}], \forall k \in \mathcal{K}$ and ${\bf{B}}=[{\rm Re}\{{\bf{b}}\}^{\rm T},{\rm Im}\{{\bf{b}}\}^{\rm T}]^{\rm T}[{\rm Re}\{{\bf{b}}\}^{\rm T},{\rm Im}\{{\bf{b}}\}^{\rm T}]$, where ${\rm Im}\{\cdot\}$ takes the imaginary part.
%
Given normalizing and power scaling factors, CPU frequencies, and data allocation, the subproblem regarding receive beamformers can be transformed to the following semidefinite programming problem:
%
	\begin{subequations}
		\label{p_6}
		\begin{eqnarray}
			\label{objective_p_6}
			{\hspace{-0.8 cm}}&\mathop {\min }\limits_{\{{\bf{V}}_{k}\},{\bf{B}}} {\hspace{-0.2 cm}} &  \sum\nolimits_{k=1}^{K} \frac{C_{9,k}}{{\tr}({\bf{V}}_{k} {\bf{H}}^{\rm D}_{k})} +\sum\nolimits_{k=1}^{K} \frac{C_{10}}{{\tr}({\bf{B}} {\bf{H}}^{\rm G}_{k})} \\
			\label{1_constraint_p_6}
			{\hspace{-0.8 cm}}&{\rm s.t.} {\hspace{-0.2 cm}} &-p_{\max} {\tr}({\bf{V}}_{k} {\bf{H}}^{\rm D}_{k}) + \zeta_{k} \le 0, \forall k \in \mathcal{K}, \\
			\label{2_constraint_p_6}
			{\hspace{-0.8 cm}}&{}& -p_{\max} {\tr}({\bf{B}} {\bf{H}}^{\rm G}_{k}) + \omega \le 0, \forall k \in \mathcal{K}, \\
			\label{3_constraint_p_6}
			{\hspace{-0.8 cm}}&{}& {\tr}({\bf{V}}_{k})=1, \forall k \in \mathcal{K}, \\
			\label{4_constraint_p_6}
			{\hspace{-0.8 cm}}&{}& {\tr}({\bf{B}})=1, \\
			\label{5_constraint_p_6}
			{\hspace{-0.8 cm}}&{}& {\bf{V}}_{k} \succeq {\bf{0}}, \forall k \in \mathcal{K}, \\
			\label{6_constraint_p_6}
			{\hspace{-0.8 cm}}&{}& {\bf{B}} \succeq {\bf{0}}, \\
			\label{7_constraint_p_6}
			{\hspace{-0.8 cm}}&{}& {\rm rank}({\bf{V}}_{k}) = 1, \forall k \in \mathcal{K}, \\
			\label{8_constraint_p_6}
			{\hspace{-0.8 cm}}&{}& {\rm rank}({\bf{B}}) = 1,
		\end{eqnarray}
	\end{subequations}
%
\noindent where $C_{9,k}=D \bar{C} \theta_{k} \zeta_{k}/[B\log(1+\zeta_k/\sigma^2)],\forall k \in \mathcal{K}$, $C_{10}=\lceil Q/M \rceil T_s \omega $, and $\tr(\cdot)$ denotes the trace.
Moreover, matrices ${\bf{H}}^{\rm D}_{k}$ and ${\bf{H}}^{\rm G}_{k}$ are defined in (\ref{define_H_DU}) and (\ref{define_H_GU}), respectively.
However, problem (\ref{p_6}) is non-convex due to the rank-one constraints (\ref{7_constraint_p_6}) and (\ref{8_constraint_p_6}).

We employ a difference-of-convex-functions (DC) programming method to address the non-convexity of the rank-one constraints~\cite{Yang2020Federated}.
Specifically, constraints (\ref{7_constraint_p_6}) and (\ref{8_constraint_p_6}) are equivalent to (\ref{V_rank_substitution}) and (\ref{B_rank_substitution}), respectively, given by
\begin{align}
	\label{V_rank_substitution}
	&\tr({\bf{V}}_{k}) - \|{\bf{V}}_{k}\|_2 = 0, \forall k \in \mathcal{K},\\
	\label{B_rank_substitution}
	&\tr({\bf{B}}) - \|{\bf{B}}\|_2 = 0,
\end{align}
where $\|\cdot\|_2$ denotes the matrix $2$-norm.
Both (\ref{V_rank_substitution}) and (\ref{B_rank_substitution}) are still non-convex due to $-\|{\bf{V}}_{k}\|_2$ and $-\|{\bf{B}}\|_2$.
We substitute $\|{\bf{V}}_{k}\|_2$ and $\|{\bf{B}}\|_2$ with their linearizations $\|{\bf{V}}^{(n_1)}_{k}\|_2+\tr(({\bf{V}}_{k}-{\bf{V}}^{(n_1)}_{k})^{\rm T}\dot{\bf{V}}^{(n_1)}_{k})$ and $\|{\bf{B}}^{(n_1)}\|_2+\tr(({\bf{B}}-{\bf{B}}^{(n_1)})^{\rm T}\dot{\bf{B}}^{(n_1)})$, respectively.
Here, ${\bf{V}}^{(n_1)}_{k}$ and ${\bf{B}}^{(n_1)}$ are obtained in the $n_1$-th DC iteration, while $\dot{\bf{V}}^{(n_1)}_{k}$ and $\dot{\bf{B}}^{(n_1)}$ denote the subgradients obtained in the $n_1$-th DC iteration.
Then, by substituting (\ref{7_constraint_p_6}) and (\ref{8_constraint_p_6}) with their linearizations, we add them to objective (\ref{objective_p_6}) as regularizers  to convexify problem (\ref{p_6}) as follows:
\begin{subequations}
	\label{p_7}
	\begin{eqnarray}
		\label{objective_p_7}
		{\hspace{-0.8 cm}}&\mathop {\min }\limits_{\{{\bf{V}}_{k}\},{\bf{B}}} {\hspace{-0.2 cm}} &  \sum\nolimits_{k=1}^{K} \frac{C_{9,k}}{{\tr}({\bf{V}}_{k} {\bf{H}}^{\rm D}_{k})} +\sum\nolimits_{k=1}^{K} \frac{C_{10}}{{\tr}({\bf{B}} {\bf{H}}^{\rm G}_{k})} \notag \\
		&{}&+\sum\nolimits_{k=1}^{K} \beta [\tr({\bf{V}}_{k})-\tr({\bf{V}}_{k}^{\rm T} \dot{\bf{V}}^{(n_1)}_{k})] \notag \\ 
		&{}&+ \beta [\tr({\bf{B}})-\tr({\bf{B}}^{\rm T}\dot{\bf{B}}^{(n_1)})] \\
		\label{1_constraint_p_7}
		{\hspace{-0.8 cm}}&{\rm s.t.} {\hspace{-0.2 cm}} & \text{(\ref{1_constraint_p_6})~--~(\ref{6_constraint_p_6})},
	\end{eqnarray}
\end{subequations}
where $\beta > 0$ denotes the penalty factor.
In view of the convexity of problem (\ref{p_7}), it can be solved using standard optimization toolkits like CVX~\cite{Grant2014CVX}.
The DC-based algorithm for solving problem (\ref{p_6}) is summarized in Algorithm~\ref{algorithm_1}.

\begin{algorithm}[t]
	\caption{A DC-Based Algorithm for Optimizing Receive Beamformers}
	\label{algorithm_1}
	\begin{algorithmic}[1]
		\STATE \textbf{Input: } Feasible receive beamformers $(\{{\bf{v}}^{(0)}_{k}\},{\bf{b}}^{(0)})$, the maximum numbers of iterations $N_1$, and $n_1=0$.
		\STATE Calculate $\{{\bf{V}}^{(0)}_{k}\}$ and ${\bf{B}}^{(0)}$.
		\REPEAT
		\STATE Update $n_1 \leftarrow n_1+1$.
		\STATE Calculate subgradients $\{\dot{\bf{V}}^{(n_1-1)}_{k}\}$ and $\dot{\bf{B}}^{(n_1-1)}$ using $\{{\bf{V}}^{(n_1-1)}_{k}\}$ and ${\bf{B}}^{(n_1-1)}$, respectively.
		\STATE Obtain $\{{\bf{V}}^{(n_1)}_{k}\}$ and ${\bf{B}}^{(n_1)}$ by solving problem (\ref{p_7}).
		\UNTIL The objective (\ref{objective_p_6}) converges or $n_1 \ge N_1$.
		\STATE Recover $\{{\bf{v}}^{*}_{k}\}$ and ${\bf{b}}^{*}$ based on $\{{\bf{V}}^{(n_1)}_{k}\}$ and ${\bf{B}}^{(n_1)}$, respectively.
		\STATE \textbf{Output: }Optimized receive beamformers $\{{\bf{v}}^{*}_{k}\}$ and ${\bf{b}}^{*}$.
	\end{algorithmic}
\end{algorithm}

\subsubsection{CPU Frequencies}
\label{CPU_frequencies_NS}
Given normalizing and power scaling factors, receive beamformers, and data allocation, by introducing an auxiliary variable $\tau_2 \ge T^{\rm ALL}$, the subproblem regarding CPU frequencies are given by
\begin{subequations}
	\label{p_8}
	\begin{eqnarray}
		\label{objective_p_8}
		{\hspace{-0.8 cm}}&\mathop {\min }\limits_{\{\hat{f}_{k}\},\tilde{f},\tau_2} {\hspace{-0.2 cm}} & \sum\nolimits_{k=1}^{K} C_{11,k} \hat{f}_{k}^2 + C_{12} \tilde{f}^2 \\
		\label{1_constraint_p_8}
		{\hspace{-0.8 cm}}&{\rm s.t.} {\hspace{-0.2 cm}} & \tau_2 - T_{\max} \le 0, \\
		\label{2_constraint_p_8}
		{\hspace{-0.8 cm}}&{} & \frac{C_{13,k}}{\hat{f}_{k}} - \tau_2 + T^{\rm G} \le 0, \forall k \in \mathcal{K}, \\
		\label{3_constraint_p_8}
		{\hspace{-0.8 cm}}&{} & \frac{C_{14}}{\tilde{f}} - \tau_2 + \max_{k \in \mathcal{K}} \{T^{\rm D}_{k}\} \le 0, \\
		\label{4_constraint_p_8}
		{\hspace{-0.8 cm}}&{} & \text{(\ref{device_CPU_constraint_p_1})~and~(\ref{BS_CPU_constraint_p_1})},
	\end{eqnarray}
\end{subequations}
where $C_{11,k}=C_{13,k}\hat{\kappa}, \forall k \in \mathcal{K}$, $C_{12}=C_{14}\tilde{\kappa}$, $C_{13,k}=D(1-\theta_k)\hat{C}_k, \forall k \in \mathcal{K}$, and $C_{14}=D\tilde{C}\sum\nolimits_{k=1}^{K} \theta_k$.
Problem (\ref{p_8}) is jointly convex with respect to $\{\hat{f}_k\}$, $\tilde{f}$, and $\tau_2$.
Hence, the closed-form solutions are derived by solving its Karush-Kuhn-Tucker (KKT) conditions, as presented in Lemma~\ref{lemma_1}.
\begin{lemma}
	\label{lemma_1}
	By solving the KKT conditions, the closed-form solution to problem (\ref{p_8}) is given by
	\begin{align}
		\label{solution_hat_f}
		&\hat{f}_{k}^{*}=\frac{C_{13,k}}{T_{\max}-T^{\rm G}}, \forall k \in \mathcal{K},\\
		\label{solution_tilde_f}
		&\tilde{f}^{*}=\frac{C_{14}}{T_{\max}-\max_{k \in \mathcal{K}}\{T^{\rm D}_{k}\}},\\
		\label{solution_tau_2}
		&\tau_2^{*} = T_{\max}.
	\end{align}
\end{lemma}
\begin{IEEEproof}
	Please refer to Appendix~\ref{proof_of_lemma_1} in~\cite{Zheng2024Appendix}.
\end{IEEEproof}

\subsubsection{Data Allocation}
\label{data_allocation_NS}
Given normalizing and power scaling factors, receive beamformers, and CPU frequencies, by introducing an auxiliary variable $\tau_3 \ge T^{\rm ALL}$, the subproblem regarding data allocation is reduced to
\begin{subequations}
	\label{p_9}
	\begin{eqnarray}
		\label{objective_p_9}
		{\hspace{-0.8 cm}}&\mathop {\min }\limits_{\{\theta_{k}\},\tau_3} {\hspace{-0.2 cm}} & \sum\nolimits_{k=1}^{K} C_{15,k} \theta_k \\
		\label{1_constraint_p_9}
		{\hspace{-0.8 cm}}&{\rm s.t.} {\hspace{-0.2 cm}} & \tau_3 - T_{\max} \le 0, \\
		\label{2_constraint_p_9}
		{\hspace{-0.8 cm}}&{} & C_{16,k}\theta_k + C_{17}\sum\nolimits_{k'=1}^{K} \theta_{k'} - \tau_3 \le 0, \forall k \in \mathcal{K}, \\
		\label{3_constraint_p_9}
		{\hspace{-0.8 cm}}&{} & -C_{18,k}\theta_k-\tau_3+T^{\rm G}+C_{18,k} \le 0, \forall k \in \mathcal{K}, \\
		\label{4_constraint_p_9}
		{\hspace{-0.8 cm}}&{} & 0 \le \theta_k \le C_{19}, \forall k \in \mathcal{K}, 
	\end{eqnarray}
\end{subequations}
where $C_{15,k}=\zeta_k C_{16,k}/|{\bf{v}}_{k}^{\rm H}{\bf{h}}^{\rm D}_{k}|^2-D\hat{C}_{k}\hat{\kappa}\hat{f}_{k}^2+D\tilde{C}\tilde{\kappa}\tilde{f}^2, \forall k \in \mathcal{K}$, $C_{16,k}=D\bar{C}/[B \log(1+\zeta_k/\sigma^2)], \forall k \in \mathcal{K}$, $C_{17}=\tilde{C} D/\tilde{f}$, $C_{18,k}=\hat{C}_{k} D/\hat{f}_{k}, \forall k \in \mathcal{K}$, and $C_{19}=\theta_{\max}$.
Problem (\ref{p_9}) is a linear programming problem regarding $\{\theta_k\}$ and $\tau_3$, which can be effectively solved using CVX.

\begin{algorithm}[t]
	\caption{Proposed Algorithm for Solving Problem (\ref{p_1})}
	\label{algorithm_2}
	\begin{algorithmic}[1]
		\STATE \textbf{Input: } A feasible solution $(\{\zeta_{k}^{(0)}\}$, $\nu^{(0)}$, $\omega^{(0)}$, ${\bf{b}}^{(0)}$, $\{{\bf{v}}^{(0)}_{k}\}$, $\{\hat{f}_{k}^{(0)}\}$, $\tilde{f}^{(0)},\{\theta_{k}^{(0)}\})$, the maximum numbers of iterations $N_2$, and $n_2=0$.
		\REPEAT
		\STATE Update $n_2 \leftarrow n_2+1$.
		\STATE Given ${\bf{b}}^{(n_2-1)}$, $\{{\bf{v}}^{(n_2-1)}_{k}\}$, $\{\hat{f}_{k}^{(n_2-1)}\}$, $\tilde{f}^{(n_2-1)}$, $\{\theta_{k}^{(n_2-1)}\}$, calculate $\nu^{(n_2)}$, $\omega^{(n_2)}$, and $\{\zeta^{(n_2)}_{k}\}$ via (\ref{solution_nu}), (\ref{solution_omega}), and (\ref{solution_zeta}), respectively.
		\STATE Given $\{\zeta_{k}^{(n_2)}\}$, $\nu^{(n_2)}$, $\omega^{(n_2)}$, $\{\hat{f}_{k}^{(n_2-1)}\}$, $\tilde{f}^{(n_2-1)}$, $\{\theta_{k}^{(n_2-1)}\!\}$, obtain ${\bf{b}}^{(n_2)}$ and $\{\!{\bf{v}}^{(n_2)}_{k}\!\}$ by using Algorithm~\ref{algorithm_1}.
		\STATE Given $\{\zeta_{k}^{(n_2)}\}$, $\nu^{(n_2)}$, $\omega^{(n_2)}$, ${\bf{b}}^{(n_2)}$, $\{{\bf{v}}^{(n_2)}_{k}\}$, $\{\theta_{k}^{(n_2-1)}\}$, calculate $\{\hat{f}_k^{(n_2)}\}$ and $\tilde{f}^{(n_2)}$ via (\ref{solution_hat_f}) and (\ref{solution_tilde_f}), respectively.
		\STATE Given $\{\zeta_{k}^{(n_2)}\}$, $\nu^{(n_2)}$, $\omega^{(n_2)}$, ${\bf{b}}^{(n_2)}$, $\{{\bf{v}}^{(n_2)}_{k}\}$, $\{\hat{f}_{k}^{(n_2)}\}$, $\tilde{f}^{(n_2)}$, obtain $\{\theta_{k}^{(n_2)}\}$ by solving problem (\ref{p_9}).
		\UNTIL The objective (\ref{objective_p_1}) converges or $n_2 \ge N_2$.
		\STATE \textbf{Output: }The optimized solution $(\{\zeta_{k}^{(n_2)}\}$, $\nu^{(n_2)}$, $\omega^{(n_2)}$, ${\bf{b}}^{(n_2)}$, $\{{\bf{v}}^{(n_2)}_{k}\}$, $\{\hat{f}_{k}^{(n_2)}\}$, $\tilde{f}^{(n_2)}$, $\{\theta_{k}^{(n_2)}\})$.
	\end{algorithmic}
\end{algorithm}

\subsubsection{Overall Algorithm for Non-Stable Region}
\label{overall_algorithm_for_non_stable_region}
The overall algorithm proposed for solving problem (\ref{p_1}) in the non-stable region $\mathcal{R}^{\rm NS}$ is summarized in Algorithm~\ref{algorithm_2}, where a variable with the superscript $(n_2)$ denotes its value obtained at the $n_2$-th iteration.
The solutions to subproblems (\ref{p_3}), (\ref{p_6}), (\ref{p_8}), and (\ref{p_9}) create a series of non-increasing objective values for (\ref{objective_p_1}).
Meanwhile, as the objective (\ref{objective_p_1}) essentially represents energy consumption, it is naturally lower bounded by zero.
Therefore, the convergence of Algorithm~\ref{algorithm_2} is ensured.
The main complexity of Algorithm~\ref{algorithm_2} is given by $\mathcal{O} ((K+2)N_2+(K+1)N_r^{4.5}N_1N_2+(k+1)N_2+(K+1)^3N_2)$.
Specifically, when invoking CVX to solve a convex problem, the standard interior-point method is adopted.
Thus, the complexity of optimizing the normalizing and power scaling factors, i.e., $\nu_t$, $\omega$, and $\{\zeta_k\}$, is $\mathcal{O}(K+2)$.
The complexity of performing Algorithm~\ref{algorithm_1} is $\mathcal{O}((K+1)N_r^{4.5}N_1)$~\cite{Luo2010Semidefinite}.
The complexities of optimizing CPU frequencies, i.e., $\{\hat{f}_k\}$ and $\tilde{f}$, and data allocation, i.e., $\{\theta_{k}\}$, can be given by $\mathcal{O} (K+1)$ and $\mathcal{O} ((K+1)^3)$~\cite{Tian2024Two}, respectively.

\vspace{-0.2 cm}
\subsection{Algorithm for Stable Region}
\label{algorithm_for_stable_region}
We address problem (\ref{p_2}) in the stable region $\mathcal{R}^{\rm S}$ by decoupling it into four subproblems as well.
The specific subproblems and their solutions are presented as follows:
\subsubsection{Normalizing and Power Scaling Factors}
\label{normalizing_and_power_scaling_factors_S}
Given receive beamformers, CPU frequencies, and data allocation, by introducing an auxiliary variable $\tau_4 \ge T^{\rm ALL}$, the subproblem regarding normalizing and power scaling factors is reduced to$\!\!\!\!$
%
\begin{subequations}
	\label{p_10}
	\begin{eqnarray}
		\label{objective_p_10}
		{\hspace{-0.8 cm}}&\mathop {\min }\limits_{\{{\zeta}_{k}\},\nu,\omega,\tau_4} & \sum\limits_{k=1}^{K} \left[C_{1,k} \frac{\zeta_k}{\log(1+\frac{\zeta_k}{\sigma^2})} + C_{2,k} \omega \right] \\
		\label{1_constraint_p_10}
		{\hspace{-0.8 cm}}&{\rm s.t.}& \tau_4 - T_{\max} \le 0, \\
		\label{2_constraint_p_10}
		{\hspace{-0.8 cm}}&{}&  \frac{C_{3,k}}{\log(1+\frac{\zeta_k}{\sigma^2})} + T^{\rm E} - \tau_4 \le 0, \forall k \in \mathcal{K}, \\
		\label{3_constraint_p_10}
		{\hspace{-0.8 cm}}&{}& \max_{k \in \mathcal{K}} \{T^{\rm L}_{k}\} + T^{\rm G} - \tau_4 \le 0, \\
		\label{4_constraint_p_10}
		{\hspace{-0.8 cm}}&{}& C_{20} \nu + C_{21} \le 0, \\
		\label{5_constraint_p_10}
		{\hspace{-0.8 cm}}&{}& C_{22} \nu -2 \sqrt{\omega}\sqrt{\nu} + \omega + C_{8} \le 0, \\
		\label{6_constraint_p_10}
		{\hspace{-0.8 cm}}&{}& \sqrt{\omega}=\sqrt{\nu}, \\
		\label{7_constraint_p_10}
		{\hspace{-0.8 cm}}&{}& \text{(\ref{4_constraint_p_3}),~and~(\ref{5_constraint_p_3})},
	\end{eqnarray}
\end{subequations}
where $C_{20}=A^2-\epsilon_3\mu(4\mu-L)/L$, $C_{21}=Q\sigma^2/2$, and $C_{22}=1-\epsilon_4 K$.
Problem (\ref{p_10}) is non-convex because of the non-convex objective (\ref{objective_p_10}) and constraint (\ref{5_constraint_p_10}).
We further decouple it into two subproblems: one involving $\nu$ and $\omega$, and the other involving $\{\zeta_{k}\}$ and $\tau_4$.

By plugging in constraint (\ref{6_constraint_p_10}), i.e., $\sqrt{\omega}=\sqrt{\nu}$, the subproblem regarding $\nu$ and $\omega$ degrades to a univariate linear programming problem of $\nu$, given by
\begin{subequations}
	\label{p_11}
	\begin{eqnarray}
		\label{objective_p_11}
		{\hspace{-0.8 cm}}&\mathop {\min }\limits_{\nu} & \left(\sum\limits_{k=1}^{K} C_{2,k} \right) \nu \\
		\label{1_constraint_p_11}
		{\hspace{-0.8 cm}}&{\rm s.t.}& \max\left\{-\frac{C_{21}}{C_{20}},\frac{C_{8}}{1-C_{22}}\right\} \le \nu \le C_{5}.
	\end{eqnarray}
\end{subequations}
Since $C_{2,k} \! \ge \! 0, \forall k \! \in \! \mathcal{K}$, the objective (\ref{p_11}) monotonously increases with $\nu$.
Thus, the optimal $\nu$ and $\omega$ are given by
\begin{align}
	\label{solution_nu_omega_S}
	\nu^{*}=\omega^{*}=\max\left\{-\frac{C_{21}}{C_{20}},\frac{C_{8}}{1-C_{22}}\right\}.
\end{align}

On the other hand, it can be verified that the subproblem involving $\{\zeta_k\}$ and $\tau_4$ has a form identical to problem (\ref{p_5}).
Hence, due to the problem similarity, the optimal $\{\zeta_k\}$ can be given by (\ref{solution_zeta}), while the optimal $\tau_4$ can be determined by
\begin{align}
	\label{solution_tau_4_S}
	\tau_4^{*} = T_{\max}.
\end{align}

\subsubsection{Receive Beamformers}
\label{receive_beamformers_S}
Given normalizing and power scaling factors, CPU frequencies, and data allocation, by using the matrices $\{{\bf{V}}_k\}$ and ${\bf{B}}$ defined in Section~\ref{receive_beamformers_NS}, one can find that the subproblem of receive beamformers in the stable region $\mathcal{R}^{\rm S}$ has the same form as problem (\ref{p_6}).
Due to the problem similarity, one can obtain the optimal receiver beamformers $\{{\bf{v}}^{*}_k\}$ and ${\bf{b}}^{*}$ by applying Algorithm~\ref{algorithm_1}.

\subsubsection{CPU Frequencies}
\label{CPU_frequencies_S}
Given normalizing and power scaling factors, receive beamformers, and data allocation, the subproblem of CPU frequencies $\{\hat{f}_k\}$ and $\tilde{f}$ is the same as problem (\ref{p_8}), which is convex.
Based on Lemma~\ref{lemma_1}, the optimal CPU frequencies $\{\hat{f}^{*}_k\}$ and $\tilde{f}^{*}$ in the stable region $\mathcal{R}^{\rm S}$ can be obtained using (\ref{solution_hat_f}) and (\ref{solution_tilde_f}), respectively.

\subsubsection{Data Allocation}
\label{data_allocation_S}
Given normalizing and power scaling factors, receive beamformers, and CPU frequencies, by introducing an auxiliary variable $\tau_5 \ge T^{\rm ALL}$, the subproblem of data allocation is reduced to
\begin{subequations}
	\label{p_12}
	\begin{eqnarray}
		\label{objective_p_12}
		{\hspace{-0.8 cm}}&\mathop {\min }\limits_{\{\theta_{k}\},\tau_5} {\hspace{-0.2 cm}} & \sum\limits_{k=1}^{K} C_{15,k} \theta_k \\
		\label{1_constraint_p_12}
		{\hspace{-0.8 cm}}&{\rm s.t.} {\hspace{-0.2 cm}} & \tau_5 - T_{\max} \le 0, \\
		\label{2_constraint_p_12}
		{\hspace{-0.8 cm}}&{} & C_{16,k}\theta_k + C_{17}\sum\limits_{k'=1}^{K} \theta_{k'} - \tau_5 \le 0, \forall k \in \mathcal{K}, \\
		\label{3_constraint_p_12}
		{\hspace{-0.8 cm}}&{} & -C_{18,k}\theta_k-\tau_5+T^{\rm G}+C_{18,k} \le 0, \forall k \in \mathcal{K}, \\
		\label{4_constraint_p_12}
		{\hspace{-0.8 cm}}&{} & C_{23} \le \theta_k \le 1, \forall k \in \mathcal{K}, 
	\end{eqnarray}
\end{subequations}
where $C_{23}=\theta_{\min}$.
Problem (\ref{p_12}) is a linear programming problem, which can be effectively solve using CVX.

\begin{algorithm}[t]
	\caption{Proposed Algorithm for Solving Problem (\ref{p_2})}
	\label{algorithm_3}
	\begin{algorithmic}[1]
		\STATE \textbf{Input: } A feasible solution $(\{\zeta_{k}^{(0)}\}$, $\nu^{(0)}$, $\omega^{(0)}$, ${\bf{b}}^{(0)}$, $\{{\bf{v}}^{(0)}_{k}\}$, $\{\hat{f}_{k}^{(0)}\}$, $\tilde{f}^{(0)},\{\theta_{k}^{(0)}\})$, the maximum numbers of iterations $N_3$, and $n_3=0$.
		\REPEAT
		\STATE Update $n_3 \leftarrow n_3+1$.
		\STATE Given ${\bf{b}}^{(n_3-1)}$, $\{{\bf{v}}^{(n_3-1)}_{k}\}$, $\{\hat{f}_{k}^{(n_3-1)}\}$, $\tilde{f}^{(n_3-1)}$, $\{\theta_{k}^{(n_3-1)}\}$, calculate $\nu^{(n_3)}$ and $\omega^{(n_3)}$ via (\ref{solution_nu_omega_S}), and calculate $\{\zeta^{(n_3)}_{k}\}$ via (\ref{solution_zeta}).
		\STATE Given $\{\zeta_{k}^{(n_3)}\}$, $\nu^{(n_3)}$, $\omega^{(n_3)}$, $\{\hat{f}_{k}^{(n_3-1)}\}$, $\tilde{f}^{(n_3-1)}$, $\{\theta_{k}^{(n_3-1)}\}$, obtain ${\bf{b}}^{(n_3)}$ and $\{{\bf{v}}^{(n_3)}_{k}\}$ by using Algorithm~\ref{algorithm_1}.
		\STATE Given $\{\zeta_{k}^{(n_3)}\}$, $\nu^{(n_3)}$, $\omega^{(n_3)}$, ${\bf{b}}^{(n_3)}$, $\{{\bf{v}}^{(n_3)}_{k}\}$, $\{\theta_{k}^{(n_3-1)}\}$, calculate $\{\hat{f}_k^{(n_3)}\}$ and $\tilde{f}^{(n_3)}$ via (\ref{solution_hat_f}) and (\ref{solution_tilde_f}), respectively.
		\STATE Given $\{\zeta_{k}^{(n_3)}\}$, $\nu^{(n_3)}$, $\omega^{(n_3)}$, ${\bf{b}}^{(n_3)}$, $\{{\bf{v}}^{(n_3)}_{k}\}$, $\{\hat{f}_{k}^{(n_3)}\}$, $\tilde{f}^{(n_3)}$, obtain $\{\theta_{k}^{(n_3)}\}$ by solving problem (\ref{p_12}).
		\UNTIL The objective (\ref{objective_p_2}) converges or $n_3 \ge N_3$.
		\STATE \textbf{Output: }The optimized solution $(\{\zeta_{k}^{(n_3)}\}$, $\nu^{(n_3)}$, $\omega^{(n_3)}$, ${\bf{b}}^{(n_3)}$, $\{{\bf{v}}^{(n_3)}_{k}\}$, $\{\hat{f}_{k}^{(n_3)}\}$, $\tilde{f}^{(n_3)}$, $\{\theta_{k}^{(n_3)}\})$.
	\end{algorithmic}
\end{algorithm}

\subsubsection{Overall Algorithm for Stable Region}
\label{overall_algorithm_for_stable_region}
The overall algorithm for solving problem (\ref{p_2}) in the stable region $\mathcal{R}^{\rm S}$ is summarized in Algorithm~\ref{algorithm_3}.
In Algorithm~\ref{algorithm_3}, a variable with the superscript $(n_3)$ refers to its value obtained in the $n_3$-th iteration.
The convergence of Algorithm~\ref{algorithm_3} can be analyzed similarly to that of Algorithm~\ref{algorithm_2}.
In addition, the complexity of Algorithm~\ref{algorithm_3} can be given by $\mathcal{O} ((K+2)N_3+(K+1)N_r^{4.5}N_1N_3+(k+1)N_3+(K+1)^3N_3)$.


\vspace{-0.2 cm}
\section{Simulation Results}
\label{simulation_results}

\begin{table}[t]
	\caption{Simulation Parameters}
	\centering
	\vspace{-0.2 cm}
	\begin{tabular}{|l|l|} \hline
		\textbf{\!\!Parameters} & \textbf{\!\!Value} \\ \hline
		\text{\!\!Maximum transmit power} & \makecell[l]{\!\!$\hat{p}_{\max}\!=\!23$ dBm} \\ \hline
		\text{\!\!Bandwidth for data uploading} & \makecell[l]{\!\!$B=10$ kHz} \\ \hline
		\text{\!\!Noise power} & \makecell[l]{\!\!$\sigma^2=-80$ dBm} \\ \hline
		\text{\!\!AirComp symbol duration\!\!} & \makecell[l]{\!\!$T_s=1$ ms} \\ \hline
		\text{\!\!Local dataset size} & \makecell[l]{\!\!$D=3\times10^{3}$} \\ \hline
		\text{\!\!Penalty factor} & \makecell[l]{\!\!$\beta=1$} \\ \hline
		\text{\makecell[l]{\!\!Number of symbols in an\\\!\!AirComp symbol}} & \makecell[l]{\!\!$M=14$} \\ \hline
		\text{\makecell[l]{\!\!Maximum CPU frequencies \\\!\!of the BS and devices}} & \makecell[l]{\!\!$\tilde{f}_{\max}=10$ GHz, \\\!\!$\hat{f}_{\max}=1$ GHz} \\ \hline
		\text{\makecell[l]{\!\!Number of CPU circles for\\\!\!processing a data sample}} & \makecell[l]{\!\!$\tilde{C}=1\times10^{8}$, \\\!\!$\hat{C}_k \! \in \! [1.5, 2.8] \!\! \times \!\! 10^{8}$} \\ \hline
		\text{\makecell[l]{\!\!Effective switched capacitance \\\!\!of the BS and devices}} & \makecell[l]{\!\!$\tilde{\kappa}=1\times10^{-28}$,\\\!\!$\hat{\kappa}=1\times10^{-28}$} \\ \hline
		\text{\makecell[l]{\!\!Maximum and minimum\\\!\!ratios of SL data}} & \makecell[l]{\!\!$\theta_{\max}=0.3$,\\\!\!$\theta_{\min}=0.2$\!\!} \\ \hline
	\end{tabular}
	\label{table_2}
	\vspace{-0.4 cm}
\end{table}

\vspace{-0.2 cm}
\subsection{Simulation Setup}
\label{simulation_setup}
We build a virtual urban square area of $100$ m $\times~100$ m for simulation, which models a real urban environment spanning from ${22.2823832}^\circ$N to ${22.2832841}^\circ$N latitude and ${114.1552437}^\circ$E to ${114.1562117}^\circ$E longitude on Earth.
There are $K=20$ single-antenna devices, represented by red icons, randomly distributed in the area, and the BS equipped with $N_r=16$ antennas is deployed at the center.
The height of the BS is $30$ m, and the height of each device is $1.5$ m.
We adopt a clustered delay line multiple-input multiple-output link-level fading channel model to generate the channel coefficient vectors for gradient aggregation and data uploading~\cite{3GPPStudy}, i.e., $\{{\bf{h}}^{\rm G}_{k}\}$ and $\{{\bf{h}}^{\rm D}_{k}\}$.
To verify learning performance, we train a MLP, a CNN, and a ResNet to classify the Fashion-MNIST~\cite{Xiao2017Fashion}, CIFAR-10, and CIFAR-100 datasets~\cite{Krizhevsky2009Learning}, respectively.
The MLP has $3$ fully-connected hidden layers, and the learning rate for training the MLP is set to $\eta=0.01$.
The CNN contains $3$ convolutional layers, $3$ max-pooling layers, and $2$ fully-connected layers, while the learning rate is set to $\eta=0.001$.
We adopt a ResNet~\cite{He2016ResNet} with customized network architecture to classify the CIFAR-100 dataset, whose architecture is demonstrated in the Appendix~\ref{architecture_of_the_adopted_ResNet} of~\cite{Zheng2024Appendix}.
The learning rate for training the ResNet is set to $\eta=0.001$.
Unless otherwise specified, other main simulation parameters are listed in Table~\ref{table_2}.
As specified in~\cite{Dahlman20205G}, a slot in the 5G NR systems typically contains $14$ symbols.
Moreover, prior work~\cite{Cao2022Transmission} have adopted AirComp blocks consisting of $14$ symbols.
Following these established configurations, we thus set $M=14$.
Most of the basic simulation parameters are determined based on our previous work~\cite{Zheng2024Retransmission} and other related studies on SemiFL~\cite{Ren2024Convergence,Huang2023Wireless,Liu2023Wireless}.
Other unique parameters specific to this work, such as the maximum and minimum ratios of SL data and the learning rate, are determined through conducting iterative pre-experiments.
Due to space limitations, additional key simulation results are provided in the Appendix~\ref{additional_simulation_results} of~\cite{Zheng2024Appendix}.

\begin{figure*}
	\begin{minipage}[t]{1 \textwidth}
		\centering
		\subfigure[Training MLP on the Fashion-MNIST dataset.]{
			\label{accuracy_vs_rounds_fig_5_MLP}
			\includegraphics[width=0.32 \textwidth]{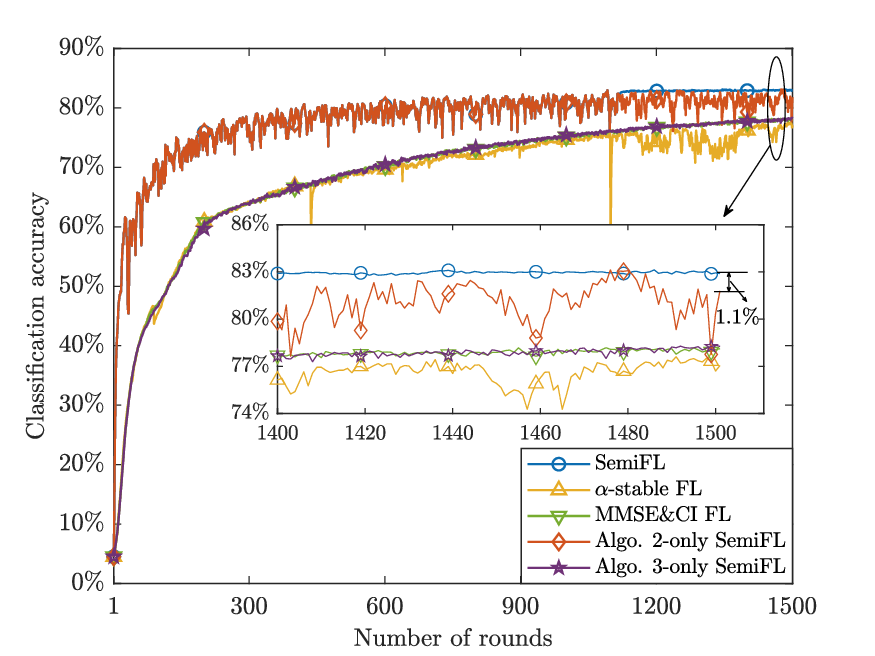}}
		\subfigure[Training CNN on the CIFAR-10 dataset.]{
			\label{accuracy_vs_rounds_fig_5_CNN}
			\includegraphics[width=0.32 \textwidth]{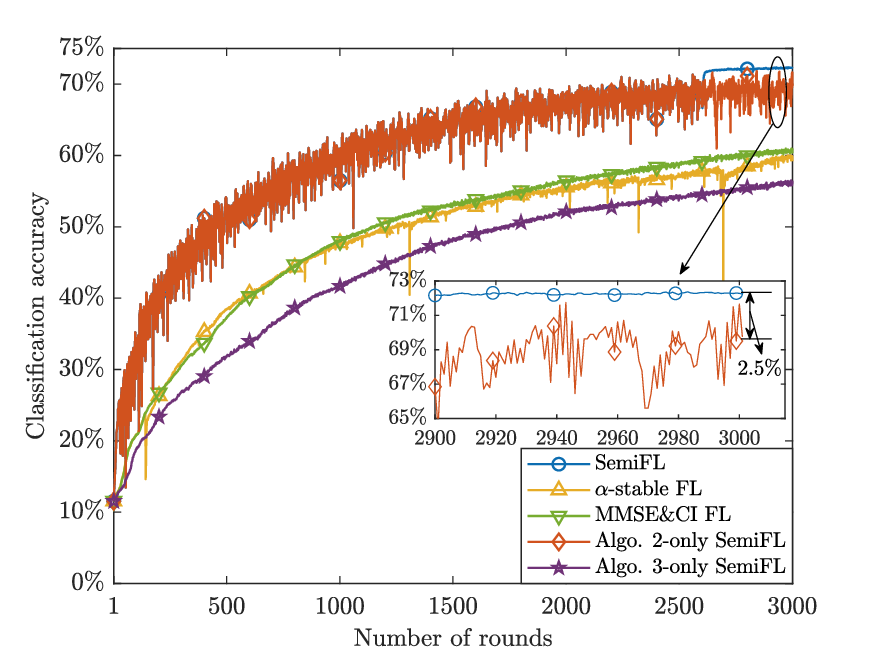}}
		\subfigure[Training ResNet on the CIFAR-100 dataset.]{
			\label{accuracy_vs_rounds_fig_5_ResNet}
			\includegraphics[width=0.32 \textwidth]{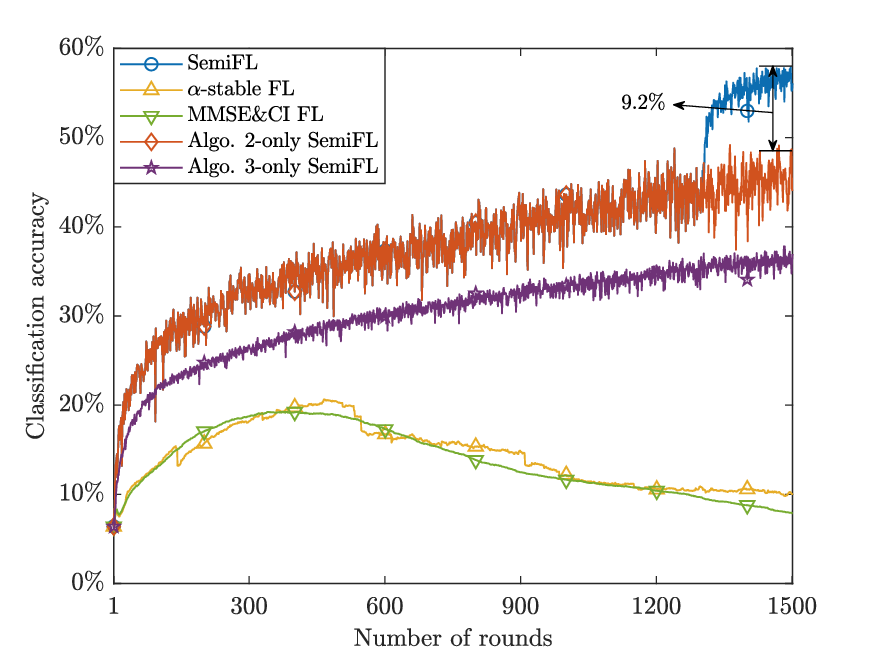}}
		\vspace{-0.3 cm}
		\caption{Learning performance comparison between SemiFL and benchmarks on the Fashion-MNIST, CIFAR-10, and CIFAR-100 datasets, where $\epsilon_1=10$, $\epsilon_2=1$, $\epsilon_3=0.8$, and $\epsilon_4=0.01$.}
		\label{accuracy_vs_rounds_fig_5}
		\vspace{-0.6 cm}
	\end{minipage}
\end{figure*}

\begin{figure*}
	\begin{minipage}[t]{1 \textwidth}
		\centering
		\subfigure[Training MLP on the Fashion-MNIST dataset.]{
			\label{accuracy_vs_rounds_fig_6_MLP}
			\includegraphics[width=0.32 \textwidth]{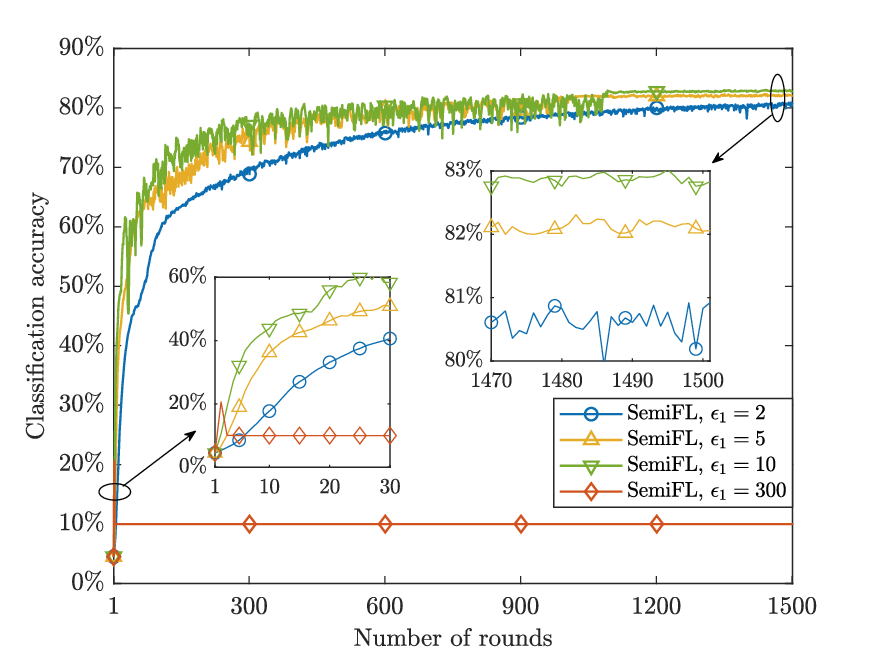}}
		\subfigure[Training CNN on the CIFAR-10 dataset.]{
			\label{accuracy_vs_rounds_fig_6_CNN}
			\includegraphics[width=0.32 \textwidth]{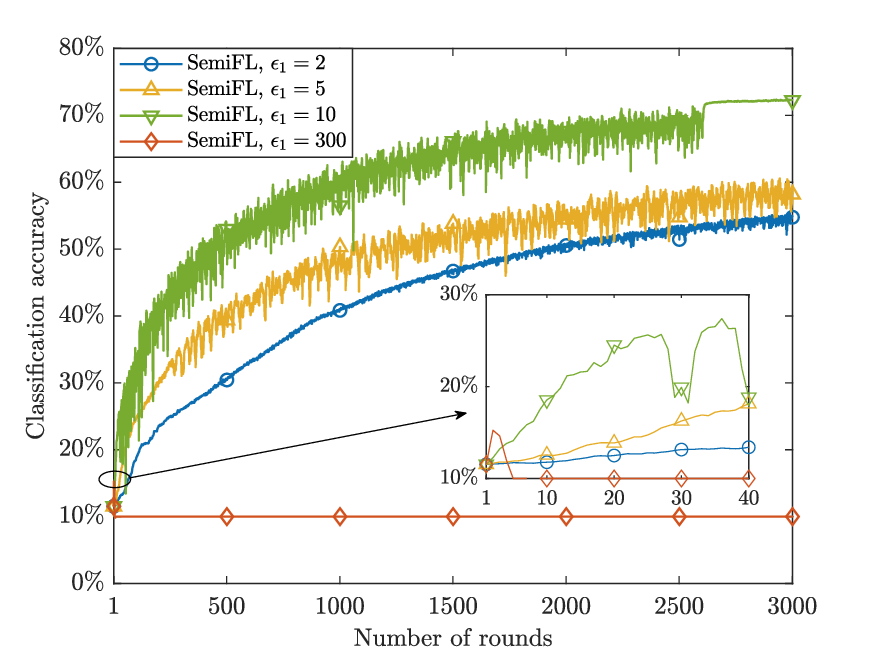}}
		\subfigure[Training ResNet on the CIFAR-100 dataset.]{
			\label{accuracy_vs_rounds_fig_6_ResNet}
			\includegraphics[width=0.32 \textwidth]{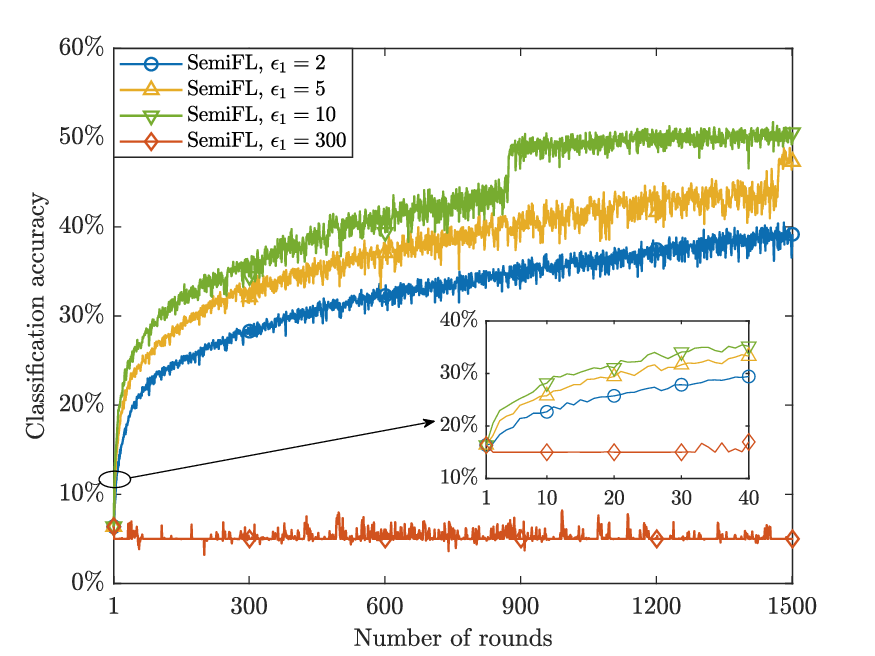}}
		\vspace{-0.3 cm}
		\caption{Learning performance comparison of the proposed SemiFL on the Fashion-MNIST, CIFAR-10, and CIFAR-100 datasets with different $\epsilon_1$ values, where $\epsilon_3\!=\!0.8$ and $\epsilon_4\!=\!0.01$. Note that we set $\epsilon_2\!=\!1$ when $\epsilon_1\!=\!2$ or $5$, and set $\epsilon_2\!=\!5$ when $\epsilon_1\!=\!10$. When $\epsilon_1\!=\!300$, a sufficiently large $\epsilon_2$ is adopted.}
		\label{accuracy_vs_rounds_fig_6}
		\vspace{-0.6 cm}
	\end{minipage}
\end{figure*}

\vspace{-0.3 cm}
\subsection{Learning Results and Analysis}
\label{simulation_results_and_analysis}
\begin{spacing}{1}
	To validate the effectiveness of over-the-air distortion in accelerating convergence of SemiFL, we consider the following state-of-the-art benchmarks for comparison:
	\begin{itemize}
		\item \textbf{FL with noise yielding $\alpha$-stable distribution ($\alpha$-stable FL)}~\cite{Yang2022Revisiting}: Devices upload only local gradients to the BS for aggregation, where the aggregation noise obeys an $\alpha$-stable distribution.
		This scheme is reported to improve model generalization.
		\item \textbf{FL with MMSE and CI (MMSE\&CI FL)}~\cite{Ni2022Federated,Liu2020Over}: All devices upload only local gradients to the BS for aggregation, where the variables are optimized using the CI\&MMSE scheme to suppress over-the-air distortion.
		\item \textbf{SemiFL with only Algorithm~\ref{algorithm_2} (Algo.~\ref{algorithm_2}-only SemiFL)}: Only Algorithm~\ref{algorithm_2} is applied in both the non-stable region $\mathcal{R}^{\rm NS}$ and the stable region $\mathcal{R}^{\rm S}$.
		\item \textbf{SemiFL with only Algorithm~\ref{algorithm_3} (Algo.~\ref{algorithm_3}-only SemiFL)}: Only Algorithm~\ref{algorithm_3} is applied in both the non-stable region $\mathcal{R}^{\rm NS}$ and the stable region $\mathcal{R}^{\rm S}$.
	\end{itemize}

	\vspace{-0.2 cm}
	Fig.~\ref{accuracy_vs_rounds_fig_5} shows the learning performance comparisons between the proposed SemiFL and state-of-the-art benchmarks on the Fashion-MNIST, CIFAR-10, and CIFAR-100 datasets.
	In Figs.~\ref{accuracy_vs_rounds_fig_5_MLP},~\ref{accuracy_vs_rounds_fig_5_CNN},~and~\ref{accuracy_vs_rounds_fig_5_ResNet}, it is seen that the convergence of SemiFL is significantly accelerated by leveraging over-the-air distortion, achieving faster convergence than benchmarks.
	Upon reaching the stable region, SemiFL exhibits stable and improved final convergence by invoking Algorithm~\ref{algorithm_3}, obtaining accuracy gains of $1.1\%$, $2.5\%$, and $9.2\%$ on the Fashion-MNIST, CIFAR-10, and CIFAR-100 datasets, respectively, compared to Algo.~\ref{algorithm_2}-only SemiFL.
	Meanwhile, it is also seen that using either Algorithm~\ref{algorithm_2} or Algorithm~\ref{algorithm_3} alone results in inferior convergence than the combined usage of them, verifying the effectiveness of the two-region MSE threshold configuration outlined in Remark~\ref{remark_4}.
	Additionally, it is observed that SemiFL with proposed algorithms outperforms the widely adopted MMSE\&CI FL scheme in all cases.
	This confirms that the conventional distortion-suppressing criterion for AirComp-based gradient aggregation is overly conservative.
	Moreover, Fig.~\ref{accuracy_vs_rounds_fig_5_ResNet} shows that both the MMSE\&CI FL and $\alpha$-stable FL fail to train the complex ResNet, whereas SemiFL with proposed algorithms keeps admirable learning performance.
\end{spacing}

\begin{spacing}{1}
	Fig.~\ref{accuracy_vs_rounds_fig_6} shows the impact of different $\epsilon_1$ values on the learning performance of SemiFL.
	In Figs.~\ref{accuracy_vs_rounds_fig_6_MLP},~\ref{accuracy_vs_rounds_fig_6_CNN}, and~\ref{accuracy_vs_rounds_fig_6_ResNet}, as $\epsilon_1$ increases from $2$ to $10$, we see that the convergence acceleration effect of over-the-air distortion is enhanced in early rounds, while higher final	accuracy is obtained in the end.
	This confirms that $\epsilon_1$, or equivalently the power scaling factor-to-normalizing factor ratio $\sqrt{\omega}/\sqrt{\nu}$, is a key factor in modulating the acceleration effect of over-the-air distortion.
	However, Figs.~\ref{accuracy_vs_rounds_fig_6_MLP},~\ref{accuracy_vs_rounds_fig_6_CNN}, and~\ref{accuracy_vs_rounds_fig_6_ResNet} also show that an excessively large $\epsilon_1$ leads to training failure for SemiFL, as excessive over-the-air distortion collapses the global model.
	This indicates that the value of $\epsilon_1$ should be carefully moderated.
\end{spacing}

\begin{figure*}
	\begin{minipage}[t]{1 \textwidth}
		\centering
		\subfigure[Training MLP on the Fashion-MNIST dataset.]{
			\label{accuracy_non_iid_MLP}
			\includegraphics[width=0.32 \textwidth]{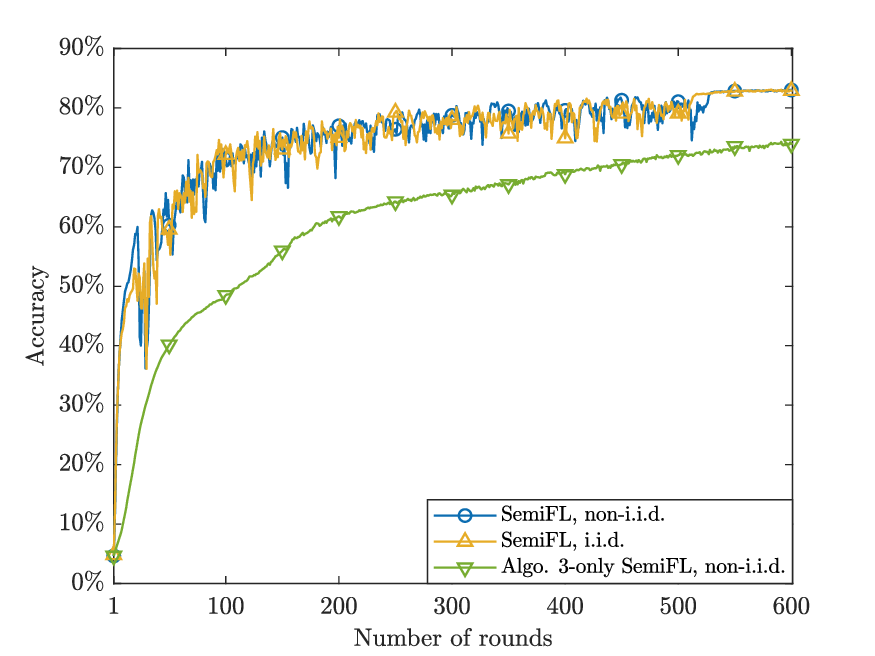}}
		\subfigure[Training CNN on the CIFAR-10 dataset.]{
			\label{accuracy_non_iid_CNN}
			\includegraphics[width=0.32 \textwidth]{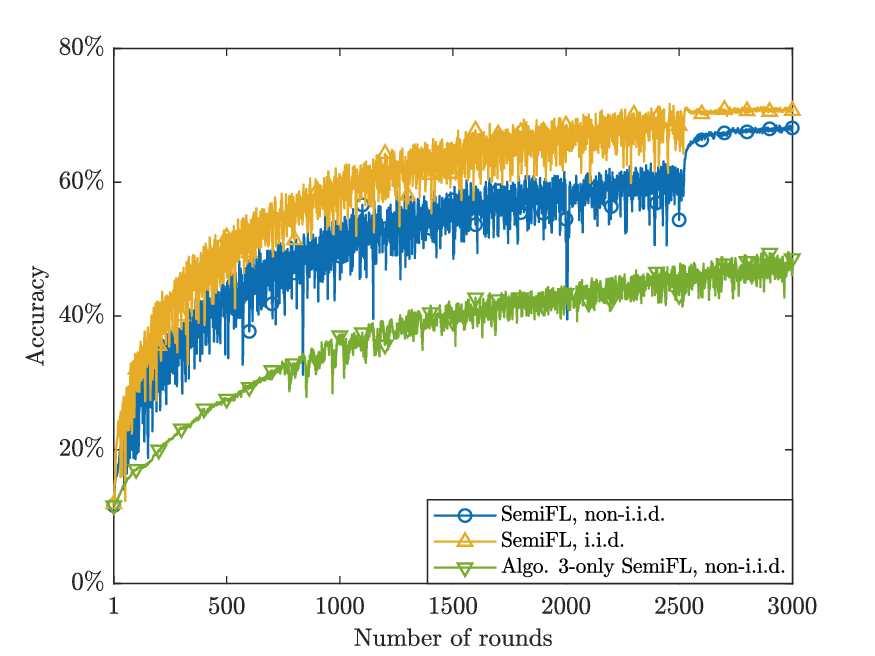}}
		\subfigure[Training ResNet on the CIFAR-100 dataset.]{
			\label{accuracy_non_iid_ResNet}
			\includegraphics[width=0.32 \textwidth]{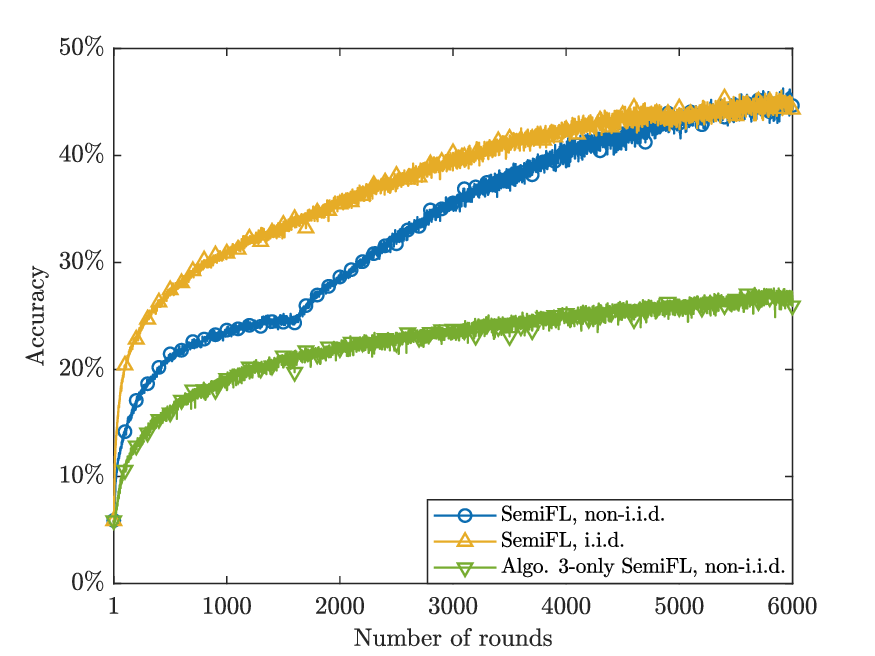}}
		\vspace{-0.3 cm}
		\caption{Learning performance with non-IID data.}
		\label{accuracy_non_iid}
		\vspace{-0.2 cm}
	\end{minipage}\\
	\begin{minipage}{1 \textwidth}
		\centering
		\subfigure[{Training MLP on the Fashion-MNIST dataset.}]{
			\label{accuracy_ablation_MLP}
			\includegraphics[width=0.32 \textwidth]{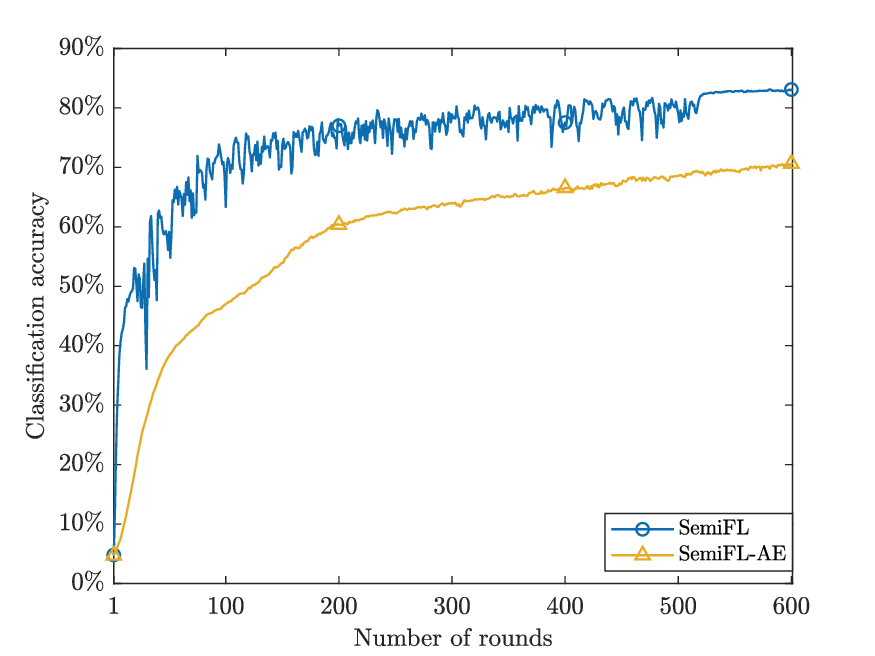}}
		\subfigure[{Training CNN on the CIFAR-10 dataset.}]{
			\label{accuracy_ablation_CNN}
			\includegraphics[width=0.32 \textwidth]{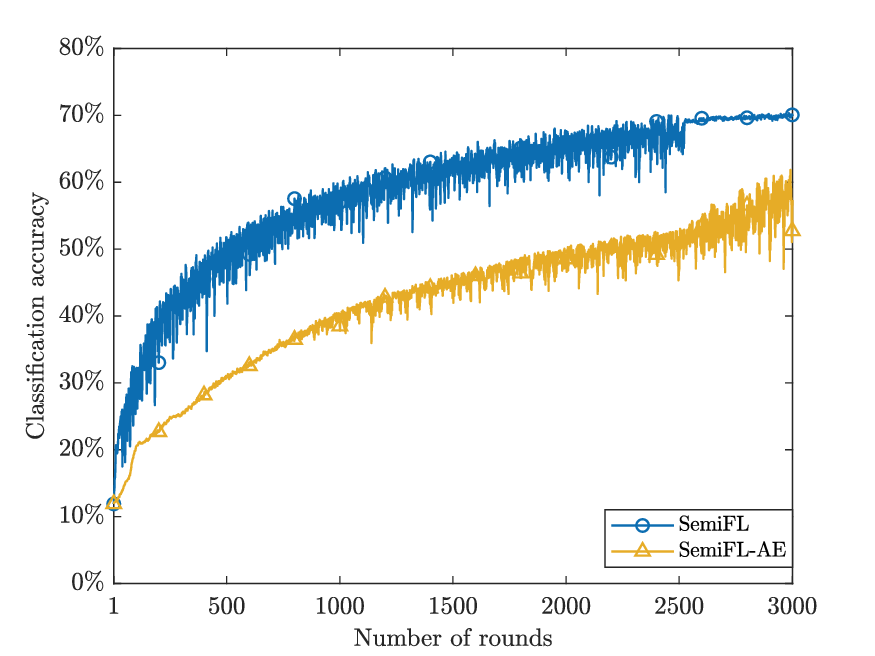}}
		\subfigure[{Training ResNet on the CIFAR-100 dataset.}]{
			\label{accuracy_ablation_ResNet}
			\includegraphics[width=0.32 \textwidth]{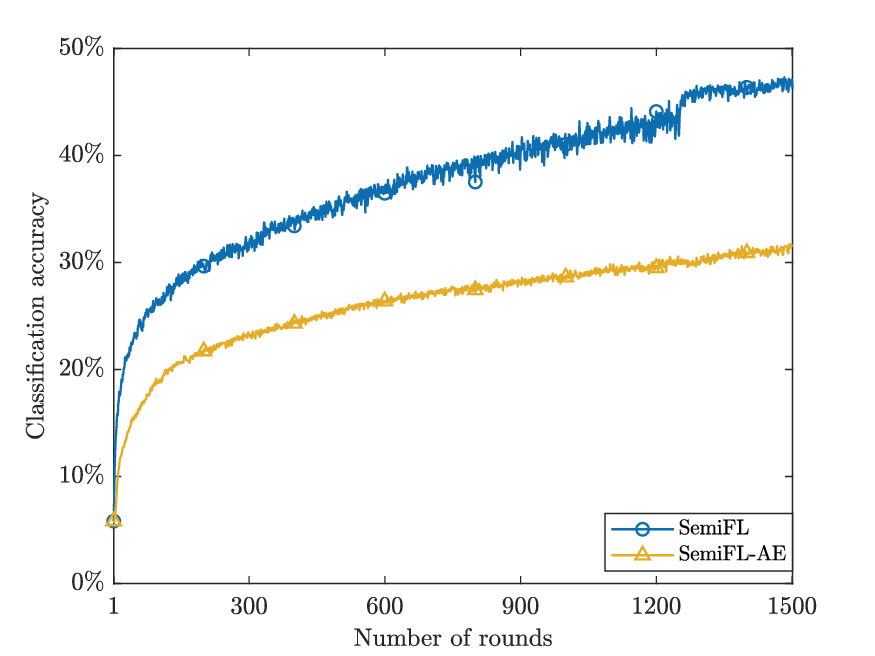}}
		\vspace{-0.3 cm}
		{\caption{Learning performance comparison in an amplitude distortion manipulation ablation experiment.}
			\label{accuracy_ablation}}
		\vspace{-0.2 cm}
	\end{minipage}\\
	\begin{minipage}[t]{1 \textwidth}
		\centering
		\subfigure[{Training MLP on the Fashion-MNIST dataset.}]{
			\label{accuracy_vs_rounds_epsilon_1_MLP}
			\includegraphics[width=0.32 \textwidth]{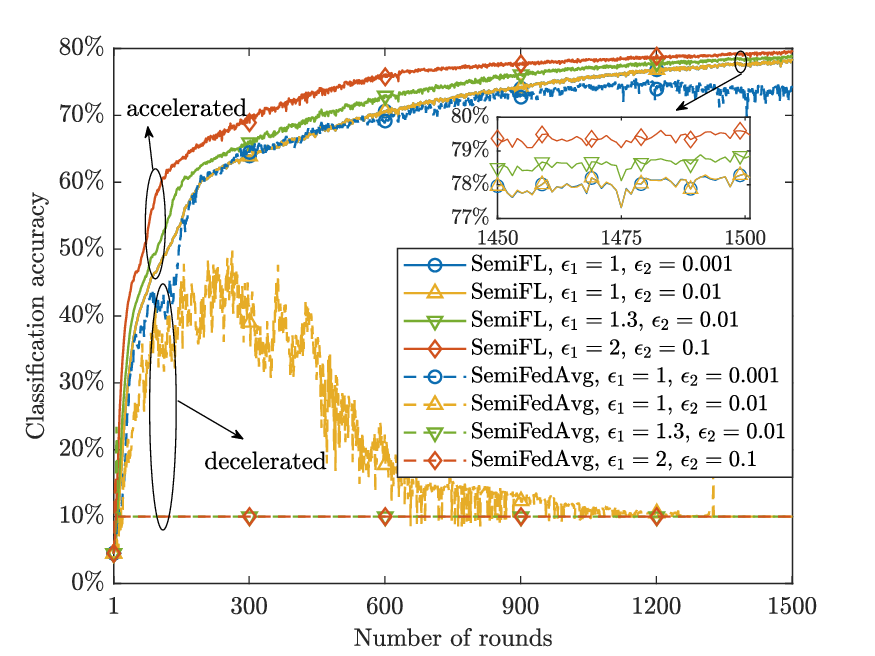}}
		\subfigure[{Training CNN on the CIFAR-10 dataset.}]{
			\label{accuracy_vs_rounds_epsilon_1_CNN}
			\includegraphics[width=0.32 \textwidth]{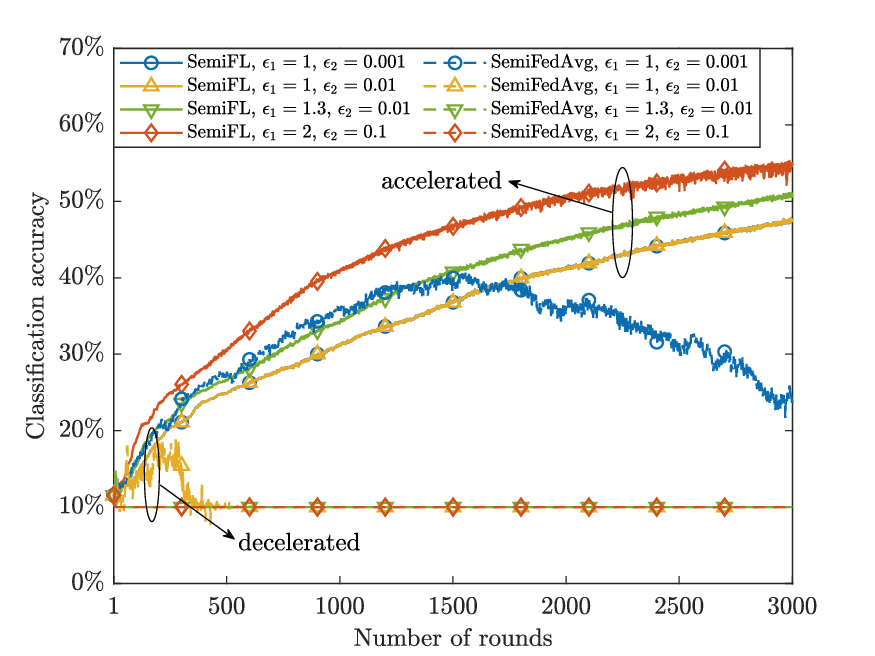}}
		\subfigure[{Training ResNet on the CIFAR-100 dataset.}]{
			\label{accuracy_vs_rounds_epsilon_1_ResNet}
			\includegraphics[width=0.32 \textwidth]{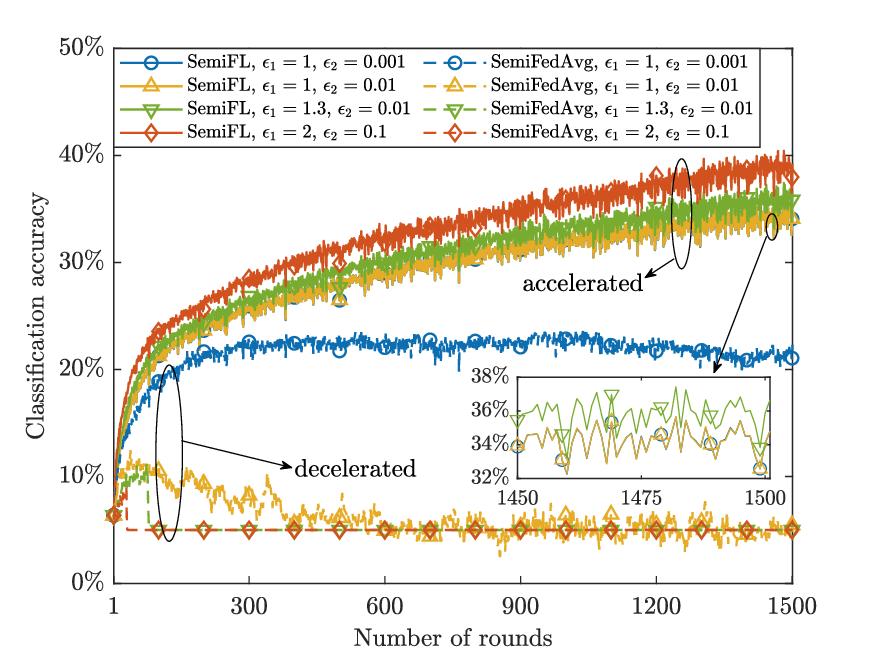}}
		\vspace{-0.3 cm}
		{\caption{Learning performance comparison between the proposed SemiFL and SemiFedAvg on the Fashion-MNIST, CIFAR-10, and CIFAR-100 datasets with different $\epsilon_1$ and $\epsilon_2$ values, where $\epsilon_3=0.8$ and $\epsilon_4=0.01$.}
			\label{accuracy_vs_rounds_epsilon_1}}
		\vspace{-0.4 cm}
	\end{minipage}
\end{figure*}

	%
	As shown in Fig.~\ref{accuracy_non_iid}, it is seen that non-IID data decelerates the convergence of SemiFL compared to IID data, especially when training CNN and ResNet.
	However, as the training process enters the stable region, SemiFL's convergence performance gradually improves as the SL component becomes dominant over FL in SemiFL, which aligns with our theoretical analysis of non-IID data.
	Moreover, even with non-IID data, it is seen that SemiFL with amplified over-the-air distortion still demonstrates faster converge than Algo. 3-only SemiFL across all cases.
	This validates the effectiveness of our proposed approach under non-IID data conditions.

%
{
	As shown in Fig.~\ref{accuracy_ablation}, the ablation experiment on SemiFL, denoted as SemiFL-AE, which removes amplitude distortion while retaining only the noise perturbation in over-the-air distortion, significantly degrades the learning performance across all three experimental settings. 
	Compared to SemiFL with our proposed scheme, SemiFL-AE not only slows down convergence but also attains a significantly lower final classification accuracy.
	This confirms that manipulating amplitude distortion is essential for accelerating convergence in the non-stable region.}

{Fig.~\ref{accuracy_vs_rounds_epsilon_1} shows the learning performance comparisons between SemiFL and a benchmark, named SemiFL with federated averaging (SemiFedAvg).
	The only difference of SemiFedAvg is that devices upload model parameters, i.e., weights and biases, to the BS for aggregation, rather than gradients.
	In Figs.~\ref{accuracy_vs_rounds_epsilon_1_MLP},~\ref{accuracy_vs_rounds_epsilon_1_CNN}, and~\ref{accuracy_vs_rounds_epsilon_1_ResNet}, it is intriguing to see that as $\epsilon_1$ and $\epsilon_2$ increase, the convergence of SemiFL is gradually accelerated, whereas the convergence of SemiFedAvg is significantly decelerated.
	This is because over-the-air distortion is directly imposed on model parameters in SemiFedAvg, fundamentally disrupting the global model.
	For the proposed SemiFL, over-the-air distortion only affects uploaded gradients, leaving the global model intact.
	This also highlights a key insight: the convergence acceleration effect of over-the-air distortion is tailored to AirComp-based gradient aggregation.
	Moreover, the figures in Fig.~\ref{accuracy_vs_rounds_epsilon_1} show that increasing $\epsilon_2$ while keeping $\epsilon_1$ constant cannot trigger the convergence acceleration effect of over-the-air distortion, as evidenced by the overlap of the yellow and blue solid lines.
	However, it should be emphasized that a large $\epsilon_2$ enables the usage of a large $\epsilon_1$, enhancing the convergence acceleration effect of over-the-air distortion.}

\vspace{-0.8 cm}
{\subsection{Energy Consumption Results and Analysis}
	\label{simulation_energy_results_and_analysis}}

{To validate the effectiveness of the proposed algorithms in conserving energy, we compare with the following baselines:
	\begin{itemize}
		\item \textbf{MMSE\&CI}~\cite{Ni2022Federated,Liu2020Over,Sun2022Dynamic}: By setting $\sqrt{\omega}=\sqrt{\nu}$ in both the non-stable region $\mathcal{R}^{\rm NS}$ and the stable region $\mathcal{R}^{\rm S}$, this scheme minimizes ${\rm MSE}$ to suppress over-the-air distortion throughout the entire SemiFL process.
		\item \textbf{SDR Beamformer}~\cite{Luo2010Semidefinite}: By dropping the rank-one constraints (44h) and (44i), receive beamformers $\{{\bf{v}}_k\}$ and ${\bf{b}}$ are solved using SDR.
		\item \textbf{Maximum transmit power (Max TP)}: By setting $\zeta_k=p_{\max}|{\bf{v}}_{k}^{\rm H}{\bf{h}}^{\rm DU}_{k}|^2, \forall k \in \mathcal{K}$ and $\omega_k=p_{\max}|{\bf{b}}^{\rm H}{\bf{h}}^{\rm GU}_{k}|^2, \forall k \in \mathcal{K}$, devices use the maximum transmit power to upload gradients and data.
		\item \textbf{Maximum CPU frequencies (Max CPU)}: By setting $\hat{f}_{k}=\hat{f}_{\max},\forall k \in \mathcal{K}$ and $\tilde{f}=\tilde{f}_{\max}$, devices and the BS use maximum CPU frequencies to perform FL and SL.
		\item \textbf{Random data allocation (RDA)}: The ratios of SL data, $\{\theta_k\}$, are randomly determined.
	\end{itemize}
	The computing energy consumption is generally orders-of-magnitude larger than that of communication.
	In addition, \textbf{MMSE\&CI}, \textbf{SDR Beamformer}, and \textbf{Max TP} benchmarks have no impact on computing energy consumption, while \textbf{Max CPU} and \textbf{RDA} benchmarks leave the uploading energy consumption unaffected.
	Thus, we separate uploading and computing energy consumption into two separate figures to highlight our algorithms' improvement clearly.}

\begin{figure}[t]
	\centering
	\subfigure[{Overall uploading energy consumption comparison.}]{
		\label{uploading_energy_comparison}
		\includegraphics[width=0.47 \textwidth]{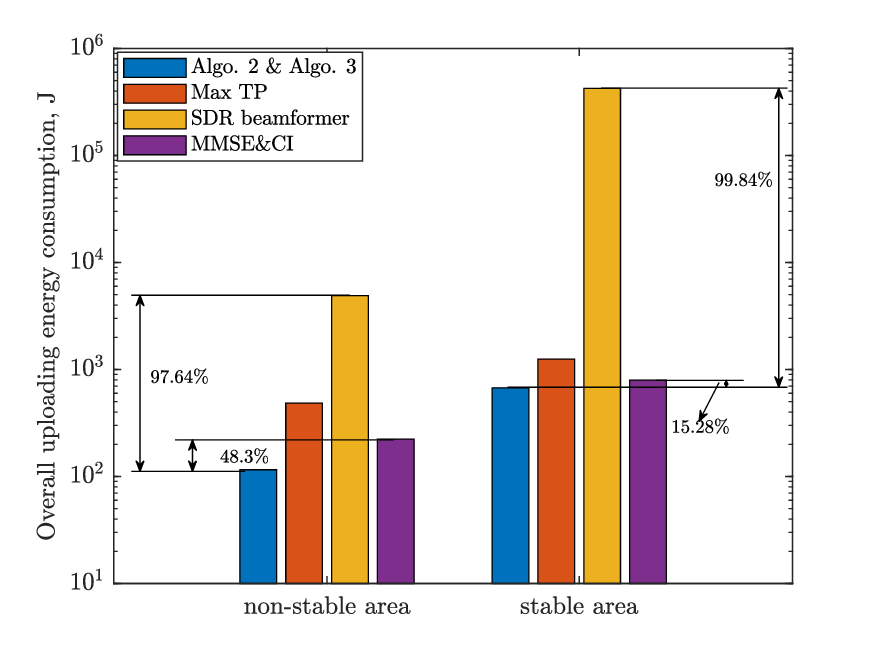}}\\
	\vspace{-0.4 cm}
	\subfigure[{Overall computing energy consumption comparison.}]{
		\label{computing_energy_comparison}
		\includegraphics[width=0.47 \textwidth]{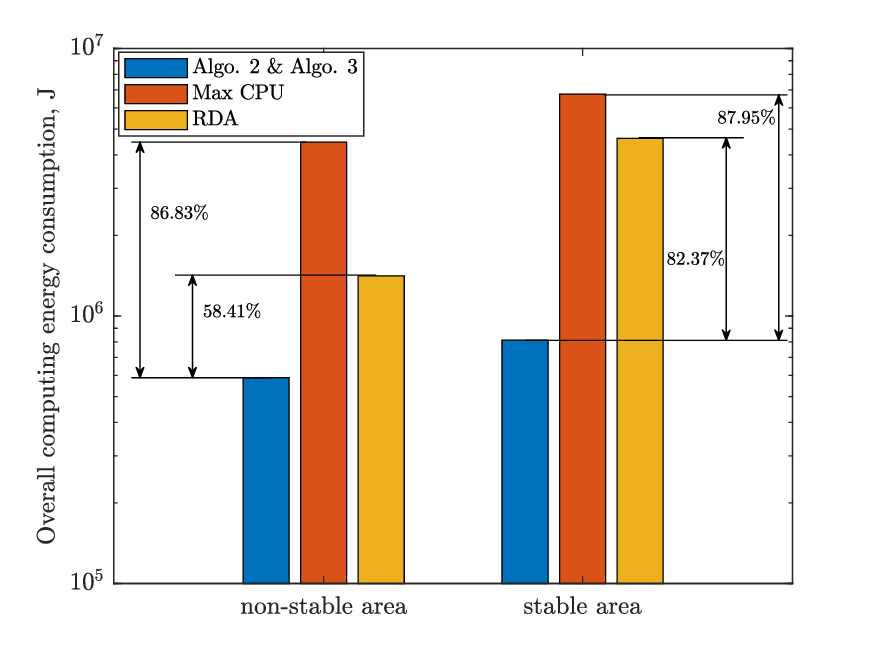}}
	\vspace{-0.3 cm}
	{\caption{Overall energy consumption comparison with $\epsilon_1=1.2$, $\epsilon_2=1$, $\epsilon_3=0.8$, and $\epsilon_4=0.01$, where $T=500$ rounds are considered in both the non-stable and stable regions.}
		\label{energy_consumption_comparison}}
	\vspace{-0.4 cm}
\end{figure}

{
	Fig.~\ref{energy_consumption_comparison} shows the energy consumption of our proposed algorithms in comparison with baselines.
	Note that $T=500$ rounds are considered in both regions.
	To show the performance gains more clearly, the overall energy consumption in objective (35a) is decomposed into two metrics: the overall uploading energy consumption, $\sum\nolimits_{t=1}^{T}\sum\nolimits_{k=1}^{K}(E^{\rm G}_{t,k}+E^{\rm D}_{t,k})$, and the overall computing energy consumption, $\sum\nolimits_{t=1}^{T}(\sum\nolimits_{k=1}^{K}E^{\rm L}_{t,k}+E^{\rm E}_{t})$.
	In Fig.~\ref{uploading_energy_comparison}, it is seen that the proposed algorithms achieve the lowest overall uploading energy consumption in both regions.
	Particularly, our proposed algorithms conserves $97.64\%$ and $48.3\%$ of uploading energy in the non-stable region, compared to SDR Beamformer and MMSE\&CI, respectively.
	Meanwhile, in the stable region, our proposed algorithms can save $99.84\%$ and $15.28\%$ of uploading energy compared to SDR Beamformer and MMSE\&CI schemes, respectively.
	In Fig.~\ref{computing_energy_comparison}, it is observed that our proposed algorithms outperform Max CPU and RDA by saving $86.83\%$ and $58.41\%$ of computing energy, respectively, in the non-stable region.
	Furthermore, our proposed algorithms save $87.95\%$ and $82.37\%$ of computing energy compared to Max CPU and RDA in the stable region, respectively.
	Additionally, one can find that the proposed algorithms consume more energy in the stable region than the stable region.
	This is because Algorithm~3 allocates higher transmit power in the stable region to suppress the over-the-air distortion, thereby reducing the optimality gap of SemiFL, as discussed in Remark~\ref{remark_3}.}

\begin{figure}[t]
	\centering
	\subfigure[{Overall energy consumption versus $T_{\max}$ in the non-stable region.}]{
		\label{overall_energy_consumption_non_stable}
		\includegraphics[width=0.47 \textwidth]{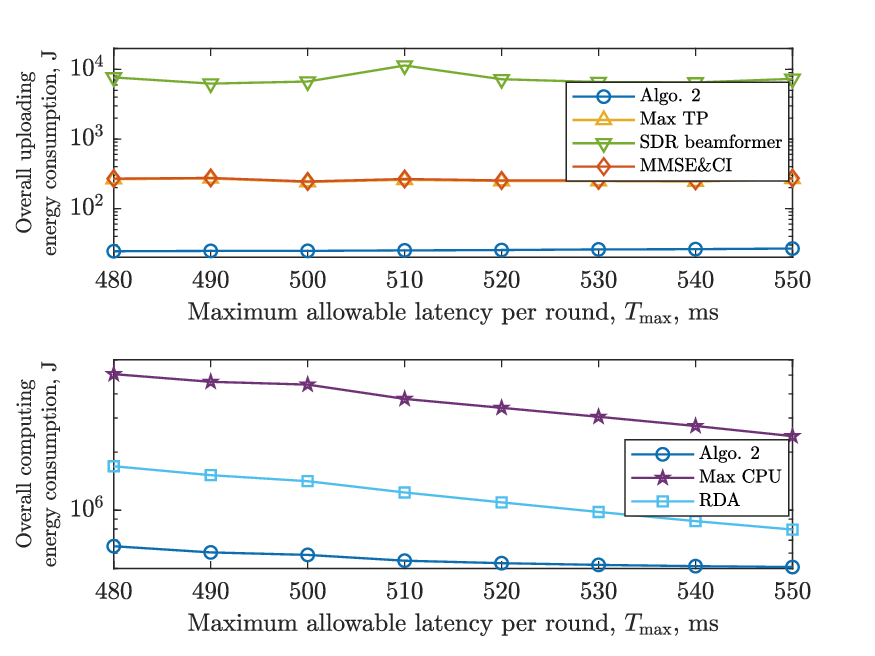}}\\
	\vspace{-0.4 cm}
	\subfigure[{Overall energy consumption versus $T_{\max}$ in the stable region.}]{
		\label{overall_energy_consumption_stable}
		\includegraphics[width=0.47 \textwidth]{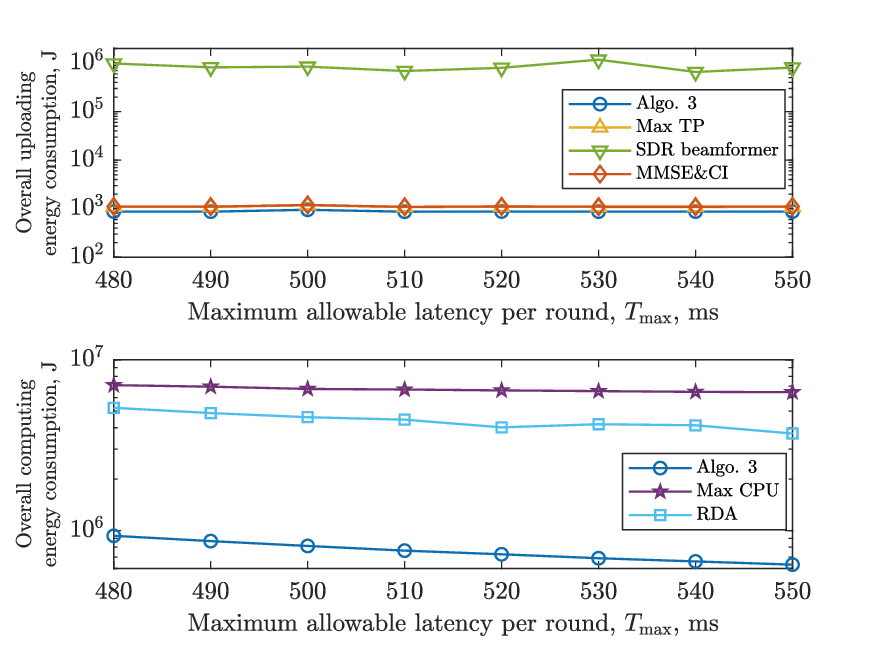}}
	\vspace{-0.2 cm}
	{\caption{Overall energy consumption versus $T_{\max}$ with $\epsilon_1=1.2$, $\epsilon_2=1$, $\epsilon_3=0.8$, and $\epsilon_4=0.01$.}
		\label{overall_energy_T_max}}
	\vspace{-0.8 cm}
\end{figure}

{
	Fig.~\ref{overall_energy_T_max} shows the overall uploading and computing energy consumption versus the maximum allowable latency per round, $T_{\max}$.
	In Fig.~\ref{overall_energy_consumption_non_stable}, it is seen that Algorithm~2 achieves lower overall uploading and computing energy consumption than all benchmarks in the non-stable region.
	Meanwhile, the overall uploading energy consumption is insensitive to changes in $T_{\max}$, whereas the overall computing energy consumption decrease as $T_{\max}$ increases.
	This is because a larger $T_{\max}$ allows devices and the BS to use lower CPU frequencies, thereby reducing computing energy consumption.
	In Fig.~\ref{overall_energy_consumption_stable}, it is seen that Algorithm~3 obtains the lowest energy consumption in the stable region.
	Since the tendencies of all curves are similar to those in the non-stable region, the same conclusion can be drawn.}

\begin{figure}[t]
	\centering
	\includegraphics[width=0.47 \textwidth]{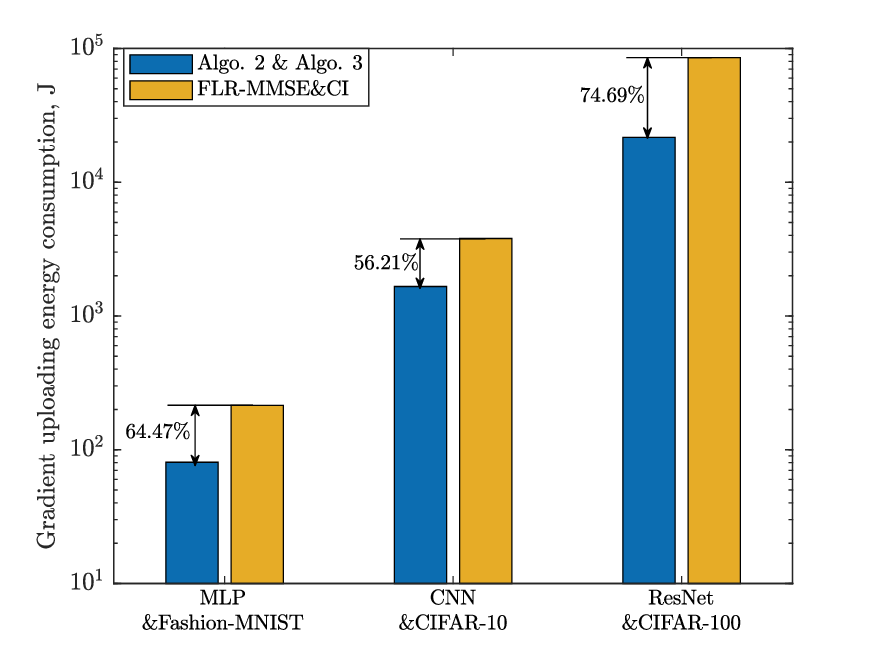}
	\vspace{-0.3 cm}
	{\caption{Gradient uploading energy consumption comparison between our approach and the FLR-MMSE\&CI scheme.}
		\label{energy_joint_or_separate}}
	\vspace{-0.4 cm}
\end{figure}

{
	Fig.~\ref{energy_joint_or_separate} demonstrates the energy consumption comparison between our proposed approach and the FLR-MMSE\&CI scheme on the three datasets.
	Note that our approach adopts a fixed learning rate $\eta_t$ and intentionally introduces the amplitude distortion $\frac{\sqrt{\omega_t}}{\sqrt{\nu_t}}$.
	The FLR-MMSE\&CI benchmark adopts a fixed learning rate while employing an MMSE\&CI scheme to minimize over-the-air distortion.
	Fig.~\ref{energy_joint_or_separate} shows that our approach reduces energy consumption for gradient uploading by $64.47\%$, $56.21\%$, and $74.69\%$ on the Fashion-MNIST, CIFAR-10, and CIFAR-100 datasets, respectively.
	This further underscores our method’s superiority in jointly optimizing convergence speed and energy efficiency, compared to FLR-MMSE\&CI which treats learning rate adjustment and distortion suppression separately.}

\begin{spacing}{0.97}
	\vspace{-0.2 cm}
	\section{Conclusion}
	\label{conclusion}
	In this paper, we proposed a novel approach that harnesses over-the-air distortion to accelerate the convergence of SemiFL.
	To preserve data privacy, the new SemiFL framework incorporated FL with SL, by which only the intermediate outputs of local model's shallow layers were uploaded to the BS.
	In the non-stable region, we amplified amplitude distortion to increase the learning rate in an energy-efficient manner, thereby accelerating SemiFL's convergence.
	In the stable region, we eliminated amplitude distortion while suppressing noise perturbation to maintain a small learning rate for improving the final convergence of SemiFL.
	We presented theoretical analyses to demonstrate the efficacy of our proposed approach across diverse regions, under both IID and non-IID data distributions.
	Furthermore, we formulated two energy consumption minimization problems, one for each region type, to implement a two-region MSE threshold configuration scheme that better leveraged over-the-air distortion.
	Then, we proposed two algorithms to solve the formulated two problems, where closed-form solutions to some optimization variables were derived.
	Extensive simulations using three AI models-dataset combinations demonstrated that under diverse network conditions and data distributions, our approach efficiently accelerated convergence of SemiFL while achieving improved final convergence.
	Meanwhile our algorithms effectively reduced the energy consumption of SemiFL by jointly optimizing communication, computation, and data allocation.
\end{spacing}
	
	\appendices
	
	{
		\section{Architecture of the Adopted ResNet}
		\label{architecture_of_the_adopted_ResNet}
		\begin{figure}[H]
			\centering
			\includegraphics[width=0.49\textwidth]{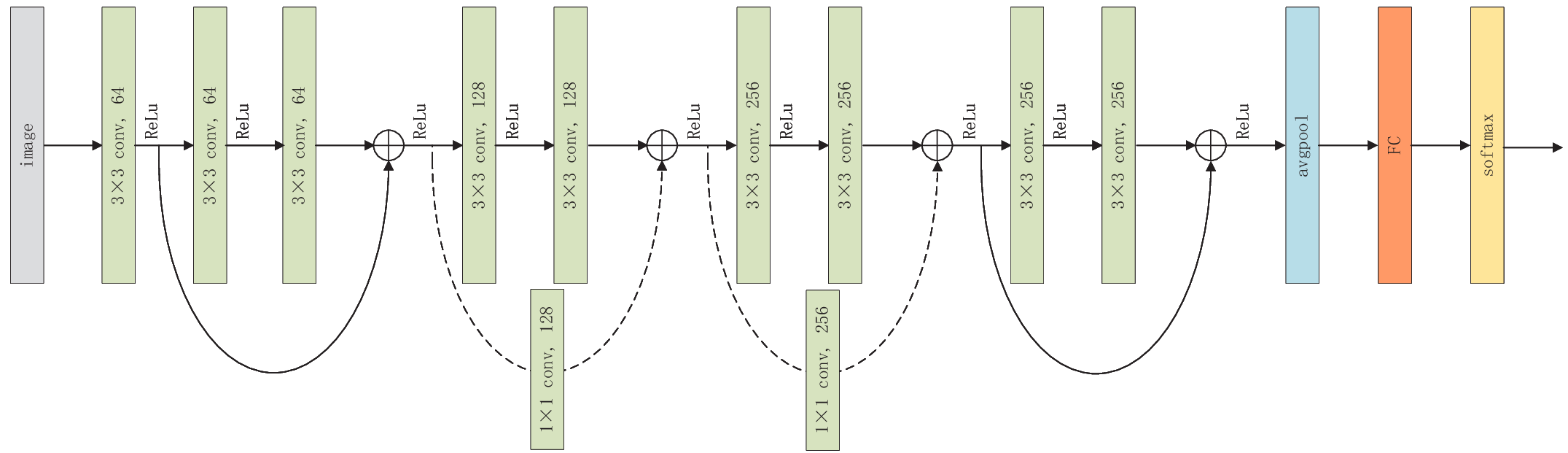}
			{\caption{An architecture demonstration of the adopted ResNet.}
				\label{ResNet_structure}}
		\end{figure}
		The architecture of the adopted ResNet is demonstrated in Fig.~\ref{ResNet_structure}.
		The adopted ResNet mainly contains four stacks of layers, where each stack contains two convolutional layers.
		For the first and fourth stacks, the input is directly added to the output of the two convolutional layers through a skip connection.
		For the second and third stacks, the input is added to the output of the aforementioned two convolutional layers after applying another convolutional layer to the skip connection.
		Finally, the four stacks of layers are sequentially followed by an average pooling layer, a fully connected layer, and a softmax layer.
	}

	{
		\section{Additional Simulation Results}
		\label{additional_simulation_results}}
	
	\begin{figure*}[t]
		\centering
		\begin{minipage}[t]{1 \textwidth}
			\centering
			\subfigure[{Training MLP on the Fashion-MNIST dataset.}]{
				\label{accuracy_joint_or_separate_MLP}
				\includegraphics[width=0.322 \textwidth]{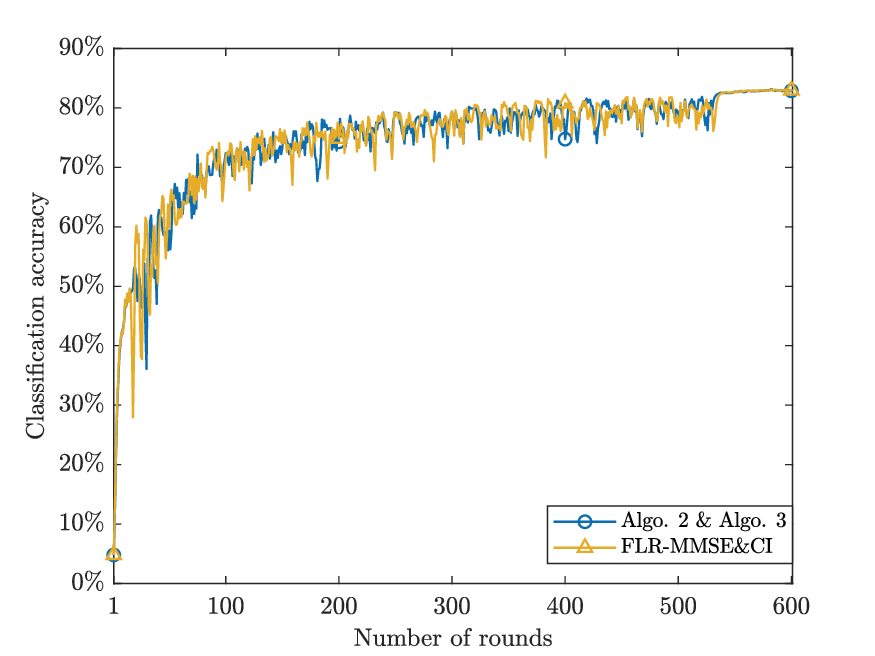}}
			\subfigure[{Training CNN on the CIFAR-10 dataset.}]{
				\label{accuracy_joint_or_separate_CNN}
				\includegraphics[width=0.322 \textwidth]{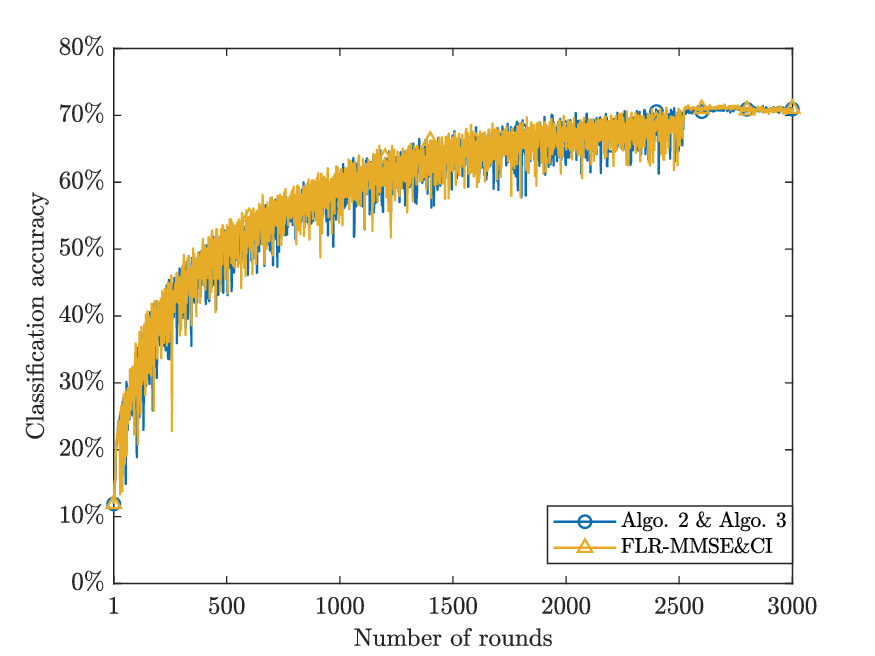}}
			\vspace{-0.3 cm}
			\subfigure[{Training ResNet on the CIFAR-100 dataset.}]{
				\label{accuracy_joint_or_separate_ResNet}
				\includegraphics[width=0.322 \textwidth]{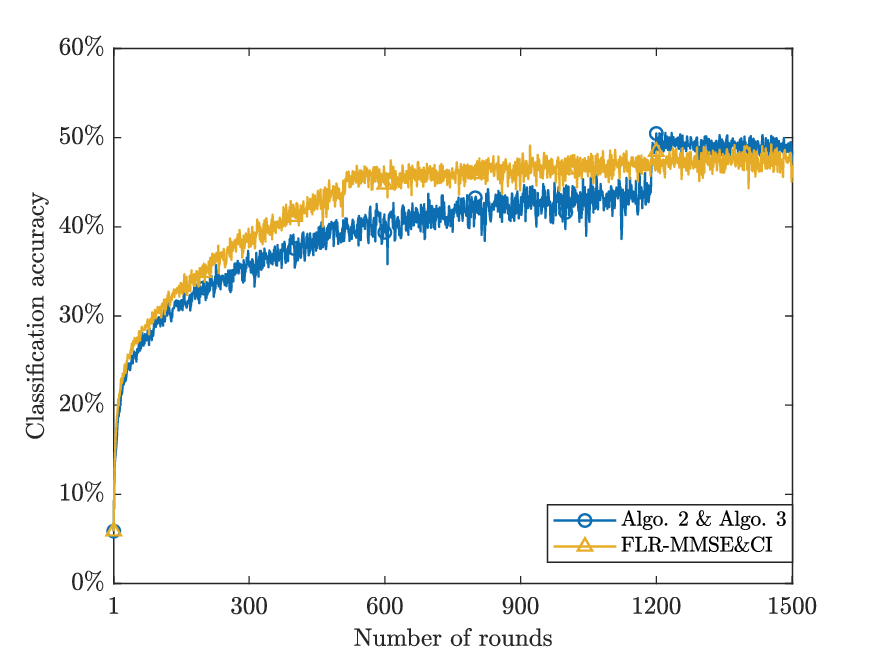}}
		\end{minipage}
		{\caption{Learning performance and energy consumption between our approach and the FLR-MMSE\&CI scheme.}
			\label{accuracy_joint_or_separate}}
		\vspace{-0.3 cm}
	\end{figure*}
	
	\begin{figure*}
		\begin{minipage}[t]{1 \textwidth}
			\centering
			\subfigure[{Training MLP on the Fashion-MNIST dataset.}]{
				\label{accuracy_diff_channel_MLP}
				\includegraphics[width=0.322 \textwidth]{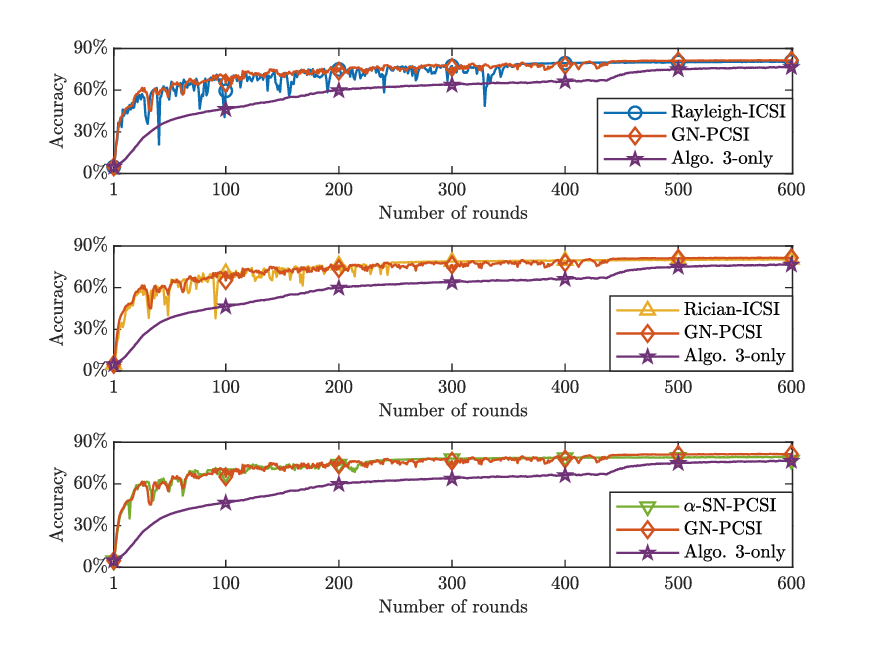}}
			\subfigure[{Training CNN on the CIFAR-10 dataset.}]{
				\label{accuracy_diff_channel_CNN}
				\includegraphics[width=0.322 \textwidth]{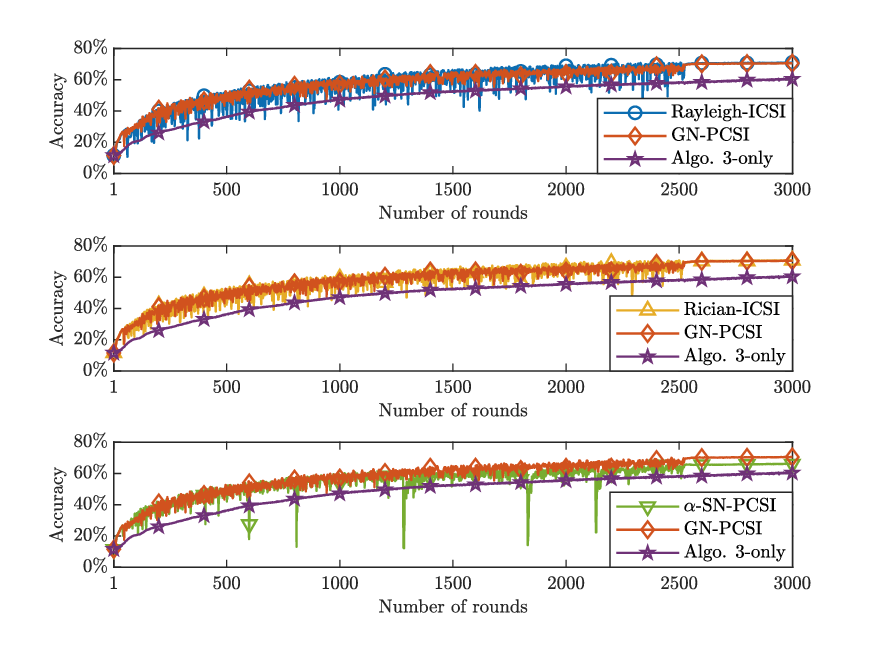}}
			\subfigure[{Training ResNet on the CIFAR-100 dataset.}]{
				\label{accuracy_diff_channel_ResNet}
				\includegraphics[width=0.322 \textwidth]{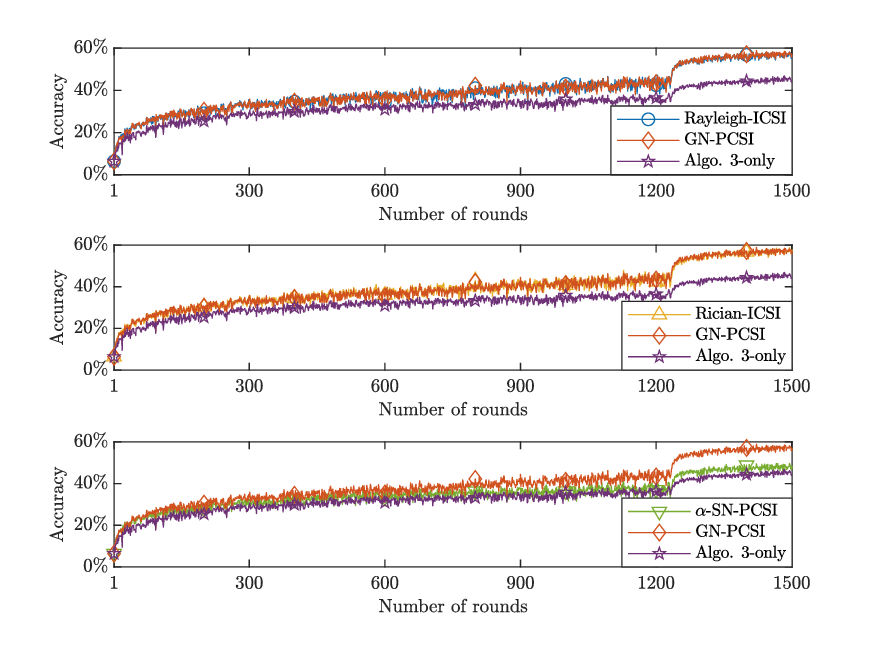}}
			\vspace{-0.1 cm}
			{\caption{Learning performance under different network conditions.}
				\label{accuracy_diff_channel}}
		\end{minipage}
		\vspace{-0.4 cm}
	\end{figure*}

	{
		Fig.~\ref{accuracy_joint_or_separate} demonstrates the learning performance comparison between our proposed approach and the FLR-MMSE\&CI scheme on the considered three datasets.
		Experiments on the Fashion-MNIST and CIFAR-10 datasets show that FLR-MMSE\&CI, which adopts a learning rate equals to $\frac{\sqrt{\omega_t}}{\sqrt{\nu_t}} \eta_t$ while employing the MMSE\&CI scheme to suppress over-the-air distortion, achieves nearly identical learning performance to our approach.
		Note that our approach adopts a fixed learning rate $\eta_t$ and intentionally introduces the amplitude distortion $\frac{\sqrt{\omega_t}}{\sqrt{\nu_t}}$.
		In addition, results on the CIFAR-100 dataset show that though FLR-MMSE\&CI converges faster than our approach in the non-stable region, our approach which utilizes over-the-air distortion still achieves better final convergence.
		These findings suggest that the observed learning performance improvements are attributable to our communication-oriented approach, i.e., amplifying over-the-air distortion, particularly amplitude distortion, in the non-stable region while suppressing it in the stable region.}
	
	{
		To examine the robustness of the proposed approach under both imperfect CSI and $\alpha$-stable noise, we consider the following setting for the channel coefficient and noise as baselines:
		\begin{itemize}
			\item \textbf{Rayleigh channel with imperfect CSI (Rayleigh-ICSI)}~\cite{Zheng2022Balancing}: The channel coefficient vector is rewritten as ${\bf{h}}^{\rm G}_{t,k}+\Delta {\bf{h}}^{\rm G}_{t,k}$, where ${\bf{h}}^{\rm G}_{t,k}$ denotes the estimated channel which follows a Rayleigh distribution, and $\Delta {\bf{h}}^{\rm G}_{t,k}$ denotes the estimation error which follows a circularly symmetric complex Gaussian (CSCG) distribution.
			We set the strength of $\Delta {\bf{h}}^{\rm G}_{t,k}$ to be $10$ times stronger than that of ${\bf{h}}^{\rm G}_{t,k}$, i.e., $\frac{\|\Delta {\bf{h}}^{\rm G}_{t,k}\|^2}{\|{\bf{h}}^{\rm G}_{t,k}\|^2}=1$.
			The noise ${\bf{n}}^{\rm G}_{t}$ follows a Gaussian distribution
			\item \textbf{Rician channel with imperfect CSI (Rician-ICSI)}~\cite{Zheng2022Balancing}: The channel coefficient vector is also rewritten as ${\bf{h}}^{\rm G}_{t,k}+\Delta {\bf{h}}^{\rm G}_{t,k}$, whereas ${\bf{h}}^{\rm G}_{t,k}$ follows a Rician distribution with a Rician factor $10$, and $\Delta {\bf{h}}^{\rm G}_{t,k}$ follows CSCG distribution as well.
			We also set $\frac{\|\Delta {\bf{h}}^{\rm G}_{t,k}\|^2}{\|{\bf{h}}^{\rm G}_{t,k}\|^2}=1$.
			The noise ${\bf{n}}^{\rm G}_{t}$ follows a Gaussian distribution.
			\item \textbf{$\alpha$-stable noise with perfect CSI ($\alpha$-SN-PCSI)}~\cite{Yang2022Revisiting}: The channel coefficient vector ${\bf{h}}^{\rm G}_{t,k}$ follows a Rayleigh distribution. 
			The noise ${\bf{n}}^{\rm G}_{t}$ follows a symmetric $\alpha$-stable distribution.
			We set the parameter $\alpha$ to $\alpha=1.4$.
			\item \textbf{Gaussian noise with perfect CSI (GN-PCSI)}:The channel coefficient vector ${\bf{h}}^{\rm G}_{t,k}$ follows a Rayleigh distribution.
			The noise ${\bf{n}}^{\rm G}_{t}$ follows a Gaussian distribution.
	\end{itemize}}

	{
		As shown in Fig.~\ref{accuracy_diff_channel}, SemiFL with the proposed approach, amplifying over-the-air distortion to accelerate convergence, works well across all considered channel and noise conditions.
		It is seen that SemiFL under the above four network conditions converges faster than Algo. 3-only SemiFL, while gradually approaching the performance of SemiFL in the case of GN-PCSI.
		This confirms the robustness and effectiveness of the proposed approach under different channel and noise types.
		Furthermore, it is also noticed that the curves for Rayleigh-ICSI, Rician-ICSI, and $\alpha$-SN-PCSI schemes exhibit more pronounced fluctuations than GN-PCSI across all experiments.
		This is because both the imperfect CSI and the $\alpha$-stable noise introduce stronger interference in gradient aggregation than perfect CSI.}
	
	\begin{figure*}
		\begin{minipage}[t]{1 \textwidth}
			\centering
			\subfigure[{Training MLP on the Fashion-MNIST dataset.}]{
				\label{loss_vs_rounds_MLP}
				\includegraphics[width=0.322 \textwidth]{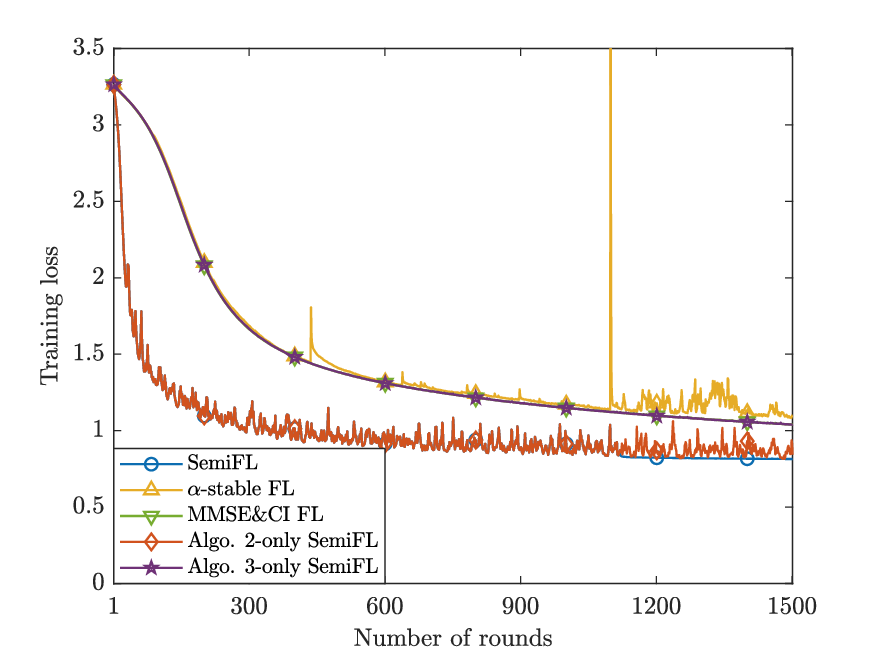}}
			\subfigure[{Training CNN on the CIFAR-10 dataset.}]{
				\label{loss_vs_rounds_CNN}
				\includegraphics[width=0.322 \textwidth]{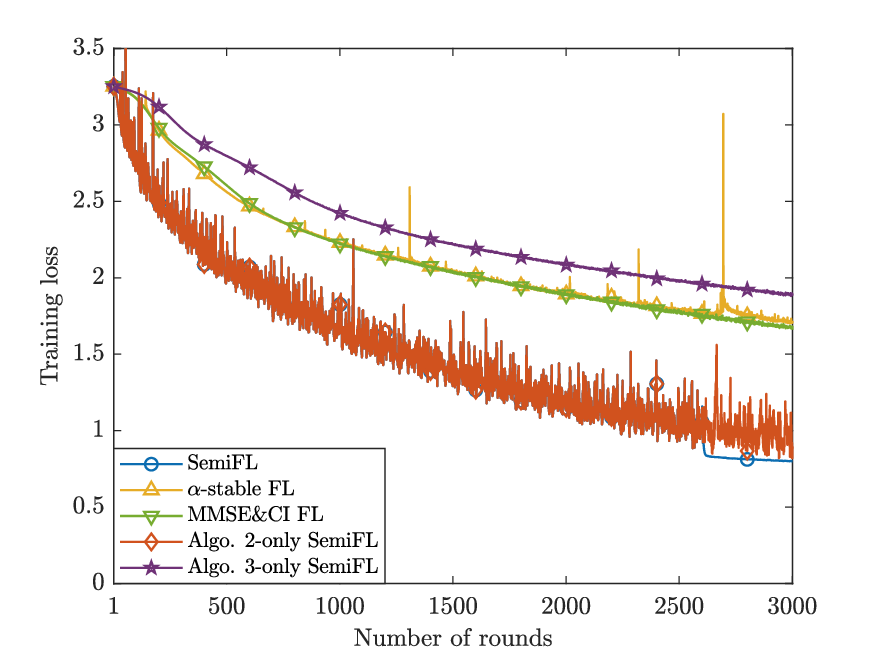}}
			\subfigure[{Training ResNet on the CIFAR-100 dataset.}]{
				\label{loss_vs_rounds_ResNet}
				\includegraphics[width=0.322 \textwidth]{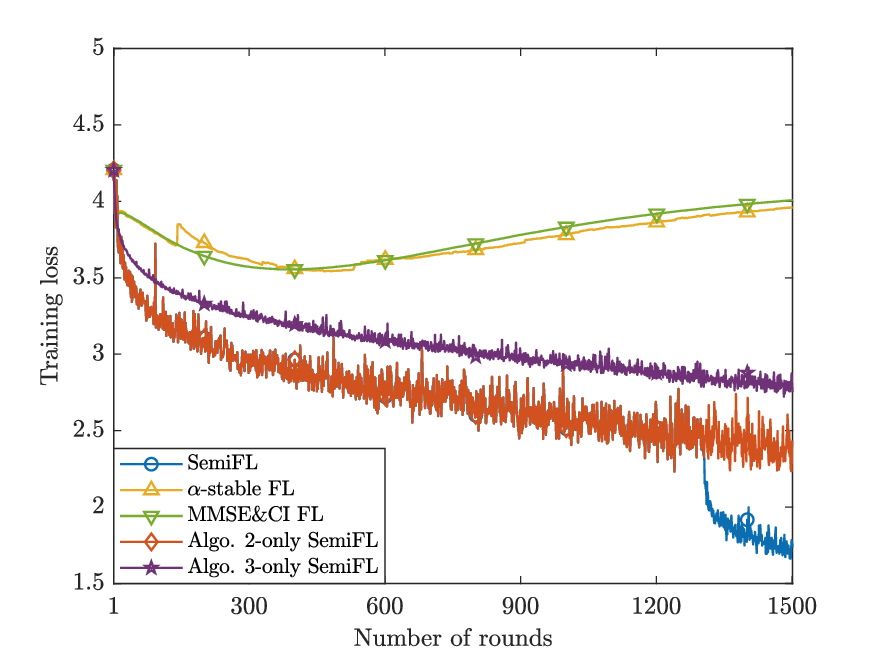}}
			\vspace{-0.2 cm}
			{\caption{Training loss comparison between SemiFL and benchmarks on the Fashion-MNIST, CIFAR-10, and CIFAR-100 datasets, where $\epsilon_1=10$, $\epsilon_2=1$, $\epsilon_3=0.8$, and $\epsilon_4=0.01$.}
				\label{loss_vs_rounds}}
		\end{minipage}
		\vspace{-0.4 cm}
	\end{figure*}
	
	\begin{figure*}
		\begin{minipage}[t]{1 \textwidth}
			\centering
			\subfigure[{Training MLP on the Fashion-MNIST dataset.}]{
				\label{loss_vs_rounds_epsilon_1_MLP}
				\includegraphics[width=0.322 \textwidth]{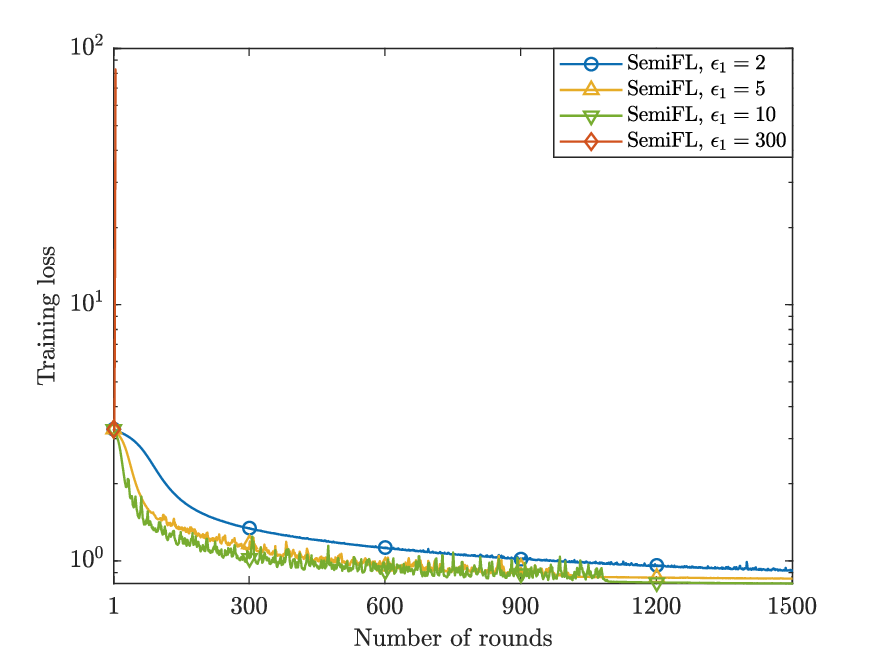}}
			\subfigure[{Training CNN on the CIFAR-10 dataset.}]{
				\label{loss_vs_rounds_epsilon_1_CIFAR-10}
				\includegraphics[width=0.322 \textwidth]{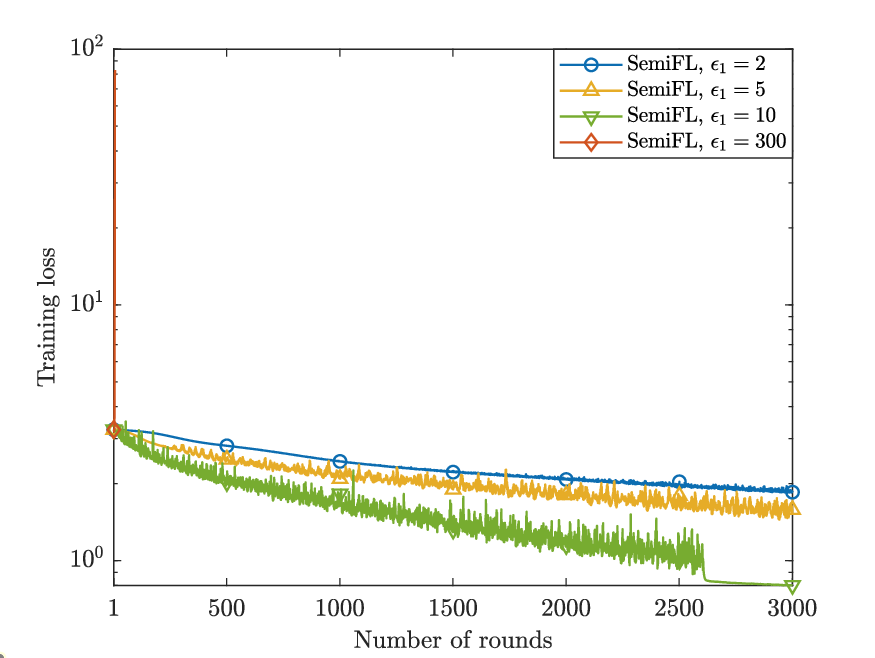}}
			\subfigure[{Training ResNet on the CIFAR-100 dataset.}]{
				\label{loss_vs_rounds_epsilon_1_CIFAR-100}
				\includegraphics[width=0.322 \textwidth]{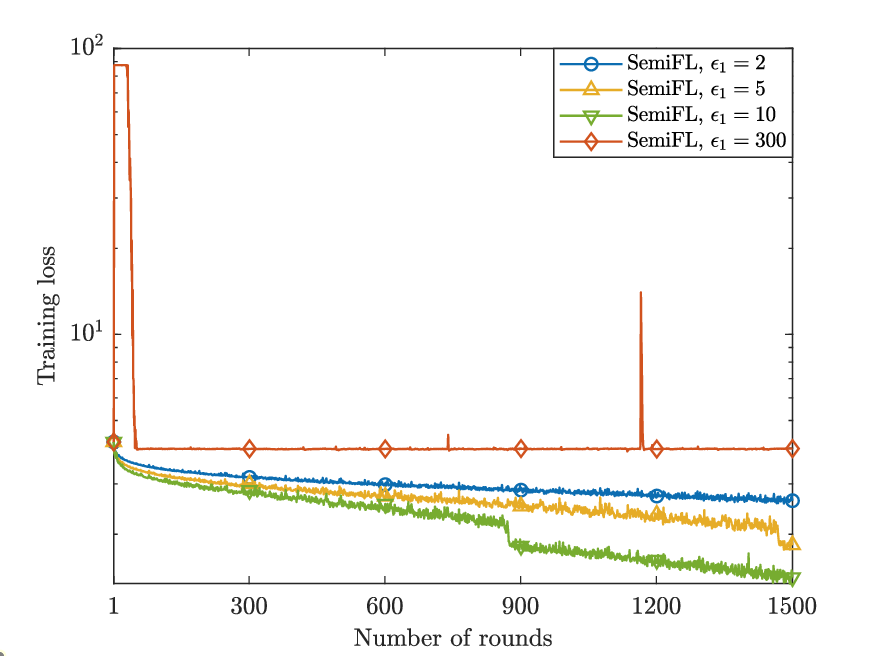}}
			{\caption{Training loss comparison of the proposed SemiFL on the Fashion-MNIST, CIFAR-10, and CIFAR-100 datasets with different $\epsilon_1$ values, where $\epsilon_3=0.8$ and $\epsilon_4=0.01$. Note that we set $\epsilon_2=1$ when $\epsilon_1=2$ or $5$, and set $\epsilon_2=5$ when $\epsilon_1=10$. When $\epsilon_1=300$, a sufficiently large $\epsilon_2$ is adopted.}
				\label{loss_vs_rounds_epsilon_1}}
		\end{minipage}
		\vspace{-0.4 cm}
	\end{figure*}

	{As shown in Fig.~\ref{loss_vs_rounds}, the proposed SemiFL with our proposed approach, i.e., increasing over-the-air distortion in the non-stable region but suppressing it in the stable region, achieves faster loss function descent than $\alpha$-stable FL, MMSE\&CI FL, and Algo. 3-only SemiFL schemes on all three datasets.
		This confirms that increasing over-the-air distortion in the non-stable region effectively increases the learning rate, thereby accelerating SemiFL's convergence.
		Moreover, it is also seen in Fig.~\ref{loss_vs_rounds} that SemiFL with our proposed approach converges to lower training loss than other schemes.
		This demonstrates that suppressing over-the-air distortion in the stable region helps maintain a small learning rate, thereby facilitating steady and improved final convergence.}

	{Fig.~\ref{loss_vs_rounds_epsilon_1} shows the impact of different $\epsilon_1$ values on the training loss of SemiFL.
		Across all datasets, it is seen that as $\epsilon_1$ increases, the decent of training loss becomes more pronounced.
		This is because a larger $\epsilon_1$ value leads to higher over-the-distortion, i.e., $\sqrt{\omega_t}/\sqrt{\nu_t}$, which increases the learning rate, thereby accelerating the convergence of SemiFL in the non-stable region.
		Meanwhile, it is observed that an excessively large $\epsilon_1$ value causes the training loss curves to vanish for Fashion-MNIST and CIFAR-10 datasets, while making a non-decreasing training loss pattern for the CIFAR-100 dateset.
		These results indicate that excessive over-the-distortion adversely affects SemiFL, learning to model collapse during training.
		This necessitates a moderate level of over-the-air distortion that accelerates convergence while maintaining model robustness for SemiFL.}

	\section{Proof of Theorem~\ref{theorem_1}}
	\label{proof_of_theorem_1}
	Based on (\ref{global_update_deep_layer}), (\ref{aggregated_FL_gradient}), and (\ref{mu_strongly_convex}) in Assumption~\ref{assumption_2}, we have
	\vspace{-0.2 cm}
	\begin{align}
		\label{appendix_A_1}
		&F({\bf{w}}_{t})-F({\bf{w}}_{t+1}) \notag \\ 
		\ge &\eta_t \nabla F({\bf{w}}_{t+1})^{\rm T} (\rho^{\rm L}_t \frac{\sqrt{\omega_t}}{\sqrt{\nu_t}}\hat{\bf{g}}^{\rm L}_{t}+\rho^{\rm L}_t \hat{\bf{n}}^{\rm G}_t+\rho^{\rm E}_t {\bf{g}}^{\rm E}_t) \notag \\ 
		&+ \frac{\mu}{2} \eta^2_t [(\rho^{\rm L}_t)^2 \|{\bf{g}}^{\rm L}_t\|^2 \!\!+\!\! (\rho^{\rm E}_t)^2 \|{\bf{g}}^{\rm E}_t\|^2 \!+\! 2\rho^{\rm L}_t\rho^{\rm E}_t({\bf{g}}^{\rm L}_t)^{\rm T}{\bf{g}}^{\rm E}_t ].
	\end{align}
	By taking the expectation on both sides, while using Assumption~\ref{assumption_4}, we have
	\begin{align}
		\label{appendix_A_2}
		\hspace{-0.2 cm}&\mathbb{E}[F({\bf{w}}_{t})-F({\bf{w}}_{t+1})] \notag \\ 
		\hspace{-0.2 cm}\ge & \eta_t (\rho^{\rm L}_t \frac{\sqrt{\omega_t}}{\sqrt{\nu_t}}+\rho^{\rm E}_t) \nabla F({\bf{w}}_{t+1})^{\rm T} \nabla F({\bf{w}}_t) \notag \\
		\hspace{-0.2 cm}&+\!\frac{\mu}{2} \eta_t^2(\rho^{\rm L}_t)^2 \mathbb{E}[\|\frac{\sqrt{\omega_t}}{\sqrt{\nu_t}}\hat{\bf{g}}^{\rm L}_t + \hat{\bf{n}}^{\rm G}_t\|^2] \notag \\
		\hspace{-0.2 cm}&+\!\frac{\mu}{2} \eta_t^2(\rho^{\rm E}_t)^2\mathbb{E}[\|{\bf{g}}^{\rm E}_t\|^2] \!+\! \mu \eta_t^2 \rho^{\rm L}_t \rho^{\rm E}_t \mathbb{E}[({\bf{g}}^{\rm L}_t)^{\rm T}{\bf{g}}^{\rm E}_t] \notag \\
		\hspace{-0.2 cm}=&\|\frac{\eta_t}{2} (\rho^{\rm L}_t \frac{\sqrt{\omega_t}}{\sqrt{\nu_t}}+\rho^{\rm E}_t) \nabla F({\bf{w}}_{t+1}) + \nabla F({\bf{w}}_t)\|^2 \notag \\
		\hspace{-0.2 cm}&-\frac{\eta_t^2}{4} (\rho^{\rm L}_t \frac{\sqrt{\omega_t}}{\sqrt{\nu_t}}+\rho^{\rm E}_t)^2 \|\nabla F({\bf{w}}_{t+1})\|^2 - \|\nabla F({\bf{w}}_t)\|^2 \notag \\
		\hspace{-0.2 cm}&+ \frac{\mu}{2}\eta_t^2(\rho^{\rm L}_t)^2(\mathbb{E}[\|\frac{\sqrt{\omega_t}}{\sqrt{\nu_t}}\hat{\bf{g}}^{\rm L}_t\|^2]+\mathbb{E}[\|\hat{\bf{n}}^{\rm G}_t\|^2]) \notag \\
		\hspace{-0.2 cm}&+\!\frac{\mu}{2} \eta_t^2(\rho^{\rm E}_t)^2\mathbb{E}[\|{\bf{g}}^{\rm E}_t\|^2] \!+\! \mu \eta_t^2 \rho^{\rm L}_t \rho^{\rm E}_t \mathbb{E}[({\bf{g}}^{\rm L}_t)^{\rm T}{\bf{g}}^{\rm E}_t].
	\end{align}
	Then, we incorporate $\|x\| \ge 0, \forall x \in \mathbb{R}$ and Assumption~\ref{assumption_3} into (\ref{appendix_A_2}).
	As a result, we have
	\vspace{-0.2 cm}
	\begin{align}
		\label{appendix_A_3}
		&\mathbb{E}[F({\bf{w}}_{t})-F({\bf{w}}_{t+1})] \notag \\ 
		\overset{}{\ge} &-\!\frac{\eta_t^2}{4} (\rho^{\rm L}_t \frac{\sqrt{\omega_t}}{\sqrt{\nu_t}}\!+\!\rho^{\rm E}_t)^2 A^2\! - \! A^2 \! + \! \frac{\mu}{2} \eta_t^2 (\rho^{\rm L}_t)^2 \frac{\omega_t}{\nu_t} \mathbb{E}[\|\hat{\bf{g}}^{\rm L}_t\|^2] \notag \\
		&+ \frac{\mu}{2} \eta_t^2 (\rho^{\rm E}_t)^2 \mathbb{E}[\|{\bf{g}}^{\rm E}_t\|^2] + \frac{\mu}{2} \eta_t^2 (\rho^{\rm L}_t)^2 \mathbb{E}[\|\hat{\bf{n}}^{\rm G}_t\|^2] \notag \\
		&+\mu \eta_t^2 \rho^{\rm L}_t \rho^{\rm E}_t \mathbb{E}[({\bf{g}}^{\rm L}_t)^{\rm T}{\bf{g}}^{\rm E}_t] \notag \\
		\overset{(a)}{\ge} &-\!\frac{\eta_t^2}{4} (\rho^{\rm L}_t \frac{\sqrt{\omega_t}}{\sqrt{\nu_t}}\!+\!\rho^{\rm E}_t)^2 A^2\! - \! A^2 \notag \\
		&+ \! \frac{\mu}{2} \eta_t^2 (\rho^{\rm L}_t)^2 \frac{\omega_t}{\nu_t} \|\nabla F({\bf{w}}_t)\|^2 \! + \! \frac{\mu}{2} \eta_t^2 (\rho^{\rm E}_t)^2 \|\nabla F({\bf{w}}_t)\|^2 \notag \\
		&+ \frac{\mu}{2} \eta_t^2 (\rho^{\rm L}_t)^2 \mathbb{E}[\|\hat{\bf{n}}^{\rm G}_t\|^2] +\mu \eta_t^2 \rho^{\rm L}_t \rho^{\rm E}_t \mathbb{E}[({\bf{g}}^{\rm L}_t)^{\rm T}{\bf{g}}^{\rm E}_t] \notag \\
		\overset{(b)}{=} &-\!\frac{\eta_t^2}{4} (\rho^{\rm L}_t \frac{\sqrt{\omega_t}}{\sqrt{\nu_t}}\!+\!\rho^{\rm E}_t)^2 A^2\! - \! A^2 \notag \\
		&+ \! \frac{\mu}{2} \eta_t^2 (\rho^{\rm L}_t)^2 \frac{\omega_t}{\nu_t} \|\nabla F({\bf{w}}_t)\|^2 \! + \! \frac{\mu}{2} \eta_t^2 (\rho^{\rm E}_t)^2 \|\nabla F({\bf{w}}_t)\|^2 \notag \\
		&+ \frac{\mu}{2} \eta_t^2 (\rho^{\rm L}_t)^2 \mathbb{E}[\|\hat{\bf{n}}^{\rm G}_t\|^2] +\mu \eta_t^2 \rho^{\rm L}_t \rho^{\rm E}_t \frac{\sqrt{\omega_t}}{\sqrt{\nu_t}} \|\nabla F({\bf{w}}_t)\|^2\notag \\
		\overset{(c)}{\ge} &-\!\frac{\eta_t^2}{4} (\rho^{\rm L}_t \frac{\sqrt{\omega_t}}{\sqrt{\nu_t}}\!+\!\rho^{\rm E}_t)^2 A^2\! - \! A^2 +\frac{\mu}{2} \eta_t^2 (\rho^{\rm L}_t \frac{\sqrt{\omega_t}}{\sqrt{\nu_t}}\!+\!\rho^{\rm E}_t)^2 \varepsilon^2 \notag \\
		&+ \frac{\mu}{2} \eta_t^2 \rho^{\rm L}_t \mathbb{E}[\|\hat{\bf{n}}^{\rm G}_t\|^2],
	\end{align}
	where $(a)$ is because $\mathbb{E}[\|\hat{\bf{g}}^{\rm L}_t\|^2]=\sum\nolimits_{q=1}^{Q} \mathbb{E}[(\hat{g}^{\rm L}_{t,q})^2] \ge \sum\nolimits_{q=1}^{Q} (\mathbb{E}[\hat{g}^{\rm L}_{t,q}])^2=\|\nabla F({\bf{w}}_t)\|^2$ and $\mathbb{E}[\|{\bf{g}}^{\rm E}_t\|^2] \! = \! \sum\nolimits_{q=1}^{Q} \mathbb{E}[({g}^{\rm E}_{t,q})^2] \! \ge \! \sum\nolimits_{q=1}^{Q} (\mathbb{E}[{g}^{\rm L}_{t,q}])^2 \! = \! \|\nabla F({\bf{w}}_t)\|^2$.
	Moreover, $(b)$ is because ${\bf{g}}^{\rm L}_t$ and ${\bf{g}}^{\rm E}_t$ are independent, while $(c)$ is because $\| \nabla F({\bf{w}}_t) \! \| \! \ge \! \varepsilon$ in the non-stable region $\mathcal{R}^{\rm NS}$.

	Recall the definition of $\hat{\bf{n}}^{\rm G}_t$ in (\ref{aggregated_FL_gradient}), one can have
	\vspace{-0.2 cm}
	\begin{align}
		\label{appendix_A_4}
		\mathbb{E}[\|\hat{\bf{n}}^{\rm G}_t\|^2]=\sum\nolimits_{q=1}^{Q} \mathbb{E}[(\hat{n}^{\rm G}_{t,q})^2]=\frac{\sigma^2Q}{2\nu_t}.
	\end{align}
	Plugging the above equation into (\ref{appendix_A_3}), we have
	\vspace{-0.2 cm}
	\begin{align}
		\label{appendix_A_5}
		&\mathbb{E}[F({\bf{w}}_{t})-F({\bf{w}}_{t+1})]\notag \\ 
		\ge & \frac{\eta_t^2}{4} (2 \mu \varepsilon^2 - A^2)(\rho^{\rm L}_t \frac{\sqrt{\omega_t}}{\sqrt{\nu_t}}\!+\!\rho^{\rm E}_t)^2 \notag \\
		&- \! A^2 \! + \! \frac{\mu\sigma^2Q\eta_t^2}{4\nu_t} (\rho^{\rm L}_t)^2.
	\end{align}
	By substituting $\rho^{\rm E}_t=1-\rho^{\rm L}_t$ into (\ref{appendix_A_5}), we can obtain (\ref{non_stable_lower_bound}).
	The proof is complete.
	
	{
		\section{Proof of Corollary~\ref{corollary_1}}
		\label{proof_of_corollary_1}}

	{
		Based on the proof of Theorem 1, one can have
		\vspace{-0.2 cm}
		\begin{align}
			\label{non-iid_non_stable_1_response_R3}
			&\mathbb{E}[F({\bf{w}}_{t})-F({\bf{w}}_{t+1})] \notag \\ 
			\ge &\eta_t \mathbb{E}[\nabla F({\bf{w}}_{t+1})^{\rm T} [\rho^{\rm L}_t \frac{\sqrt{\omega_t}}{\sqrt{\nu_t}}(\hat{\bf{g}}^{\rm L}_{t}-{\bf{g}}^{\rm L*}_{t}) \notag \\
			&+\rho^{\rm L}_t \frac{\sqrt{\omega_t}}{\sqrt{\nu_t}}{\bf{g}}^{\rm L*}_{t}+\rho^{\rm L}_t \hat{\bf{n}}^{\rm G}_t+\rho^{\rm E}_t {\bf{g}}^{\rm E}_t]] \notag \\ 
			&+ \frac{\mu}{2} \eta^2_t (\rho^{\rm L}_t)^2 \frac{\omega_t}{\nu_t} \mathbb{E}[\|\hat{\bf{g}}^{\rm L}_{t,k}\|^2] + \frac{\mu}{2} \eta^2_t (\rho^{\rm E}_t)^2 \mathbb{E}[\|{\bf{g}}^{\rm E}_{t,k}\|^2] \notag \\
			&+ \frac{\mu}{2} \eta^2_t (\rho^{\rm L}_t)^2 \mathbb{E}[\|\hat{\bf{n}}^{\rm G}_t\|^2] + \mu \eta^2 \rho^{\rm L}_t \rho^{\rm E}_t \mathbb{E}[(\hat{\bf{g}}^{\rm L}_t)^{\rm T}{\bf{g}}^{\rm E}_t] \notag
		\end{align}
		\begin{align}
			\ge & \eta_t\rho^{\rm L}_t\frac{\sqrt{\omega_t}}{\sqrt{\nu_t}} \mathbb{E}[\nabla F({\bf{w}}_{t+1})^{\rm T}[\frac{1}{K}\sum\limits_{k=1}^{K}(\hat{\bf{g}}^{\rm L}_{t,k}-{\bf{g}}^{\rm L*}_{t,k})]] \notag \\
			&+ \eta_t (\rho^{\rm L}_t\frac{\sqrt{\omega_t}}{\sqrt{\nu_t}}+\rho^{\rm E}_t) \nabla F({\bf{w}}_{t+1})^{\rm T} \nabla F({\bf{w}}_{t}) \notag \\
			&+\frac{\mu}{2} \eta_t^2\left(\rho^{\rm L}_t \frac{\sqrt{\omega_t}}{\sqrt{\nu_t}}+\rho^{\rm E}_t\right)^2 \|\nabla F({\bf{w}}_t)\|^2 + \frac{\mu \sigma^2 Q}{4\nu_t} \eta^2_t (\rho^{\rm L}_t)^2 \notag \\
			\ge & -\frac{\eta^2_t}{4}(\frac{\sqrt{\omega_t}}{\sqrt{\nu_t}})^2 \|\nabla F({\bf{w}}_{t+1})\|^2 \notag \\
			&-\mathbb{E}[\|\frac{\rho^{\rm L}_t}{K}\sum\limits_{k=1}^{K}(\hat{\bf{g}}^{\rm L}_{t,k}-{\bf{g}}^{\rm L*}_{t,k})\|^2] \notag \\
			&- \frac{\eta^2_t}{4} (\rho^{\rm L}_t\frac{\sqrt{\omega_t}}{\sqrt{\nu_t}}+\rho^{\rm E}_t)^2\|\nabla F({\bf{w}}_{t+1})\|^2 - \|\nabla F({\bf{w}}_{t})\|^2 \notag\\
			& +\frac{\mu}{2} \eta_t^2\left(\rho^{\rm L}_t \frac{\sqrt{\omega_t}}{\sqrt{\nu_t}}+\rho^{\rm E}_t\right)^2 \varepsilon^2 \notag \\
			&+ \frac{\mu \sigma^2 Q}{4\nu_t} \eta^2_t (\rho^{\rm L}_t)^2 \notag \\
			\ge &\frac{\mu}{2} \eta_t^2\left(\rho^{\rm L}_t \frac{\sqrt{\omega_t}}{\sqrt{\nu_t}}+\rho^{\rm E}_t\right)^2 \varepsilon^2 \notag \\
			&+ \frac{\mu \sigma^2 Q}{4\nu_t} \eta^2_t (\rho^{\rm L}_t)^2 \notag \\
			&- [\frac{\eta^2_t\omega_t}{4\nu_t}+\frac{\eta^2_t}{4}(\rho^{\rm L}_t \frac{\sqrt{\omega_t}}{\sqrt{\nu_t}}+\rho^{\rm E}_t)^2+1]A^2 \notag \\ 
			&-\mathbb{E}[\|\frac{\rho^{\rm L}_t}{K}\sum\limits_{k=1}^{K}(\hat{\bf{g}}^{\rm L}_{t,k}-{\bf{g}}^{\rm L*}_{t,k})\|^2].
	\end{align}}

	{
		Then, for the last term above, we have
		\vspace{-0.2 cm}
		\begin{align}
			\label{data_heterogeneity_1_response_R3}
			&-\mathbb{E}[\|\frac{\rho^{\rm L}_t}{K}\sum\limits_{k=1}^{K}(\hat{\bf{g}}^{\rm L}_{t,k}-{\bf{g}}^{\rm L*}_{t,k})\|^2]\notag \\
			=&-\frac{(\rho^{\rm L}_t)^2}{K^2} \mathbb{E} [\| \sum\limits_{k=1}^{K}\sum\limits_{c=1}^{C} (p_{t,k,c}-\frac{1}{C}) {\bf{g}}_{t,k,c} \|^2] \notag \\
			\ge&-\frac{(\rho^{\rm L}_t)^2}{K^2}\mathbb{E}[(\sum\limits_{k=1}^{K}\sum\limits_{c=1}^{C}|p_{t,k,c}-\frac{1}{C}|\|{\bf{g}}_{t,k,c}\|)^2] \notag \\
			\ge&-\frac{(\rho^{\rm L}_t)^2}{K^2} \mathbb{E}[(\sum\limits_{k=1}^{K}\sum\limits_{c=1}^{C}(p_{t,k,c}-\frac{1}{C})^2)(\sum\limits_{k=1}^{K}\sum\limits_{c=1}^{C}\|{\bf{g}}_{t,k,c}\|^2)] \notag \\
			\ge&-\frac{(\rho^{\rm L}_t A^2 C)^2}{K^2} \sum\limits_{k=1}^{K}\sum\limits_{c=1}^{C}(p_{t,k,c}-\frac{1}{C})^2
		\end{align}
		By plugging (\ref{data_heterogeneity_1_response_R3}) into (\ref{non-iid_non_stable_1_response_R3}), one can have (\ref{lower_bound_non_iid}).
		The proof is complete.}
	
	\section{Proof of Corollary~\ref{corollary_2}}
	\label{proof_of_corollary_2}

	In the non-stable region, by using (32) and $\hat{F}({\bf{w}})$ are $\delta$-strongly convex, we derive $\hat{F}({\bf{w}}_t)-\hat{F}({\bf{w}}_{t+1})$ as follows:
	\begin{align}
		\label{response_non_stable_1}
		&F({\bf{w}}_{t})\!-\! F({\bf{w}}_{t+1})\!+\!\delta (\|{\bf{w}}_{t}\|^2\!-\!\|{\bf{w}}_{t-1}\|^2)\notag \\
		\ge& (\nabla F({\bf{w}}_{t+1}) \!+\! 2 \delta {\bf{w}}_{t+1} )^{\rm T} ({\bf{w}}_{t}\!-\!{\bf{w}}_{t+1})\!+\!\frac{\delta}{2} \| {\bf{w}}_{t}\!-\!{\bf{w}}_{t+1} \|^2.
	\end{align}
	By moving the third term on the left-hand side to the right-hand side, one can have
	\begin{align}
		\label{response_non_stable_2}
		F({\bf{w}}_{t})\!-\!F({\bf{w}}_{t+1}) \!\ge\!& \nabla F({\bf{w}}_{t+1})^{\rm T}({\bf{w}}_{t}-{\bf{w}}_{t+1}) \!+\! \frac{\delta}{2}\| {\bf{w}}_{t}\!-\!{\bf{w}}_{t+1} \|^2 \notag \\ 
		&-\! \delta ( \|{\bf{w}}_{t+1}\|^2 \!+\! \|{\bf{w}}_{t}\|^2 \!-\! 2{\bf{w}}_{t+1}^{\rm T} {\bf{w}}_{t}) \notag \\
		=&\nabla F({\bf{w}}_{t+1})^{\rm T}({\bf{w}}_{t}-{\bf{w}}_{t+1}) + \frac{\mu}{2}\| {\bf{w}}_{t}-{\bf{w}}_{t+1} \|^2 \notag \\
		&-\! \frac{\delta+\mu}{2} \|{\bf{w}}_{t}-{\bf{w}}_{t+1}\|^2 \notag \\
		\overset{(a)}{\ge}& \frac{\eta_t^2}{4}(2\mu\varepsilon^2\!-\! A^2) \!\! \left[1 \! +\!\! \left(\frac{\sqrt{\omega_t}}{\sqrt{\nu_t}}\!-\!1\right) \! \rho^{\rm L}_{t}\right]^2\notag \\
		&-A^2+\frac{\mu \sigma^2 Q \eta_t^2}{4\nu_t}(\rho^{\rm L}_{t})^2 \notag \\
		&- \frac{\delta+\mu}{2} \|{\bf{w}}_{t}-{\bf{w}}_{t+1}\|^2,
	\end{align}
	where (a) results from applying the derivation process outlined in Appendix C to the first two terms in the second equality.
	The proof is complete.

	\section{Proof of Theorem~\ref{theorem_2}}
	\label{proof_of_theorem_2}
	Based on (\ref{L_smooth}) in Assumption~\ref{assumption_1}, by using ${\sqrt{\omega_t}}/{\sqrt{\nu_t}}=1$, we have
	\vspace{-0.2 cm}
	\begin{align}
		\label{appendix_B_1}
		&F({\bf{w}}_{t})-F({\bf{w}}_{t-1}) \notag \\
		\le& -\eta_{t-1} \nabla F({\bf{w}}_{t-1})^{\rm T} (\rho^{\rm L}_{t-1} {\bf{g}}^{\rm L}_{t-1} + \rho^{\rm E}_{t-1} {\bf{g}}^{\rm E}_{t-1}) \notag \\
		&+\frac{L}{2}\eta_{t-1}^2 \| \rho^{\rm L}_{t-1} \hat{\bf{g}}^{\rm L}_{t-1} + \rho^{\rm L}_{t-1} \hat{\bf{n}}^{\rm G}_{t-1} + \rho^{\rm E}_{t-1} {\bf{g}}^{\rm E}_{t-1} \|^2 \notag \\
		& -\eta_{t-1} \rho^{\rm L}_{t-1} \nabla F({\bf{w}}_{t-1})^{\rm T} \hat{\bf{n}}^{\rm G}_{t-1}.
	\end{align}
	Taking the expectation on both sides of (\ref{appendix_B_1}), we derive that
	%
	\begin{align}
		\label{appendix_B_2}
		\hspace{-0.2 cm}&\mathbb{E}[F({\bf{w}}_{t})-F({\bf{w}}_{t-1})] \notag \\
		\hspace{-0.2 cm}\le& - \eta_{t-1} \|\nabla F({\bf{w}}_{t-1})\|^2 + \frac{L}{2} \eta_{t-1}^2 (\rho^{\rm L}_{t-1})^2 \mathbb{E}[\|\hat{\bf{g}}^{\rm L}_{t-1}\|^2] \notag \\
		\hspace{-0.2 cm}&+ \frac{L}{2} \eta_{t-1}^2 (\rho^{\rm E}_{t-1})^2 \mathbb{E}[\|{\bf{g}}^{\rm E}_{t-1}\|^2] +\frac{L}{2} \eta_{t-1}^2 (\rho^{\rm L}_{t-1})^2 \mathbb{E}[\|\hat{\bf{n}}^{\rm G}_{t-1}\|^2] \notag \\
		\hspace{-0.2 cm}&+ L \eta_{t-1}^2 \rho^{\rm L}_{t-1} \rho^{\rm E}_{t-1} \mathbb{E}[(\hat{\bf{g}}^{\rm L}_{t-1})^{\rm T}{\bf{g}}^{\rm E}_{t-1}] \notag \\
		\hspace{-0.2 cm}\le & (L \eta_{t-1}^2 \rho^{\rm L}_{t-1} \rho^{\rm E}_{t-1}-\eta_{t-1})\|\nabla F({\bf{w}}_{t-1})\|^2 \notag \\
		\hspace{-0.2 cm}&+\! \frac{L}{2} \eta_{t-1}^2 A^2 [(\rho^{\rm L}_{t-1})^2 \!\!+ \!\! (\rho^{\rm E}_{t-1})^2] \!\!+\!\! \frac{L}{2} \eta_{t-1}^2 (\rho^{\rm L}_{t-1})^2 \frac{\sigma^2Q}{2\nu_t}.
	\end{align}

	Then, we employ the PL-inequality from our previous work~\cite{Zheng2023Semi}, which results in the following result:
	%
	\begin{align}
		\label{appendix_B_3}
		\|\nabla F({\bf{w}}_{t-1})\|^2 \ge 2 \mu [F({\bf{w}}_{t-1})-F({\bf{w}}^{*})].
	\end{align}
	Correspondingly, we have
	%
	\begin{align}
		\label{appendix_B_4}
		&\mathbb{E}[F({\bf{w}}_{t})-F({\bf{w}}_{t-1})] \notag \\
		\le & 2 \mu (L \eta_{t-1}^2 \rho^{\rm L}_{t-1} \rho^{\rm E}_{t-1}-\eta_{t-1}) [F({\bf{w}}_{t-1})-F({\bf{w}}^{*})] \notag \\
		&+\frac{L}{2} A^2 \eta_{t-1}^2 +\frac{L\sigma^2Q}{4\nu_t} (\rho^{\rm L}_{t-1})^2 \eta_{t-1}^2.
	\end{align}
	By setting $\eta_{t-1}$ to $\eta_{t-1}=1/\mu$ while taking the expectation on both sides, we have
	%
	\begin{align}
		\label{appendix_B_5}
		&\mathbb{E}[F({\bf{w}}_{t})-F({\bf{w}}^{*})] \notag \\
		\le & [1-2(1-\frac{L}{\mu}\rho^{\rm L}_{t-1}\rho^{\rm E}_{t-1})] \mathbb{E}[F({\bf{w}}_{t-1})-F({\bf{w}}^{*})] \notag \\
		&+ \frac{L}{2\mu^2} [A^2 + \frac{\sigma^2Q}{2\nu_t}(\rho^{\rm L}_{t-1})^2].
	\end{align}
	Based on $\rho^{\rm E}_{t-1}=1-\rho^{\rm L}_{t-1}$, it is noticed that
	%
	\begin{align}
		\label{appendix_B_6}
		&1-2(1-\frac{L}{\mu}\rho^{\rm L}_{t-1}\rho^{\rm E}_{t-1}) \notag \\
		=&-2\frac{L}{\mu} (\rho^{\rm L}_{t-1})^2 +2\frac{L}{\mu}\rho^{\rm L}_{t-1}-1 \notag \\
		\le& \frac{L}{2\mu} - 1 \triangleq \xi.
	\end{align}
	By using (\ref{appendix_B_6}) and setting $\nu_t$ to $\nu_t=\nu,\forall t \in \mathcal{T}$, we can further derive that
	\vspace{-0.2 cm}
	\begin{align}
		\label{appendix_B_7}
		&\mathbb{E}[F({\bf{w}}_{t})-F({\bf{w}}^{*})] \notag \\ 
		\overset{}{\le}& \xi \mathbb{E}[F({\bf{w}}_{t-1})-F({\bf{w}}^{*})] + \frac{L}{2\mu^2} [A^2 + \frac{\sigma^2Q}{2\nu}(\rho^{\rm L}_{t-1})^2] \notag \\
		\overset{(d)}{\le}& \xi \mathbb{E}[F({\bf{w}}_{t-1})-F({\bf{w}}^{*})] + \frac{L}{2\mu^2} (A^2 + \frac{\sigma^2Q}{2\nu}),
	\end{align}
	where $(d)$ is because $(\rho^{\rm L}_{t-1})^2\le1$.
	Recursively applying inequality (\ref{appendix_B_7}), we have
	\vspace{-0.2 cm}
	\begin{align}
		\label{appendix_B_8}
		&\mathbb{E}[F({\bf{w}}_{t})-F({\bf{w}}^{*})] \notag \\ 
		\le&  \xi^{t-1} \mathbb{E}[F({\bf{w}}_{1})\!\!-\!\!F({\bf{w}}^{*})] \!+\! \frac{1\!-\!\xi^{t-1}}{1\!-\!\xi} \frac{L}{2\mu^2} (A^2 \!+\! \frac{\sigma^2Q}{2\nu}).
	\end{align}
	Lastly, (\ref{stable_upper_bound}) can be obtained by letting the both sides of (\ref{appendix_B_8}) approach infinity.
	The proof is complete.
	
	{
		\section{Proof of Corollary~\ref{corollary_3}}
		\label{proof_of_corollary_3}}

	{
		Based on the proof of Theorem~\ref{theorem_2}, we have
		\vspace{-0.2 cm}
		\begin{align}
			\label{non-iid_stable_1_response_R3}
			&F({\bf{w}}_{t})-F({\bf{w}}_{t-1}) \notag \\
			\le& -\eta_{t-1} \nabla F({\bf{w}}_{t-1})^{\rm T} (\rho^{\rm L}_{t-1} {\bf{g}}^{\rm L}_{t-1} + \rho^{\rm E}_{t-1} {\bf{g}}^{\rm E}_{t-1}) \notag \\
			&+\frac{L}{2}\eta_{t-1}^2 \| \rho^{\rm L}_{t-1} (\hat{\bf{g}}^{\rm L}_{t-1} - {\bf{g}}^{\rm L*}_{t-1}) + \rho^{\rm L}_{t-1} {\bf{g}}^{\rm L*}_{t-1} \notag \\
			&+ \rho^{\rm L}_{t-1} \hat{\bf{n}}^{\rm G}_{t-1} + \rho^{\rm E}_{t-1} {\bf{g}}^{\rm E}_{t-1} \|^2 \notag \\
			& -\eta_{t-1} \rho^{\rm L}_{t-1} \nabla F({\bf{w}}_{t-1})^{\rm T} \hat{\bf{n}}^{\rm G}_{t-1} \notag \\
			\le&-\eta_{t-1} \nabla F({\bf{w}}_{t-1})^{\rm T} (\rho^{\rm L}_{t-1} {\bf{g}}^{\rm L}_{t-1} + \rho^{\rm E}_{t-1} {\bf{g}}^{\rm E}_{t-1}) \notag \\
			&+L\eta^2_{t-1}(\rho^{\rm L}_{t-1})^2 \|\hat{\bf{g}}^{\rm L}_{t-1}-{\bf{g}}^{\rm L*}_{t-1}\|^2 \notag \\
			&+ L\eta^2_{t-1}\| \rho^{\rm L}_{t-1} {\bf{g}}^{\rm L*}_{t-1} + \rho^{\rm L}_{t-1} \hat{\bf{n}}^{\rm G}_{t-1} + \rho^{\rm E}_{t-1} {\bf{g}}^{\rm E}_{t-1} \|^2 \notag \\
			&-\eta_{t-1} \rho^{\rm L}_{t-1} \nabla F({\bf{w}}_{t-1})^{\rm T} \hat{\bf{n}}^{\rm G}_{t-1}.
		\end{align}
		By taking the expectation on both sides, one can have
		%
		\begin{align}
			\label{non-iid_stable_2_response_R3}
			&\mathbb{E}[F({\bf{w}}_{t})-F({\bf{w}}_{t-1})] \notag \\
			\le& -\eta_{t-1} \mathbb{E}[\|\nabla F({\bf{w}}_{t-1})\|^2] + L\eta^2_{t-1} (\rho^{\rm L}_{t-1})^2 \mathbb{E}[\| {\bf{g}}^{\rm L*}_{t-1}\|^2] \notag \\
			&+ L \eta^2_{t-1} (\rho^{\rm L}_{t-1})^2 \mathbb{E}[\| \hat{\bf{n}}^{\rm G}_{t-1}\|^2] +L \eta^2_{t-1} (\rho^{\rm E}_{t-1})^2 \mathbb{E}[\| {\bf{g}}^{\rm E}_{t-1}\|^2] \notag \\
			& + 2 L \eta^2_{t-1} \rho^{\rm L}_{t-1} \rho^{\rm E}_{t-1} \mathbb{E}[({\bf{g}}^{\rm L*}_{t-1})^{\rm T} {\bf{g}}^{\rm E}_{t-1} ] \notag \\
			&+ L \eta^2_{t-1} (\rho^{\rm L}_{t-1})^2 \mathbb{E}[\|\frac{1}{K}\sum\limits_{k=1}^{K} (\hat{\bf{g}}_{t-1,k}^{\rm L}-{\bf{g}}_{t-1,k}^{\rm L*})\|^2] \notag \\
			\le&(2 L \eta^2_{t} \rho^{\rm L}_{t-1} \rho^{\rm E}_{t-1}-\eta_{t-1}) \mathbb{E}[\|\nabla F({\bf{w}}_{t-1})\|^2] \notag \\ 
			&+ L \eta^2_{t-1}[(\rho^{\rm L}_{t-1})^2+(\rho^{\rm E}_{t-1})^2] A^2 + L \eta^2_{t-1} (\rho^{\rm L}_{t-1})^2 \frac{\sigma^2 Q}{2\nu_t} \notag \\
			& + \frac{C}{K} A^2 L \eta^2_{t-1} (\rho^{\rm L}_{t-1})^2 [\sum\limits_{k=1}^{K}\sum\limits_{c=1}^{C} (p_{t-1,k,c}-\frac{1}{C})^2] \notag \\
			\le&2\mu (2 L \eta^2_{t} \rho^{\rm L}_{t-1} \rho^{\rm E}_{t-1}-\eta_{t-1}) \mathbb{E}[F({\bf{w}}_{t-1})-F({\bf{w}}^{*})] \notag \\
			&+ L \eta^2_{t-1}A^2 + L \eta^2_{t-1} (\rho^{\rm L}_{t-1})^2 \frac{\sigma^2 Q}{2\nu_t} \notag \\
			&+ \frac{C}{K} A^2 L \eta^2_{t-1} (\rho^{\rm L}_{t-1})^2 [\sum\limits_{k=1}^{K}\sum\limits_{c=1}^{C} (p_{t-1,k,c}-\frac{1}{C})^2].
		\end{align}
		By setting $\eta_t=\frac{1}{\mu}$ and subtracting $F({\bf{w}}^{*})$ from both sides, we have the following inequality after taking expectation on both sides:
		%
		\begin{align}
			\label{non-iid_stable_3_response_R3}
			&\mathbb{E}[F({\bf{w}}_{t})-F({\bf{w}}_{t-1})] \notag \\
			\le&[1-2(1-2\frac{L}{\mu}\rho^{\rm L}_{t-1}\rho^{\rm E}_{t-1})]\mathbb{E}[F({\bf{w}}_{t-1})-F({\bf{w}}^{*})] \notag \\
			&+ \frac{L}{\mu^2}[A^2+\frac{\sigma^2 Q}{2 \nu_t} (\rho^{\rm L}_{t-1})^2] \notag \\
			& + \frac{C}{K} A^2 L \eta^2_{t-1} (\rho^{\rm L}_{t-1})^2 [\sum\limits_{k=1}^{K}\sum\limits_{c=1}^{C} (p_{t-1,k,c}-\frac{1}{C})^2].
		\end{align}
		Then, we let $\hat{\xi}=1-2(1-2\frac{L}{\mu}\rho^{\rm L}_{t-1}\rho^{\rm E}_{t-1})$.
		Through using the above results, while setting $\nu_t=\nu,\forall t$, we have
		%
		\begin{align}
			\label{non-iid_stable_5_response_R3}
			&\mathbb{E}[F({\bf{w}}_{t})-F({\bf{w}}^{*})] \notag \\ 
			\le& \hat{\xi} \mathbb{E}[F({\bf{w}}_{t-1})-F({\bf{w}}^{*})] + \frac{L}{\mu^2}(A^2+\frac{\sigma^2 Q}{2\nu}) \notag \\
			&+ \frac{C A^2 L}{K \mu^2} (\rho^{\rm L}_{t-1})^2 \Delta d_{t-1}.
		\end{align}
		Recursively applying the above inequality for $t$ times, we have
		\vspace{-0.2 cm}
		\begin{align}
			\label{non-iid_stable_6_response_R3}
			\mathbb{E}[F({\bf{w}}_{t})-F({\bf{w}}^{*})] \le &\hat{\xi}^{t-1} \mathbb{E}[F({\bf{w}}_{1})-F({\bf{w}}^{*})] \notag \\
			&+ \frac{1-\hat{\xi}^{t-1}}{1-\hat{\xi}}\frac{L}{\mu^2}(A^2+\frac{\sigma^2 Q}{2\nu}) \notag \\
			&+ \frac{C A^2 L}{K \mu^2} \sum\limits_{\tau=1}^{t-1} \hat{\xi}^{t-1-\tau} (\rho^{\rm L}_{\tau})^2 \Delta d_{\tau}.
		\end{align}
		By forcing $t \rightarrow \infty$, we have (\ref{upper_bound_non_iid}).
		The poof is complete.}
	
		\section{Proof of Corollary~\ref{corollary_4}}
		\label{proof_of_corollary_4}

		In the stable region, we derive the gap incurred by the non-convex global loss function $F({\bf{w}})$ from the fact that $\hat{F}({\bf{w}})$ is $(L+2\delta)$-smooth~\cite{Yang2021Energy}, given by
		\begin{align}
			\label{response_stable_1}
			\hat{F}({\bf{w}}_{t})\!-\! \hat{F}({\bf{w}}_{t-1})\le& \nabla \hat{F}({\bf{w}}_{t+1})^{\rm T}({\bf{w}}_{t}\!-\!{\bf{w}}_{t-1})\notag\\
			&+\!\frac{L+2\delta}{2} \| {\bf{w}}_{t}\!-\!{\bf{w}}_{t+1} \|^2.
		\end{align}
		Plugging (32) into (\ref{response_stable_1}), we have
		\begin{align}
			\label{response_stable_2}
			F({\bf{w}}_{t})- F({\bf{w}}_{t-1}) \le& (\nabla F({\bf{w}}_{t-1}) \!+\! 2\delta{\bf{w}}_{t-1})^{\rm T} ({\bf{w}}_{t}\!-\!{\bf{w}}_{t-1}) \notag \\ &+\frac{L+2\delta}{2} \| {\bf{w}}_{t}-{\bf{w}}_{t-1} \|^2 \notag \\
			&+ \delta (\|{\bf{w}}_{t-1}\|^2 \!-\! \|{\bf{w}}_{t}\|^2)\notag \\
			\le&\nabla F({\bf{w}}_{t-1})^{\rm T}({\bf{w}}_{t}-{\bf{w}}_{t-1}) \notag \\
			&+ \frac{L}{2} \| {\bf{w}}_{t}-{\bf{w}}_{t-1} \|^2+ \delta \| {\bf{w}}_{t}\!-\!{\bf{w}}_{t-1} \|^2 \notag \\
			&+ 2\delta {\bf{w}}_{t-1}^{\rm T} {\bf{w}}_{t} - \delta \|{\bf{w}}_{t-1}\|^2 - \delta \|{\bf{w}}_{t}\|^2 \notag \\
			=&\nabla F({\bf{w}}_{t-1})^{\rm T}({\bf{w}}_{t}-{\bf{w}}_{t-1}) \notag \\
			&+ \frac{L}{2} \| {\bf{w}}_{t}-{\bf{w}}_{t-1} \|^2.
		\end{align}
		One can see that the above result is identical with those in Assumption~1.
		Hence, by following the derivations in Appendix~E, we have
		\begin{align}
			\label{response_stable_3}
			\mathbb{E}[F({\bf{w}}_{t})\!-\! F({\bf{w}}_{t-1})]\!\le\!&(L\eta_{t-1}^2 \rho^{\rm L}_{t-1} \rho^{\rm E}_{t-1} \!\!-\! \eta_{t-1})\|\nabla F({\bf{w}}_{t-1})\|^2\notag \\
			&\!+\!\!\frac{LA^2\eta_{t-1}^2}{2}\!+\!\!\frac{L\sigma^2Q\eta_{t-1}^2}{4\nu_t} (\rho_{t-1}^{\rm L})^2\!.\!
		\end{align}
		Based on the fact that $\hat{F}({\bf{w}})$ is $\delta$-strongly convex, we derive the PL-inequality of $\|\nabla \hat{F}({\bf{w}}_{t-1})\|^2$ as follows:
		\begin{align}
			\label{response_stable_4}
			\|\nabla F({\bf{w}}_{t-1}) + 2\delta {\bf{w}}_{t-1}\|^2 \ge& 2 \delta [F({\bf{w}}_{t-1})- F({\bf{w}}^{*}) \notag \\
			&+ \delta \|{\bf{w}}_{t-1}\|^2 - \delta \|{\bf{w}}^{*}\|^2].
		\end{align}
		By upper-bounding the left-hand side of (\ref{response_stable_4}), we have
		\begin{align}
			\label{response_stable_5}
			\|\nabla F({\bf{w}}_{t-1}) \!+\! 2\delta {\bf{w}}_{t-1}\|^2 \!=\!& \|\nabla F({\bf{w}}_{t-1}) \|^2 + 4 \delta^2\|{\bf{w}}_{t-1}\|^2 \notag \\
			&+ 4\delta \nabla F({\bf{w}}_{t-1})^{\rm T} {\bf{w}}_{t-1} \notag \\
			\le&\|\nabla F({\bf{w}}_{t-1}) \|^2 + 4 \delta^2\|{\bf{w}}_{t-1}\|^2 \notag \\
			&+ \|\nabla F({\bf{w}}_{t-1}) \|^2 + 4\delta^2\|{\bf{w}}_{t-1}\|^2 \notag \\
			=&2\|\nabla F({\bf{w}}_{t-1}) \|^2\!\!+\!8\delta^2\|{\bf{w}}_{t-1}\|^2.
		\end{align}
		Applying inequality (\ref{response_stable_5}) to the left-hand side of (\ref{response_stable_4}), we have
		\begin{align}
			\label{response_stable_6}
			\| \nabla F ({\bf{w}}_{t-1}) \|^2 \ge& \delta [F({\bf{w}}_{t-1})-F({\bf{w}}^{*})]\notag \\
			&-\delta^2\|{\bf{w}}^{*}\|^2-3\delta^2\|{\bf{w}}_{t-1}\|^2.
		\end{align}
		Substituting (\ref{response_stable_6}) into (\ref{response_stable_3}), we have
		\begin{align}
			\label{response_stable_7}
			\mathbb{E}[F({\bf{w}}_{t})\!-\! F({\bf{w}}_{t-1})]\!\le\!&(L\eta_{t-1}^2 \rho^{\rm L}_{t-1} \rho^{\rm E}_{t-1} \!-\! \eta_{t-1})\{\delta[F({\bf{w}}_{t-1})\!- \notag \\
			&F({\bf{w}}^{*})]-\delta^2\|{\bf{w}}^{*}\|^2-3\delta^2\|{\bf{w}}_{t-1}\|^2\}\notag \\
			&+\frac{LA^2\eta_{t-1}^2}{2}+\frac{L\sigma^2Q\eta_{t-1}^2}{4\nu_t} (\rho_{t-1}^{\rm L})^2 \notag \\
			=&(L\eta_{t-1}^2 \rho^{\rm L}_{t-1} \rho^{\rm E}_{t-1} - \eta_{t-1}) \delta \mathbb{E} [F({\bf{w}}_{t-1}) \notag \\
			&- F({\bf{w}}^{*})] -(L\eta_{t-1}^2 \rho^{\rm L}_{t-1} \rho^{\rm E}_{t-1} - \eta_{t-1})\notag \\
			& (\delta^2\|{\bf{w}}^{*}\|^2+3\delta^2\|{\bf{w}}_{t-1}\|^2) \notag \\
			&+\!\!\frac{LA^2\eta_{t-1}^2}{2}\!\!+\!\!\frac{L\sigma^2Q\eta_{t-1}^2}{4\nu_t} (\rho_{t-1}^{\rm L})^2\!.\!
		\end{align}
		Subtracting $F({\bf{w}}^{*})$ from both sides, while applying $\eta_{t-1}=\frac{1}{\delta}$ and $\rho^{\rm E}_{t-1}=1-\rho^{\rm L}_{t-1}$ to (\ref{response_stable_7}), we have
		\begin{align}
			\label{response_stable_8}
			\mathbb{E}[F({\bf{w}}_{t})\!-\! F({\bf{w}}^{*})]\le&\frac{L}{\delta} [\rho^{\rm L}_{t-1}\!\!-\!(\rho^{\rm L}_{t-1})^2] \mathbb{E}[F({\bf{w}}_{t-1})\!-\! F({\bf{w}}^{*})] \notag \\
			&-[\frac{L}{\delta} \rho^{\rm L}_{t-1} - \frac{L}{\delta}(\rho^{\rm L}_{t-1})^2 -1]\notag \\
			&(\delta\|{\bf{w}}^{*}\|^2+3\delta\|{\bf{w}}_{t-1}\|^2) \notag \\
			&+\frac{L}{2\delta^2}A^2 + \frac{L \sigma^2 Q}{4\delta^2 \nu} (\rho^{\rm L}_{t-1})^2 \notag \\
			\le& \frac{L}{4\delta} \mathbb{E}[F({\bf{w}}_{t-1})- F({\bf{w}}^{*})] \notag \\
			&+(\delta\|{\bf{w}}^{*}\|^2+3\delta\|{\bf{w}}_{t-1}\|^2) \notag\\
			&+\frac{L}{2\delta^2}(A^2+\frac{\sigma^2Q}{2\nu}).
		\end{align}
		By letting $\delta=\mu$, while defining $\xi'=\frac{L}{2\mu}$ and $\Delta_{t-1}=\delta\|{\bf{w}}^{*}\|^2+3\delta\|{\bf{w}}_{t-1}\|^2$, we have
		\begin{align}
			\label{response_stable_9}
			\mathbb{E}[F({\bf{w}}_{t})- F({\bf{w}}^{*})]\le& \xi' \mathbb{E}[F({\bf{w}}_{t-1})- F({\bf{w}}^{*})] + \Delta_{t-1} \notag \\
			&+ \frac{L}{2\mu^2}(A^2+\frac{\sigma^2Q}{2\nu}).
		\end{align}
		Recursively applying (\ref{response_stable_9}) for $t$ times and letting $t=T$, we finally have
		\begin{align}
			\label{response_stable_10}
			\mathbb{E}[F({\bf{w}}_{T})- F({\bf{w}}^{*})]\le& (\xi')^{T-1} \mathbb{E}[F({\bf{w}}_{1})- F({\bf{w}}^{*})] \notag \\
			&+ \sum\limits_{\tau=1}^{T-1}(\xi')^{T-1-\tau}\Delta_{\tau} \notag \\
			&+ \frac{1-(\xi')^{T-1}}{1-\xi'}\frac{L}{2\mu^2}(A^2+\frac{\sigma^2Q}{2\nu}).
		\end{align}
		As $T \rightarrow \infty$, we have
		\begin{align}
			\label{response_stable_11}
			\lim_{T \rightarrow \infty}\mathbb{E}[F({\bf{w}}_{T})- F({\bf{w}}^{*})]\le&
			\frac{L}{\mu}\frac{1}{2\mu-L}(A^2+\frac{\sigma^2Q}{2\nu}) \notag \\
			&+  \lim_{T \rightarrow \infty}\sum\limits_{\tau=1}^{T-1}(\xi')^{T-1-\tau}\Delta_{\tau}.
		\end{align}
		The proof is complete.

	\section{Proof of Lemma~\ref{lemma_1}}
	\label{proof_of_lemma_1}
	The Lagrange function of problem (\ref{objective_p_8}) is given by
	\vspace{-0.2 cm}
	\begin{align}
		\label{appendix_c_1}
		&\mathcal{L}(\{\hat{f}_{k}\},\tilde{f},\tau_2,\lambda_1,\{\lambda_{2,k}\},\lambda_3,\{\lambda_{4,k}\},\{\lambda_{5,k}\},\lambda_6,\lambda_7) \notag \\
		=& \sum\nolimits_{k=1}^{K} C_{11,k}\hat{f}_k^2+C_{12}\tilde{f}^2 \notag \\
		&+\lambda_1(\tau_2-T_{\max})+\sum\nolimits_{k=1}^{K} \lambda_{2,k} (\frac{C_{13,k}}{\hat{f}_k}-\tau_2+T^{\rm G})\notag\\
		&+\lambda_3(\frac{C_{14}}{\tilde{f}}-\tau_2+\max_{k\in\mathcal{K}}\{T^{\rm D}_{k}\}) + \sum\nolimits_{k=1}^{K} \lambda_{4,k}(-\hat{f}_k) \notag \\
		&+\! \sum\nolimits_{k=1}^{K} \lambda_{5,k} (\hat{f}_k\!\!-\!\!\hat{f}_{\max}) \!+\!\lambda_6 (-\tilde{f}) \!\!+\!\! \lambda_7 (\tilde{f}\!\!-\!\!\tilde{f}_{\max}),
	\end{align}
	where $\lambda_1$, $\{\lambda_{2,k}\}$, $\lambda_3$, $\{\lambda_{4,k}\}$, $\{\lambda_{5,k}\}$, $\lambda_6$, and $\lambda_7$ are non-negative Lagrange multipliers.
	Then, the KKT conditions of problem (\ref{objective_p_8}) are given by
	\begin{subnumcases}
		{\label{appendix_c_2}}
		\label{Lagrange_partial_hat_f}
		\frac{\partial \mathcal{L}}{\partial \hat{f}_k}=0, \forall k \in \mathcal{K}, \\
		\label{Lagrange_partial_tilde_f}
		\frac{\partial \mathcal{L}}{\partial \tilde{f}} = 0, \\
		\label{Lagrange_partial_tau_2}
		\frac{\partial \mathcal{L}}{\partial \tau_2} = 0, \\
		\label{Lagrange_complementation_1}
		\lambda_1(\tau_2-T_{\max})=0, \\
		\label{Lagrange_complementation_2}
		\lambda_{2,k} (\frac{C_{13,k}}{\hat{f}_k}-\tau_2+T^{\rm G})=0, \forall k \in \mathcal{K}, \\
		\label{Lagrange_complementation_3}
		\lambda_3(\frac{C_{14}}{\tilde{f}}-\tau_2+\max_{k\in\mathcal{K}}\{T^{\rm D}_{k}\})=0, \\
		\label{Lagrange_complementation_4}
		\lambda_{4,k}(-\hat{f}_k)=0, \forall k \in \mathcal{K}, \\
		\label{Lagrange_complementation_5}
		\lambda_{5,k} (\hat{f}_k\!-\!\hat{f}_{\max})=0, \forall k \in \mathcal{K}, \\
		\label{Lagrange_complementation_6}
		\lambda_6 (-\tilde{f})=0, \\
		\label{Lagrange_complementation_7}
		\lambda_7 (\tilde{f}\!-\!\tilde{f}_{\max})=0, \\
		\label{Lagrange_multipliers_1}
		\lambda_1 \ge 0,~\lambda_3 \ge 0,~\lambda_6 \ge 0,~\lambda_7 \ge 0, \notag \\
		\label{Lagrange_multipliers_2}
		\lambda_{2,k} \ge 0,~\lambda_{4,k} \ge 0,~\lambda_{5,k} \ge 0, \forall k \in \mathcal{K}.
	\end{subnumcases}

	It is noticed that constraint (\ref{2_constraint_p_8}) can be rewritten as
	\vspace{-0.2 cm}
	\begin{align}
		\label{appendix_c_3}
		\hat{f}_k \ge \frac{C_{13,k}}{\tau_2-T^{\rm G}}, \forall k \in \mathcal{K}.
	\end{align}
	To minimize the objective (\ref{objective_p_8}), $\hat{f}_k$ should be minimized within the feasible region.
	Moreover, the right-hand side of inequality (\ref{appendix_c_3}) monotonously decreases as $\tau_2$ increases.
	Since constraint (\ref{objective_p_8}), i.e., $\tau_2 \le T_{\max}$, should be satisfied, the right-hand side of inequality (\ref{appendix_c_3}) obtains its minimum when $\tau_2 = T_{\max}$.
	Hence, the optimal $\tau_2$ and $\hat{f}_k$ can given by (\ref{appendix_c_4}) and (\ref{appendix_c_5}), respectively, i.e., given by
	\vspace{-0.2 cm}
	\begin{align}
		\label{appendix_c_4}
		&\tau_2^{*}=T_{\max}, \\
		\label{appendix_c_5}
		&\hat{f}_k^{*} = \frac{C_{13,k}}{T_{\max}-T^{\rm G}}, \forall k \in \mathcal{K}.
	\end{align}

	similarly, constraint (\ref{3_constraint_p_8}) can be written as
	\begin{align}
		\label{appendix_c_6}
		\tilde{f} \ge \frac{C_{14}}{\tau_2-\max_{k \in \mathcal{K}}\{T^{\rm D}_k\}}.
	\end{align}
	Considering that the objective (\ref{objective_p_8}) increases as $\tilde{f}$ increases, $\tilde{f}$ should also be minimized to minimize the objective (\ref{objective_p_8}).
	In addition, the right-hand side of (\ref{appendix_c_6}) monotonously decreases as $\tau_2$ increases.
	Intuitively, $\tau_2=T_{\max}$ should be imposed to minimize the right-hand side of (\ref{appendix_c_6}), such that objective (\ref{objective_p_8}) can be minimized as well.
	Therefore, the optimal $\tilde{f}$ can be given by
	\begin{align}
		\label{appendix_c_7}
		\tilde{f}^{*} = \frac{C_{14}}{T_{\max}-\max_{k \in \mathcal{K}}\{T^{\rm D}_k\}}.
	\end{align}

	To meet constraints (\ref{2_constraint_p_8}) and (\ref{3_constraint_p_8}), it is clear that $\hat{f}_k \neq 0,\forall k \in \mathcal{K}$ and $\tilde{f} \neq 0$, so as to prevent excessive FL and CL computing latency.
	Hence, we have $\lambda_{4,k}=0,\forall k \in \mathcal{K}$ and $\lambda_6=0$.
	Additionally, by applying (\ref{appendix_c_4}), (\ref{appendix_c_5}), and (\ref{appendix_c_7}) to (\ref{Lagrange_partial_hat_f}), (\ref{Lagrange_partial_tilde_f}), and (\ref{Lagrange_partial_tau_2}), we have
	\begin{subnumcases}
		{\label{appendix_c_8}}
		\lambda_1=\sum\nolimits_{k=1}^{K} \frac{2C_{11,k}(\hat{f}_k^{*})^3}{C_{13,k}} + \frac{2C_{12}(\tilde{f}^{*})^3}{C_{14}},\\
		\lambda_{2,k}=\frac{2C_{11,k}(\hat{f}_k^{*})^3}{C_{13,k}}, \forall k \in \mathcal{K}, \\
		\lambda_3=\frac{2C_{12}(\tilde{f}^{*})^3}{C_{14}}, \\
		\lambda_{5,k}=0,\forall k \in \mathcal{K}, \\
		\lambda_7=0.
	\end{subnumcases}
	The proof is complete.
	
	\bibliographystyle{IEEEtran}
	\bibliography{reference}
	
\end{document}